%% file: main.tex
\documentclass[sigplan,10pt,screen,nonacm]{acmart-preprint}
\usepackage{booktabs}
\usepackage{array}
\usepackage{tabularx}
\usepackage{adjustbox}    % appendix-only table shrinking, see \appendix below
\usepackage{etoolbox}
\usepackage{multirow}
\usepackage{subcaption}
\usepackage{xcolor}
\usepackage{listings}
\usepackage{algorithm}
\usepackage{algpseudocode}
\usepackage{tikz}
\usepackage{pgfplots}
\pgfplotsset{compat=1.18}
\usepackage{hyperref}
\usepackage{microtype}
\usepackage{url}
\graphicspath{{figures/}}

\newcommand{\extinput}[1]{\IfFileExists{#1.tex}{\input{#1}}{}}

\newcommand{\appxref}[2]{Appendix~\ref{#1}}
\newcommand{\Appxref}[2]{Appendix~\ref{#1}}

\newcommand{\Appxrefs}{Appendices}
\newcommand{\appxitem}[2]{\ref{#1}}
\newcommand{\appxtail}{}

\newcommand{\dslref}{Appendix~\ref{app:dsl}}

\newcommand{\mk}{SyclKittens}
\newcommand{\tflops}{\text{TFLOPS}}
\newcommand{\gbps}{\text{GB/s}}
\newcommand{\us}{\text{\textmu s}}
\newcommand{\ms}{\text{ms}}
\newcommand{\pvc}{Intel Max GPU}

\begin{document}

\title{SyclKittens: A Tile Programming Model for Programmers and Coding Agents on Intel GPUs}

\author{Yehong Jiang}
\email{yehong28@mit.edu}
\orcid{0009-0006-1340-8170}
\affiliation{%
  \institution{Massachusetts Institute of Technology}
  \city{Cambridge}
  \state{Massachusetts}
  \country{USA}
}
\authornote{Work performed while the author was an intern at Intel Corporation.}

\author{Sheng Chen}
\email{sheng.s.chen@intel.com}
\orcid{0009-0000-3770-2004}
\affiliation{%
  \institution{Intel Corporation}
  \city{Shanghai}
  \country{China}
}

\author{Fangwen Fu}
\email{fangwen.fu@intel.com}
\orcid{0000-0003-3548-6066}
\affiliation{%
  \institution{Intel Corporation}
  \city{Santa Clara}
  \state{California}
  \country{USA}
}

\author{Yen-Kuang Chen}
\email{y.k.chen@intel.com}
\orcid{0000-0003-4546-9497}
\affiliation{%
  \institution{Intel Corporation}
  \city{Santa Clara}
  \state{California}
  \country{USA}
}

\author{Xinmin Tian}
\email{xinmin.tian@intel.com}
\orcid{0000-0001-6228-924X}
\affiliation{%
  \institution{Intel Corporation}
  \city{Santa Clara}
  \state{California}
  \country{USA}
}

\author{Stuart H. Sul}
\email{ssul@stanford.edu}
\orcid{0009-0006-8493-9833}
\affiliation{%
  \institution{Stanford University}
  \city{Stanford}
  \state{California}
  \country{USA}
}

\author{Simran Arora}
\email{simarora@caltech.edu}
\affiliation{%
  \institution{California Institute of Technology}
  \city{Pasadena}
  \state{California}
  \country{USA}
}

\begin{abstract}
\input{sections/abstract}
\end{abstract}




\maketitle

\input{sections/introduction}
\input{sections/section2_background_hk}
\input{sections/section3_core_hk}
\input{sections/ai_development}
\input{sections/results}
\input{sections/e2e}
\input{sections/multigpu}

\input{sections/conclusion}

\ifdefined\skanonymousediting\else
\paragraph{Contributions.} Jiang led implementation, experiments, and
writing. S.~Chen contributed implementation and debugging. Fu, Y.-K.~Chen,
and Tian contributed Intel Xe-HPC architecture expertise and baselines. Sul
contributed discussion, review, and writing. Arora provided direction,
supervision, and review.
\fi

\section*{Acknowledgments}
Generative AI tools were used to assist with kernel development and
optimization (documented in the appendices) and with editing
the manuscript. The authors reviewed and verified all resulting content,
including text, code, data, and figures, and take full responsibility
for it.

\bibliographystyle{ACM-Reference-Format}
\bibliography{references}

\appendix

% Raw result tables are wider than a sigplan column. Scoped to the appendix so
% main-paper tables keep their original typesetting.
\let\originaltabular\tabular
\let\endoriginaltabular\endtabular
\renewenvironment{tabular}[2][c]
  {\begin{adjustbox}{max width=\linewidth}\originaltabular[#1]{#2}}
  {\endoriginaltabular\end{adjustbox}}
\AtBeginEnvironment{tabularx}{%
  \let\tabular\originaltabular
  \let\endtabular\endoriginaltabular
}

\section*{Appendix overview}
These appendices separate outputs of the controlled-agent studies from the
co-designed \mk{} suite. Each experiment identifies its library baseline or
author-built reference, along with its measurement and aggregation protocol.

\paragraph{Shared measurement protocol.}
Unless stated otherwise, single-GPU experiments activate one stack per
GPU package and leave its sibling idle. Every reported configuration passes
the section-specific reference check before timing. Warmup, repetition,
aggregation, and byte/FLOP conventions are stated with each experiment.
Displayed measurements are rounded; ratios use the unrounded values.

\input{sections/supp/appx-hardware}

\section{Tile Language Mechanisms}
\label{app:dsl}
\input{sections/dsl_ext}

\input{sections/supp/appx-agent-study}

\input{sections/supp/appx-gemm}

\input{sections/supp/appx-attention-fwd}
\input{sections/supp/appx-norms}
\input{sections/supp/appx-op-fusions}

\input{sections/supp/appx-dist-gemm}

\input{sections/supp/appx-cp-attention}

\input{sections/supp/appx-collectives}

\input{sections/supp/appx-e2e}

\end{document}

%% file: sections/abstract.tex
New AI accelerators arrive before the kernels that make them fast, because
peak kernel performance requires architecture-specific expertise in operand
pipelines and data-movement techniques.
Coding agents can now write, compile, and tune kernels on their own, so they
could greatly accelerate kernel development and optimization.
What agents produce depends on the interfaces they are given. These interfaces may expose
a machine's raw capabilities or encode the known-good methods for exploiting
its hardware features efficiently. We test the performance impact of these two
interfaces on Intel GPUs through controlled experiments with five coding
models on three kernels.
With raw SYCL and execution feedback, Opus~4.8, the strongest of the three models tested with Claude Code,
writes a single-GPU GEMM kernel reaching only $57.5\%$ of Intel's tuned oneDNN
library.
We present \mk{}, a hardware-aware tile programming model for Intel GPUs that
encodes these known-good methods, so its operations place operands in
matrix-engine layouts, prefetch through the L1 cache, and communicate over the
fabric that links GPU stacks into a node.
With \mk{} under the same agent, task, and feedback budget, the agent reaches
$82.1\%$, showing that encoded methods turn hardware capabilities into
performance.
Workload-specific policies stay programmable, and engineers and agents jointly
refine these schedules in \mk{} to build a kernel suite that reaches
${\sim}96\%$ of oneDNN in geometric mean across GEMM shapes. The suite runs
Llama-3.1-8B inference $1.59\times$ faster than \texttt{torch.compile} on one
GPU and up to $2.91\times$ faster than a matched multi-GPU decode path on
Intel's oneCCL.
\mk{} is open source and available at \url{https://github.com/intel/SyclKittens}.

%% file: sections/introduction.tex
\section{Introduction}
\label{sec:introduction}

\begin{figure*}[t]
\centering
\begin{subfigure}[b]{0.455\textwidth}
\centering
\input{figures/fig1_concept}
\caption{Capabilities, methods, and policies.}
\label{fig:progression-concept}
\end{subfigure}
\hfill
\begin{subfigure}[b]{0.525\textwidth}
\centering
\input{figures/fig1_turns}
\caption{Single-GPU GEMM tuning progress.}
\label{fig:progression-turns}
\end{subfigure}
\caption{\textbf{Programming interfaces and agent tuning progress.}
(a) \mk{} supplies methods; programmers and agents choose policies.
(b) For single-GPU GEMM, \mk{} helps Opus~4.8 reach Raw SYCL's
final performance with fewer feedback calls. Curves show median best-so-far
throughput relative to oneDNN on Intel Max GPU; the dashed line is the
engineer-agent kernel. Examples come from NVIDIA ThunderKittens and AMD
HipKittens.}
\Description{Two panels. Panel (a) shows three nested rounded rectangles. The
innermost is labelled Capability and lists matrix instructions, two
dimensional block prefetch, and peer memory, tagged as exposed by raw SYCL. The
middle is labelled Method and lists matrix-engine tile layouts, cache-centric
operand delivery, and cooperative prefetch, tagged as encoded by SyclKittens.
The outermost has a dashed border, is labelled Policy, and lists partial-tile
ownership, prefetch distance, and shape-dependent schedules, tagged as searched
by engineers and agents. Panel (b) is a line chart with evaluator
feedback calls from 1 to 20 on the horizontal axis and percentage of the tuned
oneDNN baseline on the vertical axis. The raw SYCL line rises slowly from about
$3\%$ to about $50\%$ over twenty calls. The SyclKittens line starts
near $46\%$, reaches $84.5\%$ by the fourth call, and ends at
${\sim}86\%$. The SyclKittens plus examples line reaches $85.6\%$
by the third call and also ends at ${\sim}86\%$. A
dashed horizontal line marks the co-designed SyclKittens kernel at $96.2\%$.}
\label{fig:progression}
\end{figure*}
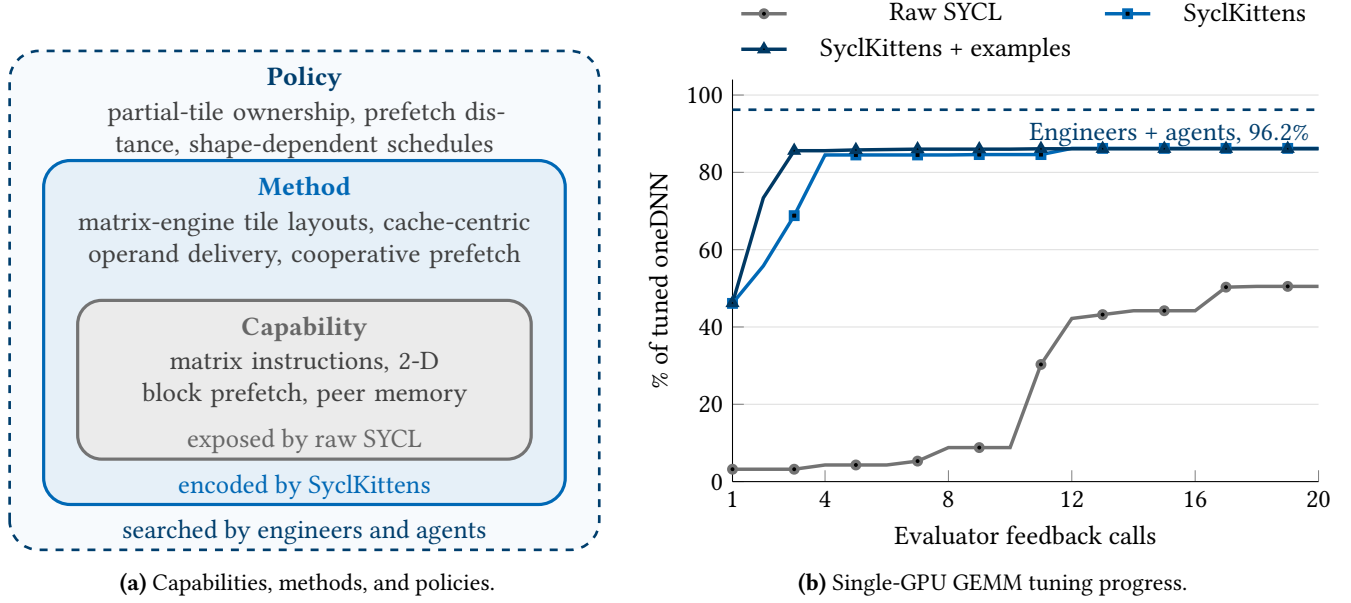

Supporting diverse AI hardware is limited by the difficulty of building
high-performance kernel ecosystems.
Tile programming models such as ThunderKittens and HipKittens package
architecture-specific methods for NVIDIA and AMD GPUs, while ParallelKittens
extends this approach for multiple GPUs~\cite{spector2025thunderkittens,
hu2024hipkittens,sul2025parallelkittens}.
Coding agents can improve programs through execution feedback,
and kernel benchmarks show that they can write, compile, benchmark, and revise
single- and multi-GPU kernels~\cite{yang2024sweagent,ouyang2025kernelbench,
chan2026parallelkernelbench}.
We ask whether a tile programming model can make Intel GPU methods reusable by
programmers and coding agents while preserving workload-specific control.

In our Claude Code experiments, three agentic-coding models produce correct Intel GPU kernels
that are substantially slower than the tuned baselines.
They use raw SYCL, a C++ programming
model for heterogeneous systems~\cite{sycl2020}. The tasks are single-GPU
matrix multiplication (GEMM), causal grouped-query attention
(GQA)~\cite{ainslie2023gqa}, and distributed GEMM. Each model receives Intel
documentation and execution feedback and works autonomously under a fixed
search budget. On single-GPU GEMM, Opus~4.8 produces the
fastest raw-SYCL kernel among the three models, but it reaches only 57.5\% of
the throughput of oneDNN, Intel's tuned deep-learning library~\cite{onednn}.
Its kernel already uses the GPU's matrix engines and fetches operands ahead of
computation. However, its thread groups do not coordinate prefetching and
repeatedly request the same operand tiles.

We introduce \mk{}, a hardware-aware tile programming model for Intel GPUs.
We target the Intel Data Center GPU Max 1550 (Intel Max GPU)~\cite{intel_max_2023,blythe2021pvc},
deployed in Aurora~\cite{aurora2025}. The tile interface must match the packed
register layouts of Xe Matrix Extensions (XMX), Intel's matrix engines, and
coordinate operand delivery through cache prefetch and 2-D block loads.
Communication operations must account for the bandwidth difference between
MDFI (Multi-Die Fabric Interconnect), the faster link between the two GPU stacks within a package, and XeLink,
the peer load/store network between packages.
Built with Intel hardware experts, \mk{} extends the tile abstractions of
ThunderKittens and dataflow algorithms from ThunderKittens and HipKittens with
three classes of Intel-specific operations.
\begin{enumerate}
\item \textbf{Matrix operand layouts and packing.} \mk{} maps register tiles to
  XMX's native execution width and fuses 2-D block loads with hardware packing
  for the column-layout right-hand matrix operand B, avoiding software packing
  shuffles. In the cumulative GEMM ablation, changing from $16\times16$ to
  $32\times32$ register tiles raises geometric-mean throughput from $37.4\%$ to
  $68.2\%$ of oneDNN across 27 shapes (\S\ref{sec:sk}).
\item \textbf{Cooperative cache prefetch.} Tile-level prefetch lets groups of
  threads prefetch disjoint regions of shared operands into the L1 cache while
  computing. In our tuned GEMM, replacing heuristic prefetch coordination with
  disjoint ownership raises geometric-mean throughput from $91.0\%$ to $96.2\%$
  of oneDNN (\S\ref{sec:sk}).
\item \textbf{Topology-aware GPU communication.} To avoid repeated XeLink
  transfers, \mk{} gathers each shard once per remote package and replicates it
  over MDFI. This yields 1.42--2.15 times the end-to-end distributed-GEMM
  throughput of the flat gather pipeline (\S\ref{sec:sk}).
\end{enumerate}

Raw SYCL exposes hardware \emph{capabilities}, including matrix instructions,
prefetch, and peer memory. \mk{} encodes \emph{methods} that coordinate these
capabilities. Programmers and coding agents choose workload \emph{policies},
including tile shape, prefetch ownership and distance, and work distribution
(Figure~\ref{fig:progression-concept}). \emph{An accelerator interface should
encode the architecture's methods as first-class operations while leaving
workload policy programmable.}

We perform three evaluations of \mk{}. Controlled coding-agent experiments
test whether encoding Intel-specific methods in the programming model helps
agents produce faster kernels. The kernel-suite evaluation tests whether these
abstractions support high performance across a broad range of operations. The
inference evaluation measures whether those kernels deliver end-to-end gains
for Llama-3.1-8B on single and multiple GPUs.

\begin{itemize}
\item For controlled coding-agent experiments, we vary programming
context across five coding models under a fixed feedback budget.
The first context, Raw SYCL (Raw),
provides Intel documentation. The second
context, \mk{}, adds its headers and interface documentation. The third
context, \mk{} + examples, provides kernel source code for NVIDIA and AMD GPUs,
or for different operations on Intel GPUs. For Opus~4.8, the best-performing
context on each task is either \mk{} or \mk{} + examples.
Single-GPU GEMM performance rises from
$57.5\%$ of oneDNN under Raw to $82.1\%$ with \mk{}. In cold iteration, where
agents start from scratch, \mk{} helps Opus reach Raw's final performance with
fewer tuning steps (Figure~\ref{fig:progression-turns}). For GQA, Opus reaches
$79.5\%$ of PyTorch's native scaled dot-product attention (SDPA) with \mk{} +
examples, up from $57.7\%$ with Raw SYCL.
All selected Opus GQA kernels use XMX
across the three contexts. The examples help the agent find more effective
execution schedules.
The GEMM gains also hold for Opus~5 and GPT-5.6-Sol. Across the five models,
GQA benefits depend on the model and programming context.

\item We build a high-performance kernel suite with \mk{} through engineer-agent
co-design. The suite reaches ${\sim}96\%$ of oneDNN in geometric mean across
all 27 GEMM shapes. On causal multi-head attention
(MHA)~\cite{vaswani2017attention}, \mk{} delivers $1.21\times$ the geometric-mean
throughput of fused SDPA with Intel Extension for PyTorch (IPEX). The suite also
includes normalization, rotary position embedding (RoPE)~\cite{su2021roformer},
matrix-vector multiplication (GEMV), multi-GPU collectives, and fused
distributed kernels.

\item We evaluate Llama-3.1-8B~\cite{grattafiori2024llama3} inference on one GPU and
on a full Intel Max node with six two-stack packages (12 GPUs). (i) On one GPU,
\mk{} works alongside \texttt{torch.compile}, improving full-request
prefill-plus-decode throughput by $1.59\times$ in geometric mean over
\texttt{torch.compile} alone. \mk{} also runs configurations where
\texttt{torch.compile} alone runs out of memory. (ii) At node scale, \mk{} combines
tensor-parallel projections~\cite{shoeybi2019megatron} with context-parallel
attention. Its decode implementation outperforms the tuned inference path built
on Intel's oneCCL collective-communication library, by up to $2.91\times$ on four GPUs, and serves longer contexts than
the baseline can fit in memory (\S\ref{sec:e2e}).
\end{itemize}

This paper makes three contributions.
\begin{enumerate}
\item We build and open-source \mk{}, an Intel GPU tile programming model that encodes
  architectural methods while preserving workload control, and quantify their
  performance contributions through GEMM ablations (\S\ref{sec:sk}).
\item We provide a controlled coding-agent study that evaluates how programming
  context, example kernels, and execution feedback affect kernel
  performance and workload-policy search (\S\ref{sec:ai}).
\item We present a co-designed Intel GPU kernel suite and evaluate its
  performance in individual kernels and Llama-3.1-8B inference
  (\S\ref{sec:kernels}, \S\ref{sec:e2e}).
\end{enumerate}
\input{sections/intro_ext}

%% file: figures/fig1_concept.tex
% Figure 1(a): capability subset method subset policy, drawn as containment.
% The three item lists must stay word-for-word with the defining paragraph in
% sections/introduction.tex. Ring colours match the lines in panel (b), namely
% black!55 = raw SYCL, intelblue = SyclKittens, intelblue!60!black = engineers.
% All text is at caption size, so the box geometry is hand-fitted to the wraps.
\definecolor{intelblue}{HTML}{0068B5}
\begin{tikzpicture}
% --- policy, outermost. Dashed because no interface encodes it. ---
\draw[rounded corners=11pt, draw=intelblue!60!black, dashed, line width=1.0pt,
      fill=intelblue!3]
  (0,0) rectangle (7.80,6.60);
\node[anchor=north, font=\normalsize\bfseries, text=intelblue!60!black]
  at (3.90,6.52) {Policy};
\node[anchor=north, font=\normalsize, text=black!75, align=center,
      text width=7.0cm]
  at (3.90,6.05) {partial-tile ownership, prefetch distance, shape-dependent schedules};
\node[anchor=north, font=\normalsize, text=intelblue!60!black]
  at (3.90,0.52) {searched by engineers and agents};

% --- method ---
\draw[rounded corners=10pt, draw=intelblue, line width=1.2pt, fill=intelblue!10]
  (0.45,0.60) rectangle (7.35,5.15);
\node[anchor=north, font=\normalsize\bfseries, text=intelblue]
  at (3.90,5.07) {Method};
\node[anchor=north, font=\normalsize, text=black!75, align=center,
      text width=6.1cm]
  at (3.90,4.60) {matrix-engine tile layouts, cache-centric operand delivery, cooperative prefetch};
\node[anchor=north, font=\normalsize, text=intelblue]
  at (3.90,1.12) {encoded by \mk{}};

% --- capability, innermost ---
\draw[rounded corners=9pt, draw=black!55, line width=1.2pt, fill=black!8]
  (0.90,1.20) rectangle (6.90,3.30);
\node[anchor=north, font=\normalsize\bfseries, text=black!60]
  at (3.90,3.22) {Capability};
\node[anchor=north, font=\normalsize, text=black!75, align=center,
      text width=5.2cm]
  at (3.90,2.75) {matrix instructions, 2-D block prefetch, peer memory};
\node[anchor=north, font=\normalsize, text=black!60]
  at (3.90,1.72) {exposed by raw SYCL};
\end{tikzpicture}

%% file: figures/fig1_turns.tex
% Figure 1(b): what feedback buys, per interface. Opus 4.8, cold-iterative,
% single-stack BF16 GEMM 4096^3. x is the evaluator call, y is the best
% held-out-correct kernel so far as a percentage of the co-measured oneDNN
% baseline, median over three independent trials.
% Raw   sgv1-prod-20260731d      iterative/gemm_opus_C_raw_r{1,2,3}
% MK    sgv1-iter-c0c2-20260803a iterative/gemm_opus_C0_r{1,2,3}
% MK+ex sgv1-iter-c0c2-20260803a iterative/gemm_opus_C2_r{1,2,3}
% Per-call values come from evaluations/dev_call_NNN/result.json.
% Colours match the rings in panel (a). The legend sits above the axis so it
% cannot collide with the raw SYCL curve.
\definecolor{intelblue}{HTML}{0068B5}
\begin{tikzpicture}
\begin{axis}[
  width=\linewidth, height=6.9cm,
  scale only axis=false,
  xmin=1, xmax=20, ymin=0, ymax=104,
  xtick={1,4,8,12,16,20},
  ytick={0,20,40,60,80,100},
  tick label style={font=\normalsize},
  xlabel={Evaluator feedback calls}, ylabel={\% of tuned oneDNN},
  label style={font=\normalsize},
  ymajorgrids, grid style={gray!25}, axis lines*=left,
  legend style={font=\normalsize, at={(0.5,1.02)}, anchor=south,
    legend columns=2, draw=none, fill=none, column sep=10pt, row sep=1pt},
  every axis plot/.append style={line width=1.3pt},
]
\addplot[draw=intelblue!60!black, dashed, line width=1.0pt, forget plot]
  coordinates {(1,96.2) (20,96.2)};
\node[font=\normalsize, anchor=north east, intelblue!60!black]
  at (axis cs:20,95.4) {Engineers + agents, $96.2\%$};

\addplot[draw=black!55, mark=*, mark size=1.4pt, mark repeat=2]
  coordinates {(1,3.2) (2,3.2) (3,3.2) (4,4.3) (5,4.3) (6,4.3) (7,5.3) (8,8.8)
  (9,8.8) (10,8.8) (11,30.3) (12,42.2) (13,43.2) (14,44.2) (15,44.2) (16,44.2)
  (17,50.3) (18,50.5) (19,50.5) (20,50.5)};
\addlegendentry{Raw SYCL}

\addplot[draw=intelblue, mark=square*, mark size=1.4pt, mark repeat=2]
  coordinates {(1,46.1) (2,55.8) (3,68.8) (4,84.5) (5,84.5) (6,84.5) (7,84.5)
  (8,84.5) (9,84.6) (10,84.6) (11,84.6) (12,86.2) (13,86.2) (14,86.2) (15,86.2)
  (16,86.2) (17,86.2) (18,86.2) (19,86.2) (20,86.2)};
\addlegendentry{\mk{}}

\addplot[draw=intelblue!55!black, mark=triangle*, mark size=1.8pt, mark repeat=2]
  coordinates {(1,46.2) (2,73.4) (3,85.6) (4,85.6) (5,85.8) (6,85.9) (7,86.0)
  (8,86.0) (9,86.0) (10,86.0) (11,86.1) (12,86.1) (13,86.1) (14,86.1) (15,86.1)
  (16,86.1) (17,86.1) (18,86.1) (19,86.1) (20,86.1)};
\addlegendentry{\mk{} + examples}
\end{axis}
\end{tikzpicture}

%% file: sections/section2_background_hk.tex
\section{Background and Motivation}
\label{sec:background}

\subsection{Tile Programming and Coding Agents}

Tile programming models expose fixed-shape register and shared-memory tiles,
layouts that address larger tensors in memory, and bulk movement, matrix,
reduction, and synchronization operations.
They sit between tensor compilers, which choose most schedules, and
low-level GPU programming, which exposes individual lanes and
instructions. This middle layer separates tile-level algorithms from the
architecture-specific layouts, movement, and scheduling that realize them.
ThunderKittens~\cite{spector2025thunderkittens} and
HipKittens~\cite{hu2024hipkittens} provide hardware-aware
interfaces on NVIDIA and AMD, respectively, and their shared primitives
provide the conceptual starting point for \mk{}. A prior IWOCL
study ported ThunderKittens directly to SYCL for Intel kernels on a single
GPU~\cite{jiang2026iwocl}, establishing SIMD16 mappings, larger register
tiles, L1 operand delivery, and cooperative prefetch. The present work
retains that foundation and adds explicit partial-tile prefetch control,
multi-GPU coordination, controlled coding-agent studies, and full-model
inference evaluation. Section~\ref{sec:sk} defines the partial-tile operation
and evaluates its ownership schedules.

Other systems divide control between the programmer and the compiler
differently. CuTe/CUTLASS provides
fine-grained layout and thread-participation control through C++
templates~\cite{nvidia_cute2024,nvidia_cutlass2025}, while TileLang separates
tile dataflow from schedule selection~\cite{wang2025tilelang}. For Intel GPUs,
programmers can use portable SYCL~\cite{sycl2020}, explicit SIMD (ESIMD)
vectors~\cite{intel_esimd2023}, compiler-selected Triton schedules through
Intel's XPU backend~\cite{tillet2019triton,intel_triton_xpu}, or template
libraries with native hardware control, such as XeTLA~\cite{intel_xetla2024}
and SYCL-TLA~\cite{sycltla2024}. Intel's oneAPI toolchain also provides
\texttt{joint\_matrix}, a portable SYCL extension that drives matrix engines
while hiding the operand layouts \mk{} makes explicit.
SK retains the TK/HK tile-dataflow style with Intel block loads and cache
prefetch, while SYCL-TLA follows CuTe/CUTLASS template composition.
Section~\ref{app:gemm-tla} compares their GEMM throughput under a common
measurement protocol.

At node scale, ParallelKittens~\cite{sul2025parallelkittens} extends
ThunderKittens to multiple GPUs and composes communication with computation
through NVIDIA interconnects. Intel oneCCL provides tuned collective
operations~\cite{intel_oneccl2025}. \mk{} exposes peer-memory operations and
synchronization within kernels so that programmers can coordinate data
movement with computation across Intel GPU stacks and packages. These
operations let kernels overlap communication with attention in the manner of
ring attention~\cite{liu2023ringattention}.

AI coding agents can generate and iteratively improve programs, but for GPU
kernels, correctness and performance are separate objectives.
KernelBench~\cite{ouyang2025kernelbench} evaluates fast, correct single-GPU
kernel generation over 250 PyTorch workloads.
ParallelKernelBench~\cite{chan2026parallelkernelbench} extends the
setting to multi-GPU PyTorch+NCCL baselines and fine-grained CUDA or DSL
kernels. Both establish that execution and profiling feedback help.
Our controlled study varies programming context and example kernels under
fixed models, tasks, and feedback budgets to examine how tile programming
affects kernel generation on Intel GPUs (\S\ref{sec:ai}).

\subsection{Intel GPU Model}
\label{sec:intel-fundamentals}

We study the Intel Data Center GPU Max
1550 (\pvc{})~\cite{intel_max_2023,blythe2021pvc,jiang2022pvc,aurora2025}.
Each package is one PCIe device built from two \emph{stacks}, Intel's term for
its physical GPU tiles. We run the flat device hierarchy, so each stack is
exposed as its own SYCL device with 56 Xe cores and 64~GB of HBM2e, and a node
presents 12 stacks across 6 packages. In workload and performance
descriptions, we call each exposed stack a \emph{GPU}; we reserve
\emph{stack} and \emph{package} for the physical topology. Throughout this
paper, \emph{tile} refers only to a register or shared tile in the language,
and not to a partition of the hardware.
Figure~\ref{fig:intel-platform} summarizes the stack and node hierarchy, and the three
hardware properties discussed below shape the \mk{} design in \S\ref{sec:sk}.

\input{figures/fig_topology}

\paragraph{Execution and registers.}
Intel software addresses register operands as linear vectors along the SIMD
direction, one element per lane, whereas NVIDIA tensor-core interfaces
use matrix fragments distributed across a warp. The shared tile abstractions
therefore require different physical register layouts on the two architectures.
A subgroup is the SIMD execution group within a workgroup. The \pvc{} supports
16- or 32-lane subgroups, while its Xe Matrix Extensions (XMX) matrix engines
only consume 16-lane subgroups. When all 16 lanes move
one four-byte register element, the subgroup access spans exactly one 64-byte
cache line. These execution and transaction widths determine the register-tile
mapping in \S\ref{sec:exec-width}.

\paragraph{Operand delivery.}
The \pvc{} supports direct matrix-operand delivery through the cache hierarchy.
A hardware 2-D prefetch warms an L1 cache of up to 512~KB per Xe core, and a
2-D demand load then moves the cached operand into subgroup registers. Shared
local memory (SLM) remains available. The \mk{} matrix path studied here uses
L1 prefetch and 2-D demand loads without intermediate SLM staging
(Figure~\ref{fig:intel-platform}(a)). This differs from the asynchronous
global-to-shared-memory staging used by ThunderKittens.

\paragraph{Node fabric.}
The Multi-Die Fabric Interconnect (MDFI) links the two GPU stacks within a package.
XeLink, Intel's peer load/store network between packages, connects packages
all-to-all through independent even- and odd-stack planes. Together, these
fabrics allow any pair of stacks to communicate directly
(Figure~\ref{fig:intel-platform}(b)). 
Multicast and reduction are composed in software over these peer paths.
XeLink also provides per-link copy engines for explicit transfers. 
However, peer-to-peer communications over these fabrics
differ in bandwidth and the number of transfers they
can sustain concurrently. MDFI provides substantially higher bandwidth than
XeLink.
Same-plane traffic uses a direct XeLink path, while cross-plane traffic
traverses an additional MDFI link. 
In distributed GEMM, a
topology-aware gather that crosses XeLink once per remote package and replicates
over MDFI delivers $1.42$--$2.15\times$ the
end-to-end throughput of the flat gather pipeline (\S\ref{sec:sk:multigpu}).
\Appxref{app:hardware-map}{Hardware and
Programming-Interface Map} reports measurement methods, full values, and a
cross-vendor terminology map.

The tile interface separates algorithmic operations from their
architecture-specific lowering.
Intel's upcoming Crescent Island data center GPU
illustrates which hardware properties a new
lowering must consider. It is based on the Intel Xe3p architecture, which
retains matrix engines and an L1 operand path~\cite{mohan2026cri}. Its
scale-up network uses PCIe switches instead of XeLink but remains a load/store
fabric. Porting \mk{} could preserve client tile dataflow and schedule
structure while requiring library changes to subgroup mappings, operand
packing, load and prefetch lowering, peer synchronization, and topology
plans. Validation on the new hardware would be required; the present
performance measurements are on Max 1550.

\subsection{What Raw-SYCL Agents Implement}
\label{sec:raw-sycl-motivation}

We inspect the Claude Code cold-iteration GEMM kernels in \S\ref{sec:ai} to identify which
matrix and communication methods raw-SYCL agents implement.
In single-GPU bfloat16 (BF16) GEMM, Sonnet and Haiku produce correct kernels
that stage operands in SLM without using XMX. Their best correct kernels reach only
$5.7\%$ and $1.6\%$ of Intel's tuned oneDNN baseline.
Opus builds an XMX pipeline from SPIR-V
operations, including a Dot Product Accumulate Systolic (DPAS) matrix
instruction, 2-D block loads, a $B$ operand in the interleaved VNNI layout that
DPAS requires, and prefetch. However, its thread groups do not coordinate
prefetching and repeatedly request the same operand tiles. Its best correct
kernel reaches $57.5\%$ of that baseline. oneDNN generates Intel GPU assembly through
nGEN~\cite{onednn_gpu_jit} and selects GEMM strategies from a tuned kernel
database~\cite{onednn_gemm_kernel_catalog}. The comparison is therefore against
oneDNN's assembly-generating implementation.

In the distributed GEMM task, input matrix $A$ is row-sharded, $B$ is replicated, and every
GPU must return the full $C=AB$. A kernel can gather $A$ before computing the
full output, or compute its local rows of $C$ and then gather the output.

All nine raw-SYCL cold-iteration trials produce correct kernels, but the models
combine these choices differently. Haiku computes local output rows but uses
vector fused multiply-add (FMA) operations. Sonnet uses XMX but
gathers the input and computes the full output on every GPU. Opus uses XMX
and overlaps copies across compute and copy queues, but keeps the same
replicated-compute decomposition and ignores the MDFI/XeLink topology.
Across these trials, no raw-SYCL candidate combines local-output
computation, XMX, and topology-aware communication. The best reaches $42.3\%$
of our frozen, co-measured, topology-aware SyclKittens hand reference.

These kernels show that access to matrix instructions and peer memory leaves
operand coordination and work distribution to the kernel author. \mk{}
encodes operand layouts, cooperative cache prefetch, and topology-aware
communication as reusable operations. Programmers and coding agents use those
operations while choosing tile shapes, prefetch ownership and distance, and
the distribution of computation. Section~\ref{sec:sk} defines these operations,
and \S\ref{sec:ai} evaluates their effect on agent-written kernels.

%% file: figures/fig_topology.tex
% Stack hierarchy and full-node two-plane XeLink topology.
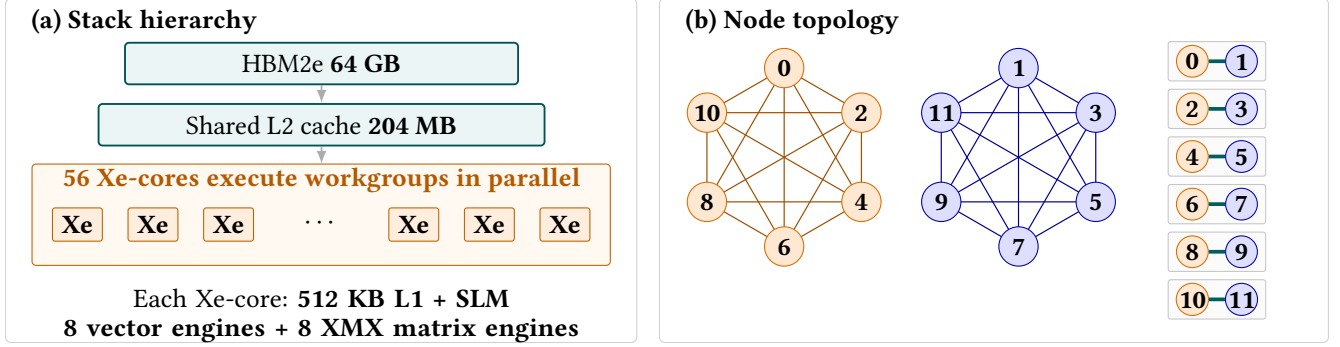
\begin{figure*}[t]
\centering
\begin{tikzpicture}[
  font=\normalsize,
  memory/.style={draw=teal!65!black, fill=teal!8, rounded corners=1.5pt,
                 minimum height=0.55cm, align=center, line width=0.7pt},
  core/.style={draw=orange!75!black, fill=orange!13, rounded corners=1pt,
               minimum width=0.66cm, minimum height=0.44cm, inner sep=1pt,
               font=\bfseries},
  even/.style={circle, draw=orange!85!black, fill=orange!18,
               minimum size=0.52cm, inner sep=0pt, font=\bfseries},
  odd/.style={circle, draw=blue!70!black, fill=blue!13,
              minimum size=0.52cm, inner sep=0pt, font=\bfseries},
  evenlink/.style={draw=orange!62!black, line width=0.45pt},
  oddlink/.style={draw=blue!58!black, line width=0.45pt},
  mdfi/.style={draw=teal!75!black, line width=1.4pt},
  flow/.style={-latex, draw=gray!65, line width=0.7pt},
]
  \draw[gray!35, rounded corners=2pt] (0,0) rectangle (8.35,4.55);
  \draw[gray!35, rounded corners=2pt] (8.65,0) rectangle (17.5,4.55);
  \node[font=\bfseries, anchor=west] at (0.22,4.24) {(a) Stack hierarchy};
  \node[font=\bfseries, anchor=west] at (8.87,4.24) {(b) Node topology};

  % Panel (a): memory hierarchy feeding 56 parallel Xe-cores.
  \node[memory, minimum width=5.2cm] (hbm) at (4.18,3.69)
    {HBM2e \textbf{64 GB}};
  \node[memory, minimum width=5.9cm] (l2) at (4.18,2.88)
    {Shared L2 cache \textbf{204 MB}};
  \draw[flow] (hbm.south) -- (l2.north);

  \draw[orange!75!black, fill=orange!5, rounded corners=1.5pt]
    (0.35,1.02) rectangle (8.00,2.36);
  \node[font=\bfseries, text=orange!70!black] at (4.18,2.15)
    {56 Xe-cores execute workgroups in parallel};
  \foreach \x in {0.95,1.95,2.95,5.40,6.40,7.40}{
    \node[core] at (\x,1.56) {Xe};
  }
  \node at (4.18,1.56) {$\cdots$};
  \draw[flow] (l2.south) -- (4.18,2.36);
  \node[align=center, text width=7.6cm] at (4.18,0.36)
    {Each Xe-core: \textbf{512 KB L1 + SLM}\\[-1pt]
     \textbf{8 vector engines + 8 XMX matrix engines}};

  % Panel (b): independent K6 meshes for even and odd stack IDs.
  \begin{scope}[shift={(10.30,2.45)}]
    \foreach \i/\lab/\ang in {0/0/90,1/2/30,2/4/-30,3/6/-90,4/8/-150,5/10/150}{
      \coordinate (E\i) at (\ang:1.18);
    }
    \foreach \i in {0,...,5}{
      \foreach \j in {0,...,5}{
        \ifnum\i<\j
          \draw[evenlink] (E\i) -- (E\j);
        \fi
      }
    }
    \foreach \i/\lab in {0/0,1/2,2/4,3/6,4/8,5/10}{
      \node[even] at (E\i) {\lab};
    }
  \end{scope}

  \begin{scope}[shift={(13.40,2.45)}]
    \foreach \i/\lab/\ang in {0/1/90,1/3/30,2/5/-30,3/7/-90,4/9/-150,5/11/150}{
      \coordinate (O\i) at (\ang:1.18);
    }
    \foreach \i in {0,...,5}{
      \foreach \j in {0,...,5}{
        \ifnum\i<\j
          \draw[oddlink] (O\i) -- (O\j);
        \fi
      }
    }
    \foreach \i/\lab in {0/1,1/3,2/5,3/7,4/9,5/11}{
      \node[odd] at (O\i) {\lab};
    }
  \end{scope}

  % Each outline is one GPU package; MDFI connects its two stacks.
  \foreach \y/\a/\b in {3.72/0/1,3.09/2/3,2.46/4/5,1.83/6/7,1.20/8/9,0.57/10/11}{
    \begin{scope}[shift={(0,\y)}]
      \draw[gray!45, fill=gray!3, rounded corners=1pt]
        (15.38,-0.26) rectangle (16.67,0.26);
      \node[even, minimum size=0.42cm] (M\a) at (15.70,0) {\a};
      \node[odd, minimum size=0.42cm] (M\b) at (16.35,0) {\b};
      \draw[mdfi] (M\a) -- (M\b);
    \end{scope}
  }
\end{tikzpicture}
\caption{\textbf{Intel Max GPU at stack and node scale.}
  (a) HBM2e memory and shared L2 cache feed the Xe-cores within a stack.
  (b) Six two-stack packages form two independent all-to-all XeLink
  planes, one for even stacks and one for odd stacks. MDFI connects
  the two stacks within each package and is the only link between
  the planes.}
\Description{A two-panel schematic of the Intel Max GPU. Panel (a) is a block
diagram of one stack: HBM2e memory and a shared L2 cache feeding an array of
Xe-cores, each drawn with its own L1/SLM and its vector and XMX engines. Panel
(b) is a node-level connectivity diagram of six packages of two stacks each,
showing MDFI links inside a package and all-to-all XeLink planes between
packages.}
\label{fig:intel-platform}
\end{figure*}

%% file: sections/section3_core_hk.tex
\section{SyclKittens}
\label{sec:sk}

\mk{} is a header-only C++ tile programming model for Intel GPUs.
Kernels declare typed register or shared-memory tiles, address
tensors through global layouts, and express dataflow through tile movement,
matrix and reduction operations, and synchronization. The backend lowers these
operations to SYCL or SPIR-V and selects their physical lane mappings.
Programmers retain C++ control flow and direct access to SYCL or SPIR-V
intrinsics. Distributed layouts, peer memory, and GPU-resident synchronization
extend the same programming model across a node.

The Intel-specific operations encode matrix operand layouts and packing,
cooperative cache prefetch, and topology-aware communication. Tile shape,
prefetch ownership and distance, and work distribution remain programmable.
In the cumulative single-GPU GEMM ablation, changes to operand layouts,
prefetch, and workload-specific schedules raise the geometric mean from
$37.4\%$ to $96.3\%$ of tuned oneDNN. The distributed movement ablation
reaches $82$--$90\%$ of communication-free GEMM throughput across twelve GPUs.
\S\ref{sec:tile-interface} defines the interface.
\S\ref{sec:exec-width} and \S\ref{sec:prefetch} explain operand layouts,
packing, and cooperative prefetch, and \S\ref{sec:sk:multigpu} describes
communication within a node.

\subsection{Tile Interface}
\label{sec:tile-interface}

The register-tile type \texttt{rt<T,M,N,layout,shape>} represents a logical
$M{\times}N$ two-dimensional array of elements of type \texttt{T}. The
\texttt{layout} parameter selects row- or column-major organization, while
\texttt{shape} selects a static physical base tile. The GEMM ablation compares
$16{\times}16$ and $32{\times}32$ base tiles; rectangular shapes are also
supported. Sub-tile coordinates address rectangular regions within a tile and support
cooperative prefetch in \S\ref{sec:prefetch}.
Bulk rectangular movement and matrix or reduction operations are shared with
ThunderKittens. Kernels choose the ordering of load, compute, and store stages
over workgroup tiles.

A Global Layout (GL) pairs a device-memory pointer
with logical batch, depth, row, and column dimensions. Tile movement operations
use its coordinates rather than computing byte offsets in each kernel. A
Parallel Global Layout (PGL)~\cite{sul2025parallelkittens} extends this
coordinate space across devices. Additional API and lowering details appear in
\dslref{}.

\subsection{Matrix Operand Layouts and Packing}
\label{sec:exec-width}

\mk{} maps logical row- and column-major tiles onto linear vectors along the SIMD
direction, using the 16-lane subgroups that XMX consumes.
This differs from matrix fragments distributed across a 32-lane warp in
ThunderKittens' NVIDIA mapping. Fixing the subgroup width
at 16 avoids a 32-to-16 operand reorganization inherited from that mapping.
For BF16 data, a 32-element tile row matches one 64-byte L1 transaction. A
16-element row consumes only half the cache line, leaving the other half for a later
access. The $32{\times}32$ base tile therefore halves L1 transactions per 32
values and consumes each line immediately. In the cumulative GEMM ablation,
this change raises the geometric mean from $37.4\%$ to $68.2\%$ of oneDNN
(Table~\ref{tab:gemm-ablation}).

\paragraph{Direct B-operand load and packing.}
The \mk{} API supports two paths from global memory into registers. One stages
through a shared tile in SLM, and the other moves directly from a Global Layout
into a register tile. ThunderKittens and HipKittens use shared-memory staging
followed by \texttt{ldmatrix} or \texttt{ds\_read\_b64\_tr\_b16}, respectively.
The \pvc{} has no equivalent SLM matrix-operand load, so loading a packed B
operand from SLM requires multiple ordinary reads. The \mk{} matrix path
instead loads operands directly through L1 cache, avoiding this load sequence
and the additional staging and synchronization. For a column-layout
right-hand matrix operand B,
\texttt{load\_transform} lowers to \texttt{LoadTransformINTEL}, a hardware
operation that fuses a global 2-D block load with VNNI packing for DPAS. An
ordinary load followed by software shuffles is measurably slower; \dslref{}
reports the packing implementation and its effect.

\subsection{Cooperative Cache Prefetch}
\label{sec:prefetch}
\label{sec:sk:onestack}

\mk{} separates operand delivery into a prefetch phase and a demand phase.
\texttt{prefetch\_load} issues a 2-D prefetch hint that warms L1 cache without
writing the tile's registers. A later 2-D demand load transfers the cached
operands into subgroup registers. This split allows kernels to prefetch
future operands while computing with the current tiles, without allocating a
shared-memory staging buffer.

\texttt{prefetch\_load\_coop} takes sub-tile offsets that identify the operand
region to prefetch. Kernels assign these offsets to computing subgroups and
choose the prefetch distance. The operation provides partial-tile prefetch;
the schedule determines ownership and lookahead.
The complete prologue and steady-state example in
\S\ref{app:gemm-coop-loop} shows these responsibilities in client code.
The same operations are available to programmers and agents; the interface
does not prescribe a separate agent-only syntax or an inferred schedule.

In the cumulative GEMM ablation (Table~\ref{tab:gemm-ablation}),
selective subgroup-masked prefetch raises the
geometric mean to $88.8\%$ of oneDNN. Loop unrolling and snake traversal, which
keeps neighboring workgroups within a compact L2 reuse region, raise it to
$91.0\%$.
The masked schedule still warms shared operand regions more than once.
With partial-tile de-duplication,
each subgroup prefetches a disjoint
region of the shared operands using the sub-tile coordinates from
\S\ref{sec:tile-interface}.
This ownership assignment warms each shared cache region once and raises the mean to
$96.2\%$.
A final per-shape selection of prefetch distance closes the sequence at
$96.3\%$ of oneDNN.
\Appxref{app:gemm}{BF16 GEMM} reports the protocol, errors, and per-shape
results for this GEMM ablation.

\begin{table}[t]
\centering
\caption{Six-node, 27-shape cumulative BF16-input/BF16-output GEMM ablation on
the \pvc{}. Values are geometric means of each shape's median \mk{}/oneDNN ratio
across nodes; each row adds one change to the row above.}
\label{tab:gemm-ablation}
\small
\setlength{\tabcolsep}{4pt}
\begin{tabular}{@{}p{0.58\columnwidth}rr@{}}
\toprule
\textbf{Configuration} & \textbf{\% oneDNN} & \textbf{vs. prior} \\
\midrule
Direct $16{\times}16$ register tiles, no prefetch & 37.4\% & -- \\
\quad + $32{\times}32$ register tiles & 68.2\% & $1.822\times$ \\
\quad + selective subgroup-masked L1 prefetch & 88.8\% & $1.301\times$ \\
\quad + explicit unroll-32 & 89.1\% & $1.003\times$ \\
\quad + snake traversal & 91.0\% & $1.022\times$ \\
\quad + partial-tile de-duplication & 96.2\% & $1.056\times$ \\
\quad + shape-specific prefetch distance & 96.3\% & $1.001\times$ \\
\bottomrule
\end{tabular}
\end{table}

\subsection{Topology-Aware GPU Communication}
\label{sec:sk:multigpu}

\texttt{make\_pgl} combines per-device GLs and peer pointers, and
\texttt{pgl.dev\_ptrs[peer]} exposes remote high-bandwidth memory (HBM) to
ordinary kernel loads and stores. The GPUs share a SYCL context, allowing
kernels to dereference peer pointers directly.

GPUs synchronize through peer-visible counters without returning to the host.
After writing peer memory, the writing GPU publishes completion with a
system-scope release atomic (\texttt{sycl::atomic\_ref}). The reading GPU waits
with a system-scope acquire atomic. Observing the matching release makes the
preceding writes visible to that GPU. \texttt{signal}, \texttt{wait}, and
\texttt{barrier\_all} implement this protocol.

The topology API groups adjacent stacks into GPU packages with
\texttt{topo::group\_pvc} and assigns transfer ownership with
\texttt{topo::plan\_all\_gather}. A hierarchical gather sends each shard over
XeLink once per remote package, then replicates it to the partner stack over
MDFI. This reduces repeated XeLink crossings relative to a flat gather.

PGL permits both kernel-driven peer access and explicit copy-engine transfers.
Kernels choose how to divide computation and schedule these transfers.
Table~\ref{tab:dist-gemm-ablation} holds the GEMM core fixed and varies
input-gather schedules to isolate communication changes. Each GPU gathers the
missing rows of $A$ and computes the full $C$. The agent study in
\S\ref{sec:ai} also explores computing local rows of $C$ and then gathering
the output.

\begin{table}[t]
\centering
\caption{Distributed GEMM movement ablation on 12 GPUs, reported as
end-to-end throughput relative to the same GEMM core measured without
communication in the same run.}
\label{tab:dist-gemm-ablation}
\small
\setlength{\tabcolsep}{4pt}
\begin{tabularx}{\columnwidth}{@{}>{\raggedright\arraybackslash}Xrr@{}}
\toprule
\textbf{Movement method} & \textbf{$6144^3$} & \textbf{$12288^3$} \\
\midrule
Flat kernel or copy-engine gather & 38\% & 63\% \\
Pipelined gather and GEMM & 40\% & 62\% \\
Fused gather and GEMM & 47\% & 62\% \\
Topology-aware gather and GEMM & \textbf{82\%} & \textbf{90\%} \\
\bottomrule
\end{tabularx}
\end{table}

The topology-aware pipeline achieves $2.15\times$ and $1.42\times$ the
end-to-end distributed-GEMM throughput of the flat gather pipeline at the
smaller and larger shapes, respectively. The same primitives support
collectives and ring attention.
\Appxrefs{}~\appxitem{app:dist-gemm}{Distributed GEMM},
\appxitem{app:cp-attn}{Context-Parallel Attention (Ring and All-Gather)}, and
\appxitem{app:collectives}{Multi-GPU Collectives}\appxtail{} report full sweeps.

%% file: sections/ai_development.tex
\section{Controlled Coding-Agent Evaluation}
\label{sec:ai}

We ask how programming context affects agent-written kernel performance,
when example kernels help, and how execution feedback improves kernels
started from scratch or from a correct implementation. We evaluate five
coding models on single-GPU GEMM, distributed GEMM, and causal grouped-query
attention (GQA). The \mk{} contexts supply its complete interface, including
cooperative prefetch, while agents choose the kernel schedule. The
co-designed kernels evaluated in \S\ref{sec:sk} are not supplied to agents.
Tables~\ref{tab:agent-gemm}, \ref{tab:agent-newer-gemm},
and~\ref{tab:agent-gqa} report the controlled results.

\subsection{Experimental Design}
\label{sec:ai:protocol}

\paragraph{Models and tasks.}
We use Claude Code for Opus~4.8, Sonnet~5, and Haiku~4.5, and GitHub Copilot
for Opus~5 and GPT-5.6-Sol. The tasks are $4096^3$ BF16 GEMM on one GPU,
row-sharded $6144^3$ all-gather GEMM on 12 GPUs, and causal GQA at
$B{=}1$, $H{=}32$, $H_{KV}{=}8$, $N{=}2048$, and $D{=}128$.
Comparisons vary programming context within each model and controller;
the two controller groups are not a controlled comparison of model capability.

\paragraph{Programming contexts.}
Raw supplies Intel SYCL/SPIR-V documentation. \mk{} adds headers and
API documentation. \mk{} + examples additionally supplies 
task-specific kernel source code. For single-GPU GEMM, the examples are
ThunderKittens and HipKittens GEMMs. For
distributed GEMM, the context also adds a two-GPU SyclKittens
all-reduce that shows Intel peer access but not an all-gather GEMM or 12-GPU
schedule. For GQA, it adds two HipKittens GQA kernels. None
provides an Intel implementation of the target kernel.

\paragraph{Hardware access and launch control.}
All contexts permit direct SYCL/SPIR-V intrinsics and forbid oneAPI
\texttt{joint\_matrix}. All GQA conditions use a byte-identical raw-pointer
harness and candidate function signature. Candidates control launch geometry
and submissions in every context.
Every GQA condition also forbids ESIMD, a separate Intel-specific SIMD
programming model that neither raw SYCL nor \mk{} builds on.
All contexts use the SYCL toolchain. Triton-XPU is evaluated as an autotuned
kernel-performance baseline in \S\ref{sec:kernels}.

\paragraph{Experimental coverage.}
Each evaluated model-context pair receives three independent starts.
All five models undergo cold-iterative evaluation. The Claude Code group
also undergoes single-shot and warm-iterative evaluation and spans large,
mid-size, and small model tiers. The Copilot group covers Raw and \mk{}
on both GEMM tasks and all three contexts on GQA.
Raw versus \mk{} measures access to the interface and its encoded expertise
together; \mk{} versus \mk{} + examples measures the additional examples.

\paragraph{Feedback and selection.}
We use a custom Claude Code harness rather than the mini-SWE-agent
implementation used by
ParallelKernelBench~\cite{chan2026parallelkernelbench,yang2024sweagent}.
Both protocols use compilation, correctness, and performance feedback, while
ours fixes condition-specific staged files and one evaluator.
In \emph{single-shot} runs, agents can inspect and edit staged files but
cannot compile or execute candidates.
\emph{Cold iterative} starts from the context's blank task stub and receives
at most 20 evaluator calls. \emph{Warm iterative} starts from an eligible
single-shot kernel, retains that source as checkpoint~0, and
receives the same feedback budget. 
Under feedback, the harness selects the development-correct candidate with
the highest co-measured baseline ratio.
After development, we grade the selected candidate on five unseen seeds, with
inputs and results withheld from the agent. Every task-specific
correctness criterion must hold on every seed.

\paragraph{Numerical contracts.}
Single-GPU GEMM and GQA candidates take BF16 inputs and write FP32 output,
whereas their \texttt{torch.mm} and SDPA baselines return BF16 output.
Programming-context comparisons keep the candidate output contract fixed.
Vendor-relative percentages are reference throughput ratios, not
precision-matched efficiency measurements.

\paragraph{Performance reporting.}
Only protocol-admissible, held-out-correct kernels receive performance scores.
\texttt{peak@3} selects the fastest absolute-throughput result among the
three starts. For Claude Code, single-GPU single-shot and cold-iterative
throughput uses a fixed 197.5-TFLOP/s oneDNN baseline reached through
\texttt{torch.mm} on Intel XPU. Copilot results use each trial's co-measured
oneDNN baseline and report the percentage belonging to the selected endpoint.
Warm-iterative results use oneDNN co-measured with checkpoint~0 and the final
kernel on the assigned GPU.
Distributed throughput uses our frozen, co-measured SyclKittens hand reference,
which gathers the input with a topology-aware schedule and computes the full
output on every GPU. Claude Code GQA uses native PyTorch XPU SDPA with
\texttt{enable\_gqa=True} and \texttt{is\_causal=True}, co-measured on the
same GPU and trial without importing IPEX. 
Figure~\ref{fig:progression}(b) instead reports median best-so-far trajectories
against co-measured oneDNN. Table~\ref{tab:agent-gemm} reports
endpoints from correct kernels under the baseline rules above, so the values
differ.
\Appxref{app:agent-study}{Controlled Coding-Agent Studies} reports exact
correctness gates, distributions, and protocol deviations.

\begin{table*}[t]
\caption{Controlled BF16 GEMM with Claude Code on Intel Max GPUs, with $4096^3$ on one GPU and
$6144^3$ on 12 GPUs. Single-shot and cold-iterative cells
report held-out-correct trials / 3 and \texttt{peak@3}.
Warm-iterative cells report held-out-correct finals / eligible sources
and the fastest correct final. Warm rows here report Opus. The single-GPU
denominator is oneDNN. The distributed denominator is our frozen, co-measured
topology-aware input-gather SyclKittens hand reference. Programming contexts
are defined in \S\ref{sec:ai:protocol}. Single-GPU paired details are in
\appxref{app:agent-single-stack-endpoint}{Controlled Coding-Agent Studies}, and
the distributed all-model study is in
\appxref{app:agent-study-ag-gemm}{Controlled Coding-Agent Studies}. A dash means
no correct or applicable cell.}
\label{tab:agent-gemm}
\centering
\footnotesize
\setlength{\tabcolsep}{2.5pt}
\renewcommand{\arraystretch}{1.12}
\begin{tabular*}{\textwidth}{@{\extracolsep{\fill}}llcccccc@{}}
\toprule
\multirow{2}{*}{\textbf{Protocol}} & \multirow{2}{*}{\textbf{Model}} &
\multicolumn{3}{c}{\textbf{Single-GPU GEMM (\% oneDNN)}} &
\multicolumn{3}{c}{\textbf{Distributed GEMM (\% SK ref.)}} \\
\cmidrule(lr){3-5}\cmidrule(lr){6-8}
 & & \textbf{Raw} & \textbf{\mk{}} & \textbf{+ examples} &
\textbf{Raw} & \textbf{\mk{}} & \textbf{+ examples} \\
\midrule
\multirow{3}{*}{Single-shot}
 & Opus   & 1/3; 6.6 & 3/3; 63.4 & 3/3; 81.2 &
			  0/3; --  & 2/3; 36.7 & 3/3; 59.4 \\
 & Sonnet & 0/3; --  & 3/3; 53.4 & 3/3; 31.0 &
			  3/3; 5.7 & 3/3; 5.4 & 3/3; 38.8 \\
 & Haiku  & 2/3; 0.8 & 2/3; 47.5 & 0/3; -- &
			  2/3; 0.5 & 1/3; 0.5 & 1/3; 0.5 \\
\midrule
\multirow{3}{*}{Cold iterative}
 & Opus   & 3/3; 57.5 & 3/3; 82.1 & 3/3; 82.6 &
			  3/3; 42.3 & 3/3; 46.8 & 3/3; 44.0 \\
 & Sonnet & 3/3; 5.7 & 3/3; 85.2 & 3/3; 84.8 &
			  3/3; 25.8 & 3/3; 50.9 & 3/3; 41.1 \\
 & Haiku  & 3/3; 1.6 & 3/3; 0.6 & 3/3; 50.3 &
			  3/3; 17.2 & 3/3; 0.7 & 2/3; 1.5 \\
\midrule
Warm iterative
 & Opus & 1/1; 14.2 & 3/3; 87.8 & 3/3; 87.7 &
			  -- & 2/2; 106.6 & 3/3; 61.5 \\
\bottomrule
\end{tabular*}
\end{table*}

\begin{table*}[t]
\centering
\caption{Controlled cold-iterative GEMM through GitHub Copilot. Cells give
admissible held-out-correct trials / 3; median / \texttt{peak@3} performance
in percent. Single-GPU percentages use each trial's co-measured oneDNN.
Distributed percentages use the node-matched frozen SK reference.
\texttt{peak@3} selects the fastest correct absolute-throughput endpoint;
the reported percentage belongs to that endpoint.}
\label{tab:agent-newer-gemm}
\small
\setlength{\tabcolsep}{7pt}
\begin{tabular}{@{}llcc@{}}
\toprule
\textbf{Task} & \textbf{Model} & \textbf{Raw} & \textbf{\mk{}} \\
\midrule
Single-GPU GEMM & Opus 5 & 1/3; 84.5 / 84.5 & 3/3; 84.7 / 94.2 \\
 & GPT-5.6-Sol & 3/3; 9.9 / 11.0 & 3/3; 85.2 / 87.9 \\
12-GPU all-gather GEMM & Opus 5 & 3/3; 39.1 / 48.4 & 3/3; 50.5 / 92.8 \\
 & GPT-5.6-Sol & 3/3; 8.8 / 8.9 & 3/3; 38.6 / 87.3 \\
\bottomrule
\end{tabular}
\end{table*}

\begin{table*}[t]
\caption{Controlled cold-iterative causal GQA at
$B{=}1$, $H{=}32$, $H_{KV}{=}8$, $N{=}2048$, and $D{=}128$ on one
Intel Max GPU. Cells report admissible held-out-correct trials / 3;
median / \texttt{peak@3} throughput in TFLOP/s among correct trials.
Controller groups follow the experimental design in \S\ref{sec:ai:protocol}.
Normalized, replicate-level, and single-shot results appear in
\appxref{app:agent-study-gqa}{Controlled Coding-Agent Studies}.}
\label{tab:agent-gqa}
\centering
\footnotesize
\setlength{\tabcolsep}{6pt}
\begin{tabular}{@{}llccc@{}}
\toprule
\textbf{Controller} & \textbf{Model} & \textbf{Raw} & \textbf{\mk{}} & \textbf{+ examples} \\
\midrule
Claude Code & Haiku 4.5 & 3/3; 0.45 / 0.56 & 3/3; 0.29 / 0.73 & 3/3; 0.40 / 1.00 \\
 & Sonnet 5 & 3/3; 14.47 / 15.36 & 3/3; 3.33 / 3.48 & 3/3; 3.41 / 6.82 \\
 & Opus 4.8 & 3/3; 26.38 / 42.30 & 3/3; 24.84 / 34.41 & 3/3; 38.04 / 59.95 \\
\midrule
GitHub Copilot & Opus 5 & 3/3; 45.99 / 57.11 & 2/3; 49.78 / 50.54 & 3/3; 70.54 / 75.55 \\
 & GPT-5.6-Sol & 3/3; 4.80 / 8.95 & 3/3; 24.08 / 24.57 & 3/3; 30.37 / 44.52 \\
\bottomrule
\end{tabular}
\end{table*}

\subsection{Results}
\label{sec:ai:results}

\paragraph{Encoded methods improve GEMM performance.}
In cold iteration, all nine Opus single-GPU trials are correct, and
\texttt{peak@3} rises from $57.5\%$ with Raw to $82.1\%$ with \mk{} and
$82.6\%$ with \mk{} + examples. Sonnet also improves from $5.7\%$ with Raw
to $85.2\%$ with \mk{}. The tile programming model supplies XMX computation,
operand layout, and data movement as operations that agents can compose while
tuning schedules. Without feedback, \mk{} raises Opus correctness from 1/3 to
3/3 and
\texttt{peak@3} from $6.6\%$ to $63.4\%$ on one GPU, while raising distributed
GEMM correctness from 0/3 to 2/3.

\label{sec:ai:newer}
The gains also hold for Opus~5 and GPT-5.6-Sol
(Table~\ref{tab:agent-newer-gemm}). On single-GPU GEMM, Opus~5 rises from
84.5\% to 94.2\% of oneDNN, and GPT-5.6-Sol rises from 11.0\% to 87.9\%.
On distributed GEMM, their endpoints rise from 48.4\% to 92.8\% and from
8.9\% to 87.3\% of the SK reference. Opus~5 produces one admissible correct
Raw single-GPU kernel and three with \mk{}; the other two Raw starts
violate the model-tool policy.

\paragraph{Context effects vary across models and tasks.}
All 27 Claude Code cold-iterative GQA trials are correct, but their
\texttt{peak@3} ordering changes by model. Examples raise Opus~4.8 from
Raw's 42.30 to 59.95~TFLOP/s, whereas Raw leads for Sonnet at 15.36,
above examples at 6.82 and \mk{} at 3.48~TFLOP/s. Haiku remains at or
below 1.00~TFLOP/s. Sonnet's reversal from GEMM to GQA shows that the
context's benefit also depends on the task. Examples also lead for Opus~5
and GPT-5.6-Sol, reaching 75.55 and 44.52~TFLOP/s, respectively, whereas
\mk{} without examples does not beat Raw's Opus~5 peak.
Every selected Opus~4.8 GQA kernel
uses XMX through direct subgroup DPAS, so its gain from examples reflects
schedule search rather than access to a different low-level mechanism.
\Appxref{app:agent-study-gqa}{Controlled Coding-Agent Studies} reports
the per-trial results.

\paragraph{Example kernels help in some settings.}
Without feedback, examples raise Opus \texttt{peak@3} from $63.4\%$ to $81.2\%$
on one GPU and from $36.7\%$ to $59.4\%$ on distributed GEMM, where correctness
also rises from 2/3 to 3/3. In cold iteration, Opus reaches about $82\%$
on single-GPU GEMM in both \mk{} contexts, and Sonnet and Haiku vary by task
and protocol. The examples' measured benefit therefore depends on the model,
task, and starting condition.

\paragraph{Agents consistently refine correct kernels.}
Warm iterative refinement retains the original correct kernel while
feedback improves its schedule. All seven eligible Opus single-GPU GEMM sources
and all six Opus GQA sources remain correct and improve, reaching $87.8\%$ and
$83.4\%$ of their respective baselines. All thirteen eligible distributed
sources across Opus, Sonnet, and Haiku in the two \mk{} contexts also remain
correct and improve. The strongest reaches $106.6\%$ of our co-measured
reference by computing local output-row shards before gathering the output,
avoiding redundant full-GEMM work on every GPU
(\appxref{app:agent-study-ag-gemm}{Controlled Coding-Agent Studies}).

\paragraph{Source inspection identifies implementation choices.}
As \S\ref{sec:raw-sycl-motivation} shows, selected Raw kernels either bypass
XMX or fail to coordinate the full operand pipeline. Every context exposes the
same SPIR-V and XMX instructions, but the strongest no-feedback Opus kernel in
\mk{} + examples combines \texttt{mma\_AB}, 2-D loads, VNNI packing, a
$32{\times}32$ register base shape, and three-tile prefetch, while the correct
single-shot Raw kernel only uses vector FMA. The \mk{} interface supplies the
matrix and movement operations, while the generated kernel chooses the tile shape
and prefetch distance. Full source inspection
appears in \appxref{app:agent-study}{Controlled Coding-Agent Studies}.

Source inspection shows how the contexts influence operand delivery.
The fastest Opus~5 SK-context kernel uses direct intrinsics with guarded,
disjoint prefetch ownership, whereas its Raw counterpart repeats operand
requests. The fastest GPT SK-context kernel uses \texttt{mma\_AB}, packed
2-D loads, and prefetch, while its Raw counterparts explicitly pack scalar
loads without prefetch. The SK context therefore supports both composition
through its API and direct use of the methods it encodes.

\paragraph{Encoded methods reach high throughput in fewer calls.}
The median Opus~4.8 \mk{} trajectory in Figure~\ref{fig:progression}(b)
passes Raw's final level on the second evaluator call and reaches 80\% of
co-measured oneDNN at call four.
Across all three single-GPU starts, the median calls to 80\% of co-measured
oneDNN are five for Opus~5 with SK and four for GPT with SK. Raw reaches
that threshold at a median of ten calls for Opus~5; fewer than two GPT Raw
starts reach it within the budget. Search progress uses development-correct
checkpoints across all starts; endpoint performance uses held-out grading.
\Appxref{app:agent-newer}{GitHub Copilot Protocol and Results} reports the
individual traces, threshold computation, numerical sensitivity, and
controller-time and token measurements.

\begin{figure*}[t]
\centering
\input{figures/fig1_tasks}
\caption{\textbf{Selected implementations from different development settings.}
Bars compare selected correct Opus kernels with engineer-agent kernels in
TFLOP/s. Annotations identify cold or warm iteration, remeasurement, and
co-design. Raw single-GPU GEMM
uses \texttt{peak@3}; the two \mk{} contexts use remeasured kernels from warm
iteration. GQA reports \texttt{peak@3} after remeasuring all kernels with the
same harness. Distributed GEMM uses a single-node Raw measurement and
node-aggregated measurements for the other kernels. Programming
contexts are defined in \S\ref{sec:ai:protocol}. These bars are separate from
the controlled endpoint summaries in Tables~\ref{tab:agent-gemm}
and~\ref{tab:agent-gqa}.}
\Description{A grouped bar chart with three task groups for single-GPU GEMM,
causal GQA, and distributed GEMM. Each group has four TFLOP/s bars for Raw
SYCL, SyclKittens, SyclKittens plus examples, and engineers plus agents. The
bars also identify cold and warm iteration, remeasured kernels, and co-design.
The programming contexts therefore do not have a common development setting.
The single-GPU GEMM values are 113.6, 163.1, 168.3, and 187.6 TFLOP/s. The causal
GQA values are 43.1, 71.1, 60.2, and 77.3 TFLOP/s. The distributed GEMM values
are 49.8, 123.2, 123.8, and 125.7 TFLOP/s.}
\label{fig:tasks}
\end{figure*}
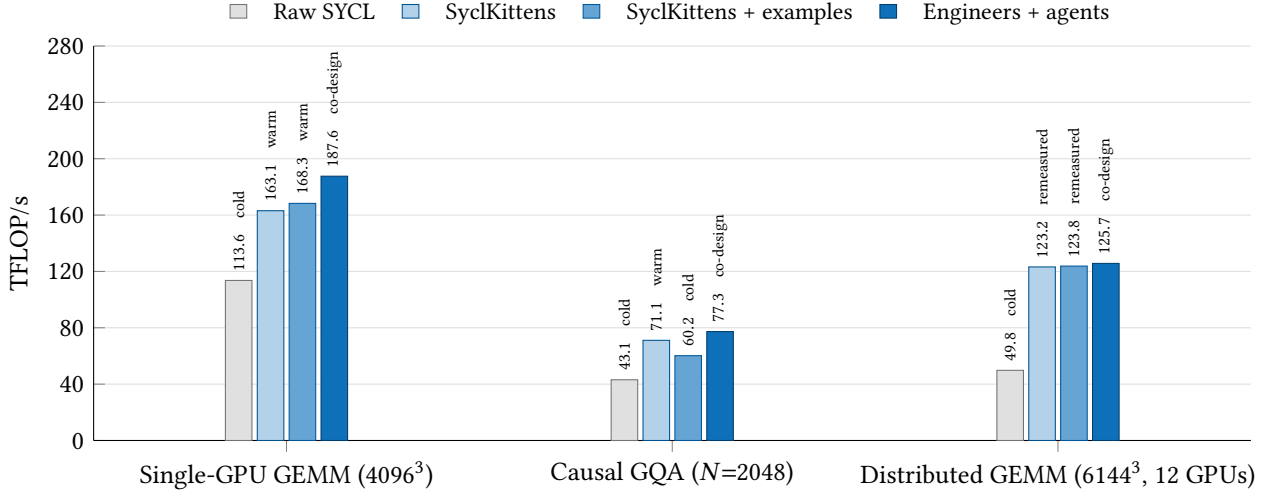

% \subsection{Engineer-Agent Kernel Co-Design}
% \label{sec:ai:handoff}

% The inference kernel suite is built through engineer-agent co-design, distinct
% from the autonomous trials above. Using the programming model developed with
% Intel hardware experts, engineers and coding agents jointly refine tile shapes,
% prefetch ownership and distance, and work distribution under correctness and
% performance checks set by engineers. Opus's warm-refined single-GPU GEMM reaches
% $87.8\%$ of co-measured oneDNN at $4096^3$. The co-designed suite reaches
% $96.3\%$ in geometric mean across 27 GEMM shapes. \S\ref{sec:kernels}
% evaluates this suite across inference operations.

%% file: figures/fig1_tasks.tex
% Figure 3: Opus 4.8, selected correct kernels from controlled and co-developed
% campaigns. All bars report absolute throughput at the task shape shown.
% GEMM  Raw 113.577 (sgv1-prod-20260731d); MK 163.103 and donors 168.312
%       (sgv1-warm-c0c2-20260803a); engineers+agents 187.5915, the median of
%       six 4096^3 node medians (skautotuned_final_ab_20260802a, policy).
% GQA   Raw 43.14, MK 71.10, donors 60.16, and hand 77.31. All four kernels
%       were replayed on one harness and node (posthoc_gqa_hand).
% Dist  Raw 49.766 (ag_gemm_raw_randn_allocHJ31_20260831_020434); MK C0
%       123.207 (matched three-way study); donors C2 123.839 and policy 14
%       125.724 (ag_gemm_c2_final_compare_allocHJ31_c2_p14_final_20260831_010233).
%       Inputs use the common Box-Muller random-normal harness. The Raw result
%       is single-node; the other results are medians of node medians.
\definecolor{intelblue}{HTML}{0068B5}
\begin{tikzpicture}
\begin{axis}[
  width=0.95\textwidth, height=6.8cm, ybar, bar width=10pt,
  ymin=0, ymax=280, ytick={0,40,80,120,160,200,240,280},
  ylabel={TFLOP/s},
  symbolic x coords={gemm,gqa,dist}, xtick={gemm,gqa,dist},
  xticklabels={{Single-GPU GEMM ($4096^3$)},{Causal GQA ($N{=}2048$)},{Distributed GEMM ($6144^3$, 12 GPUs)}},
  enlarge x limits=0.25, ymajorgrids, grid style={gray!25}, axis lines*=left,
  point meta=explicit symbolic,
  nodes near coords={\pgfmathprintnumber[fixed,fixed zerofill,precision=1]{\pgfkeysvalueof{/data point/y}}\quad\pgfplotspointmeta},
  % Rotated, because adjacent bars in the distributed group differ by ~1% and
  % their horizontal labels overlap.
  nodes near coords style={font=\tiny, /pgf/number format/fixed,
    /pgf/number format/fixed zerofill, /pgf/number format/precision=1,
    rotate=90, anchor=west},
  legend style={font=\small, at={(0.5,1.03)}, anchor=south,
    legend columns=4, draw=none, fill=none, column sep=8pt},
  legend image code/.code={\draw[#1] (0cm,-0.07cm) rectangle (0.22cm,0.11cm);},
]
\addplot[draw=black!55, fill=black!12] coordinates {(gemm,113.6) [cold] (gqa,43.1) [cold] (dist,49.8) [cold]};
\addlegendentry{Raw SYCL}
\addplot[draw=intelblue!80!black, fill=intelblue!30] coordinates {(gemm,163.1) [warm] (gqa,71.1) [warm] (dist,123.2) [remeasured]};
\addlegendentry{\mk{}}
\addplot[draw=intelblue!85!black, fill=intelblue!60] coordinates {(gemm,168.3) [warm] (gqa,60.2) [cold] (dist,123.8) [remeasured]};
\addlegendentry{\mk{} + examples}
\addplot[draw=intelblue!60!black, fill=intelblue!95] coordinates {(gemm,187.6) [co-design] (gqa,77.3) [co-design] (dist,125.7) [co-design]};
\addlegendentry{Engineers + agents}
\end{axis}
\end{tikzpicture}

%% file: sections/results.tex
\section{\mk{} Inference Kernel Suite}
\label{sec:experiments}

Engineers and coding agents co-designed the \mk{} inference kernel suite.
Engineers defined correctness and performance checks, selected disjoint
prefetch regions, and aligned their phases. Agents integrated the primitives
and collaborated on tile shapes, prefetch distance, and work distribution,
using Intel VTune to guide optimization. This section evaluates the resulting
kernels against tuned vendor paths.

\subsection{Experimental Setup}
\label{sec:eval:setup}

PyTorch~2.10 with IPEX~2.10.10~\cite{ipex}
dispatches operator baselines to oneDNN~\cite{onednn,pytorch_xpu_mm_210}, fused
SDPA, or oneCCL. Our standalone decode GEMV baseline calls Intel's optimized
oneMKL math library directly through
\texttt{oneapi::mkl::blas::gemm}~\cite{intel_onemkl2025}. Fusion and end-to-end
results use \texttt{torch.compile}.
For the Triton comparison, we use identical shapes, correctness gates,
and measurement windows with Triton-XPU~3.6's standard GEMM implementation,
autotuned.

Every sweep warms up before timing and co-measures each \mk{} candidate
with its baseline in the same thermal window. We report geometric means of
per-shape speedups. \mk{} GEMM requires relative Frobenius error below
0.02. \Appxref{app:gemm}{BF16 GEMM} reports per-shape data and
additional correctness gates.
The GEMM comparisons interleave implementations after initialization, with
10 warmups and 30 timed operations per implementation in each trial.
Operands are reused without cache flushing. Compilation and configuration
selection precede timing.

\subsection{Kernel Results}
\label{sec:kernels}

\paragraph{Dense GEMM across shapes and nodes.}
\label{sec:kernels:gemm}
The GEMM of \S\ref{sec:sk:onestack} reaches $96.3\%$ of oneDNN in geometric
mean across the 27 shapes in Figure~\ref{fig:production-suite}(a).
Across six independently measured nodes
at $8192^3$, the per-node \mk{}/oneDNN relative performance
ranges from $95.5\%$ to $96.3\%$.
An interleaved six-node sweep finds that the final cooperative revision exceeds
autotuned Triton-XPU~3.6 on all 27 shapes, with $1.087\times$ geometric-mean
throughput. \Appxref{app:gemm}{BF16 GEMM} reports the direct comparison, exact
tables, Stream-K results, and ablations.

\paragraph{Comparison with SYCL-TLA.}
The cooperative \mk{} implementation also achieves throughput comparable
to SYCL-TLA, a native SYCL tile library. Across 27 shapes on three nodes,
\mk{} reaches $0.999\times$ oneDNN and $1.016\times$ SYCL-TLA in geometric
mean. On the largest square shapes, throughput reaches
$1.08$--$1.11\times$ SYCL-TLA.
\Appxref{app:gemm-tla}{SYCL-TLA Comparison} reports the configurations,
per-shape ratios, and confidence intervals.

\paragraph{Host submission and device utilization.}
Host submission explains the small-shape performance difference between
the \mk{} and \texttt{torch.mm} paths. At $1024^3$, \texttt{torch.mm}
submission costs about 34--39~\us{} versus 8--9~\us{} for SK, although
oneDNN's device kernel is faster. At $4096^3$, submission is hidden and
both implementations are device-bound. On three large shapes, SK reaches
80--96\% of the architectural single-stack matrix ceiling at each run's
measured loaded clock. \Appxref{app:gemm-diagnosis}{Host and Device
Performance} gives the timing decomposition and roofline analysis.

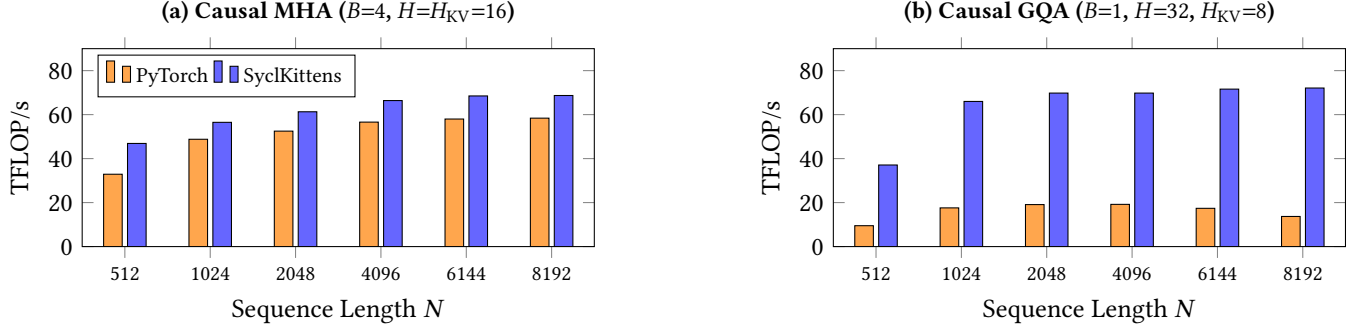
\begin{figure*}[t]
\centering
\begin{tikzpicture}
\begin{axis}[
    ybar,
    bar width=7pt,
  width=0.47\textwidth,
    height=4.2cm,
    ylabel={TFLOP/s},
    symbolic x coords={512,1024,2048,4096,6144,8192},
    xtick=data,
    x tick label style={font=\footnotesize},
  xlabel={Sequence Length $N$},
    ymin=0, ymax=90,
    legend style={at={(0.03,0.97)},anchor=north west,font=\footnotesize},
    legend columns=2,
    legend cell align={left},
    title style={font=\small\bfseries},
  title={(a) Causal MHA ($B{=}4$, $H{=}H_{\mathrm{KV}}{=}16$)},
]
\addplot[fill=orange!70] coordinates
  {(512,32.9) (1024,48.8) (2048,52.5) (4096,56.6) (6144,58.0) (8192,58.4)};
\addplot[fill=blue!60] coordinates
  {(512,46.9) (1024,56.5) (2048,61.3) (4096,66.4) (6144,68.5) (8192,68.7)};
\legend{PyTorch, \mk}
\end{axis}
\end{tikzpicture}
\hfill
\begin{tikzpicture}
\begin{axis}[
    ybar,
    bar width=7pt,
  width=0.47\textwidth,
    height=4.2cm,
    ylabel={TFLOP/s},
    symbolic x coords={512,1024,2048,4096,6144,8192},
    xtick=data,
    x tick label style={font=\footnotesize},
    xlabel={Sequence Length $N$},
    ymin=0, ymax=90,
    title style={font=\small\bfseries},
    title={(b) Causal GQA ($B{=}1$, $H{=}32$, $H_{\mathrm{KV}}{=}8$)},
]
\addplot[fill=orange!70] coordinates
  {(512,9.5) (1024,17.6) (2048,19.1) (4096,19.2) (6144,17.4) (8192,13.7)};
\addplot[fill=blue!60] coordinates
  {(512,37.1) (1024,66.0) (2048,69.8) (4096,69.8) (6144,71.6) (8192,72.1)};
\end{axis}
\end{tikzpicture}
\caption{\textbf{Forward causal attention on one Intel Max GPU.}
  (a) MHA reaches $1.21\times$ geometric mean over fused SDPA.
  (b) GQA reaches $4.0\times$ in geometric mean over the baseline's repeat-KV path, which
  dispatches GQA through MHA. Both panels use the same scale.}
\Description{Grouped bar charts of attention throughput in TFLOP/s against
sequence lengths from 512 to 8192, comparing SyclKittens with the fused SDPA
baseline.}
\label{fig:results-attn}
\end{figure*}

\paragraph{Attention reuses the HipKittens algorithm.}
\label{sec:kernels:gqa}
The attention kernel reuses the complete HipKittens dataflow algorithm
through the tile interface. For each query tile, the GQA
kernel retains HipKittens' sequence of tiled $QK^\top$, running-maximum and
normalizer updates, output rescaling, and $PV$ accumulation. The dataflow
therefore needs no algorithmic change. \mk{} re-implements its tile
operations with Intel's matrix, movement, and prefetch instructions.
Figure~\ref{fig:results-attn} reports the resulting MHA and GQA throughput. The
GQA baseline uses a repeat-KV fallback that expands eight KV heads to match
the 32 query heads and dispatches through its MHA path.

\paragraph{Decode-critical operators.}
The remaining \mk{} kernels outperform their vendor baselines on the operations that
dominate autoregressive decode. These operations either run briefly, so launch
overhead matters, or stream weights from memory.
Fused root mean square layer normalization
(RMSNorm)~\cite{zhang2019rmsnorm}, layer normalization
(LayerNorm)~\cite{ba2016layernorm}, and RoPE each use one launch.
Split-K GEMV matters most because
the five $M{=}1$ Llama projections account for about $64\%$ of per-token decode
latency. \mk{} maps each output-row reduction across cooperating subgroups so
the GPU stays occupied as tensor-parallel shards shrink. Figure~\ref{fig:production-suite}(b)
reports these gains before \S\ref{sec:e2e:decode} measures them inside a full
decode loop. \Appxrefs{}~\appxitem{app:norms}{Memory-Bound Kernels:
Norms, RoPE, and Decode GEMV} and \appxitem{app:op-fusions}{Op-Fusion
Microbenchmarks}\appxtail{} report the full grids.

\input{figures/fig_production_suite}

\input{sections/results_ext}

%% file: figures/fig_production_suite.tex
\begin{figure*}[t]
\centering
\begin{tikzpicture}
\begin{axis}[
    width=0.47\textwidth,
    height=4.8cm,
    xlabel={Canonical shape index},
    ylabel={\mk{} / oneDNN},
    xmin=0.5, xmax=27.5,
    ymin=0.84, ymax=1.18,
    xtick={1,5,9,13,17,21,25,27},
    ytick={0.85,0.90,0.95,1.00,1.05,1.10,1.15},
    tick label style={font=\footnotesize},
    grid=major,
    grid style={gray!20},
    title style={font=\small\bfseries},
    title={(a) BF16 GEMM},
]
\addplot[blue!70!black, thick, mark=*, mark size=1.7pt] coordinates {
  (1,1.165) (2,0.881) (3,0.999) (4,0.962) (5,0.940) (6,0.947)
  (7,0.944) (8,0.959) (9,0.989) (10,0.948) (11,0.96) (12,0.972)
  (13,0.964) (14,0.977) (15,0.960) (16,0.953) (17,0.924) (18,0.926)
  (19,0.925) (20,0.935) (21,0.884) (22,0.941) (23,0.909) (24,1.042)
  (25,0.962) (26,1.039) (27,0.976)
};
\addplot[black, dashed, thin, forget plot] coordinates {(0.5,1) (27.5,1)};
\draw[gray!65, dashed] (axis cs:13.5,0.84) -- (axis cs:13.5,1.18);
\node[font=\footnotesize, anchor=south] at (axis cs:7,0.845) {square};
\node[font=\footnotesize, anchor=south] at (axis cs:20.5,0.845) {rectangular};
\end{axis}
\end{tikzpicture}
\hfill
\begin{tikzpicture}
\begin{axis}[
    ybar,
    bar width=9pt,
    width=0.47\textwidth,
    height=4.8cm,
    ylabel={Speedup over baseline},
    symbolic x coords={RMSNorm,LayerNorm,RoPE,GEMV},
    xtick={RMSNorm,LayerNorm,RoPE,GEMV},
    tick label style={font=\footnotesize},
    ymin=0, ymax=2.9,
    grid=major,
    grid style={gray!20},
    legend style={at={(0.5,0.98)},anchor=north,legend columns=3,
      font=\footnotesize,draw=none},
    legend cell align={left},
    title style={font=\small\bfseries},
    title={(b) Decode-critical operators},
]
\addplot[fill=teal!45, bar shift=-5pt] coordinates
  {(RMSNorm,1.12) (LayerNorm,1.42) (RoPE,1.09)};
\addplot[fill=blue!60, bar shift=5pt] coordinates
  {(RMSNorm,2.44) (LayerNorm,2.49) (RoPE,1.98)};
\addplot[fill=orange!70, bar shift=0pt] coordinates
  {(GEMV,1.36)};
\draw[black, dashed, thin]
  (axis cs:RMSNorm,1) -- (axis cs:GEMV,1);
\legend{HBM-scale, Few-token, GEMV}
\end{axis}
\end{tikzpicture}
\caption{\textbf{\mk{} kernel suite.}
  (a) The \mk{} GEMM reaches $96.3\%$ of the
  co-measured tuned oneDNN library path in geometric mean across 27 GEMMs, exceeds it on
  three shapes, and peaks at $1.17\times$. Shape indices 1--13 are square
  and 14--27 rectangular.
  (b) Fused normalization and RoPE are near the \texttt{torch.compile}
  HBM-scale path but lead it by $1.98$--$2.49\times$ for few-token decode.
  GEMV has one bar because weight streaming is bandwidth-bound even in decode.
  \mk{}'s wide-load split-K GEMV reaches
  $1.36\times$ oneMKL in geometric mean and $1.54\times$ on $W_o$.}
\Description{Two bar charts of SyclKittens kernel performance relative to Intel
library baselines. Panel (a) plots the GEMM ratio to tuned oneDNN across 27
shape indices, square shapes first and rectangular shapes after. Panel (b) plots
fused normalization and RoPE against torch.compile, and GEMV against oneMKL.}
\label{fig:production-suite}
\end{figure*}
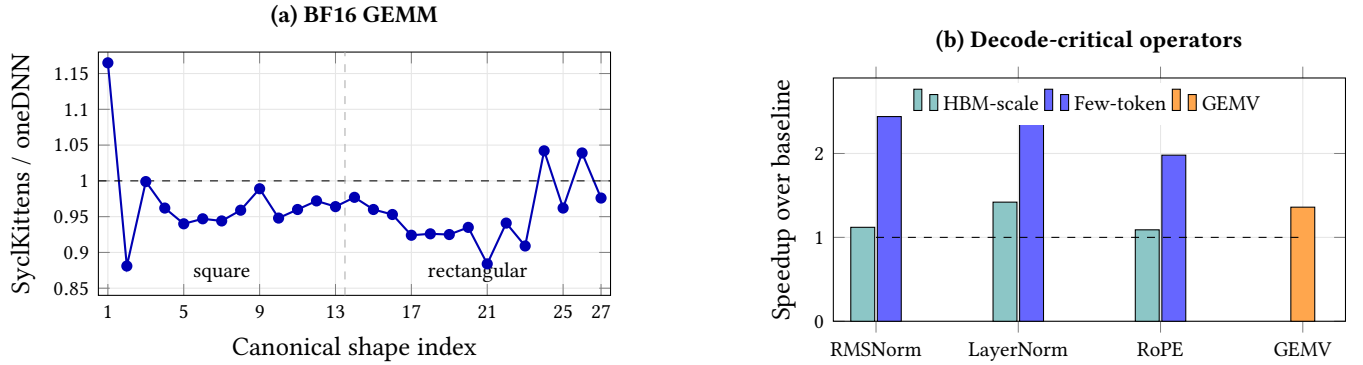

%% file: sections/e2e.tex
\section{End-to-End Inference}
\label{sec:e2e}

We measure how the kernel gains translate to Llama-3.1-8B
inference~\cite{grattafiori2024llama3} on one GPU and across a node.
Composed with \texttt{torch.compile}, \mk{} improves full-request throughput
by $1.59\times$ in geometric mean over \texttt{torch.compile} alone on the
single-GPU configurations. Across twelve GPUs,
\mk{} also serves a 256K context.
That memory-capacity result is separate
from the matched-GPU-count speedups below.

\subsection{Framework Composition on One GPU}
\label{sec:e2e:arch}

\paragraph{Operator ownership.}
\label{sec:e2e:policy}
We compare \mk{} composed with \texttt{torch.compile} with
\texttt{torch.compile} alone.
\texttt{torch.compile} generates Triton for the pointwise, standalone
normalization, and epilogue regions it fuses, but graph-breaks at attention on
XPU. Across this boundary, \mk{} supplies a FlashAttention-style
kernel~\cite{dao2022flashattention}, fused Q/K rotary, fused QKV projection,
and residual-plus-RMSNorm fusion. The kernels are exposed as
PyTorch custom operations and installed by replacing the Llama decoder layer
at runtime; HuggingFace source and \texttt{model.generate()} remain unchanged.

\begin{figure*}[t]
\centering
\begin{tikzpicture}
	\node[inner sep=0, outer sep=0] (prefill) {\includegraphics[width=0.485\textwidth]{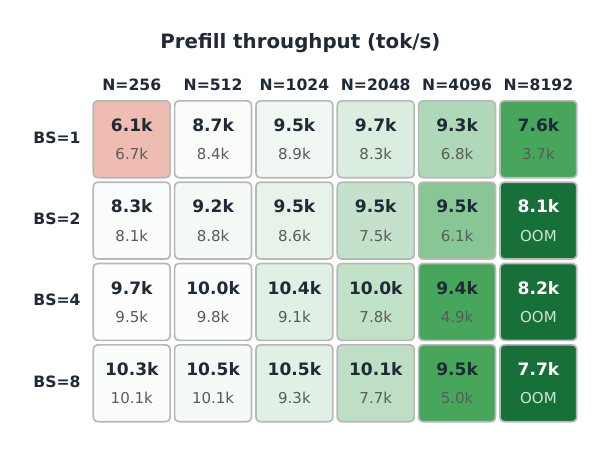}};
	\node[font=\bfseries, anchor=north west, xshift=1.5mm, yshift=-1.5mm]
		at (prefill.north west) {(a)};
\end{tikzpicture}
\hfill
\begin{tikzpicture}
	\node[inner sep=0, outer sep=0] (decode) {\includegraphics[width=0.485\textwidth]{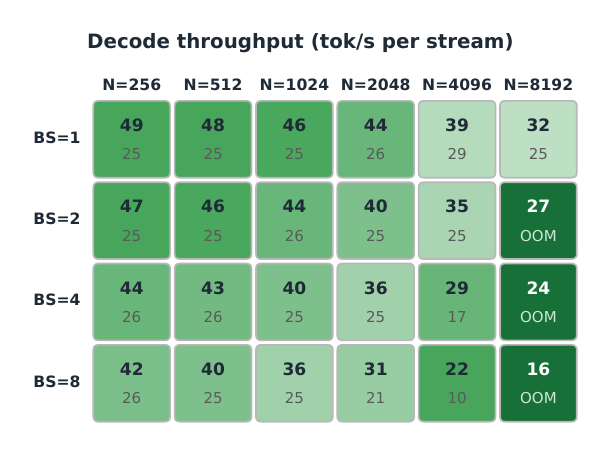}};
	\node[font=\bfseries, anchor=north west, xshift=1.5mm, yshift=-1.5mm]
		at (decode.north west) {(b)};
\end{tikzpicture}
\caption{\textbf{Llama-3.1-8B composition on one Intel Max GPU.}
	(a) Prefill and (b) decode throughput across batch size and prompt length.
	Each cell reports \mk{} composed with \texttt{torch.compile} (bold, top)
	and \texttt{torch.compile} alone (bottom); color encodes their ratio.
	Ratios use the 21 cells where both systems fit; in the three OOM cells the
	baseline runs out of memory while \mk{} serves the request.}
\Description{Two heatmaps over batch size and prompt length for Llama-3.1-8B on
a single GPU. Each cell gives throughput for SyclKittens composed with
torch.compile and for torch.compile alone, with color encoding the ratio between
them; panel (a) covers prefill and panel (b) covers decode.}
\label{fig:e2e-heatmaps}
\end{figure*}

\paragraph{Prefill and decode.}
\label{sec:e2e:decode}
Figure~\ref{fig:e2e-heatmaps} reports Llama-3.1-8B inference throughput over four
batch sizes and six context lengths. Each point times one synchronized,
deterministic \texttt{model.generate()} request with up to 128 new tokens.
The reported full-request rate counts batch-wide prompt tokens plus the
generated sequence length of one stream, as defined in
\appxref{app:e2e-composition}{End-to-End Inference}.
Geometric means use the 21
configurations in which both systems fit. The composed path improves prefill
by $1.22\times$, decode by $1.65\times$, and
full-request prefill-plus-decode throughput by $1.59\times$
(range $1.29$--$1.98\times$). Prefill gains grow with context
because the \mk{} FlashAttention-style kernel avoids the baseline's quadratic score
materialization; decode gains come from eliminating launches around that
attention boundary and from the projection path in \S\ref{sec:kernels}.
Neither system uses SYCL Graph, Intel's equivalent of CUDA graph capture,
in these measurements.

\input{sections/e2e_ext}

%% file: sections/multigpu.tex
\subsection{Node-Scale Composition}
\label{sec:multigpu}

At node scale, decode performance depends on communication latency. In the evaluated
batch-one decode, each tensor-parallel all-reduce carries one 4096-element
BF16 hidden vector (8~KiB), so launch, process, and synchronization latency
matter more than peak bulk bandwidth. \mk{} keeps all GPUs in one process and
one shared SYCL context; a host thread launches each GPU's stream, and inline
peer movement with system-scope device barriers keeps Python and cross-process
dispatch off the per-token path.

\input{figures/fig_node_scale}

\paragraph{Inference-scale collectives.}
\label{sec:multigpu:fused}
At 24 KiB, \mk{} outperforms oneCCL on all four collectives by
$1.81$--$3.31\times$. Reduce-scatter reaches $2.20\times$ at 0.19 MiB and remains
$1.10\times$ faster at 96 MiB. Destination-split all-to-all reaches
$3.49\times$ at 1.5 MiB and remains $1.20\times$ faster at 96 MiB. At that largest
payload, oneCCL leads all-reduce and all-gather, where \mk{} reaches
$0.94\times$ and $0.92\times$ of oneCCL.

\paragraph{Context-parallel attention.}
\label{sec:multigpu:ring}
Long contexts also require splitting attention across GPUs. When the full KV cache
fits in each GPU's HBM, \mk{} gathers it once and executes one flash pass.
When it does not, a ring rotates one KV shard per step and merges the online
softmax state. Copy-engine overlap leads the blocking PyTorch
\texttt{sendrecv} ring at every context we measure, by $1.21$--$2.46\times$
(Figure~\ref{fig:node-scale}(a)). At 49K total tokens, the overlapped
non-causal ring reaches 60.7~TFLOP/s per GPU, versus 42.0 for the blocking
baseline. All-gather exceeds the sequential ring across the
measured context sweep when the gathered KV cache fits in HBM; the ring
provides the memory-scaling policy beyond that point.

\subsection{Multi-GPU Decode}
\label{sec:e2e:mgpu-decode}

The Llama decoder composes tensor-parallel projections and inline
all-reduces with context-parallel attention. Figure~\ref{fig:node-scale}(b)
shows that, at matched GPU count, the \mk{} decode path outperforms the
oneCCL-based path at every context, by up to $2.91\times$ at 32K on four GPUs. This
comparison includes both the source-partitioned collective and \mk{}'s
single-process execution model. The standalone results above measure complete
collective calls without the surrounding decoder, while retaining each
implementation's process model.

The best layout changes with context. Flat four-way tensor parallelism wins
through 32K. At 64K and 128K, $\mathrm{TP}2{\times}\mathrm{CP}3$ uses six
GPUs to shard the growing KV scan, and at 256K,
$\mathrm{TP}4{\times}\mathrm{CP}3$ uses all twelve GPUs and serves at
73.3~ms/token.
The last point is a capacity result, since the two-GPU tuned oneCCL path is
out of memory there and the \mk{} configuration uses more GPUs.
\Appxrefs{}~\appxitem{app:cp-attn}{Context-Parallel
Attention (Ring and All-Gather)}, \appxitem{app:collectives}{Multi-GPU
Collectives}, and \appxitem{app:e2e}{End-to-End
Llama-3.1-8B}\appxtail{} provide per-shape tables, integration and correctness
details, per-layer collective costs, the ring-attention step breakdown,
timing-tail summaries, and layout sweeps.

%% file: figures/fig_node_scale.tex
\begin{figure*}[t]
\centering
\begin{tikzpicture}
\begin{axis}[
    width=0.47\textwidth,
    height=5.0cm,
    xmode=log,
    log basis x=2,
    xlabel={Total context length},
    ylabel={TFLOP/s per GPU},
    xmin=1300, xmax=110000,
    ymin=0, ymax=75,
    xtick={1536,6144,12288,24576,49152,98304},
    xticklabels={1.5K,6K,12K,24K,49K,98K},
    tick label style={font=\footnotesize},
    grid=major,
    grid style={gray!20},
    legend style={at={(0.5,-0.28)},anchor=north,legend columns=3,
      font=\footnotesize,draw=none,
      /tikz/every even column/.append style={column sep=5pt}},
    legend cell align={left},
    title style={font=\small\bfseries},
    title={(a) Six-package ring attention},
    every axis plot/.append style={thick},
]
\addplot[orange!75!black, mark=square*] coordinates
  {(1536,0.9) (6144,6.5) (12288,12.9) (24576,24.6)
   (49152,42.0) (98304,56.9)};
\addplot[blue!55!black, dashed, mark=triangle*] coordinates
  {(1536,1.6) (6144,11.1) (12288,20.6) (24576,34.4)
   (49152,47.7) (98304,59.6)};
\addplot[blue!75!black, mark=*] coordinates
  {(1536,1.7) (6144,15.2) (12288,31.7) (24576,53.6)
   (49152,60.7) (98304,69.0)};
\legend{PyTorch, \mk{} sequential, \mk{} overlap}
\end{axis}
\end{tikzpicture}
\hfill
\begin{tikzpicture}
\begin{axis}[
    width=0.47\textwidth,
    height=5.0cm,
    xmode=log,
    log basis x=2,
    xlabel={Context length},
    ylabel={TPOT (ms/token)},
    xtick={1024,4096,16384,65536,262144},
    xticklabels={1K,4K,16K,64K,256K},
    tick label style={font=\footnotesize},
    ymin=0, ymax=430,
    grid=major,
    grid style={gray!20},
    legend style={at={(0.03,0.97)},anchor=north west,font=\footnotesize,
      draw=none},
    legend cell align={left},
    title style={font=\small\bfseries},
    title={(b) Llama-3.1-8B decode},
    every axis plot/.append style={thick},
]
\addplot[orange!75!black, mark=diamond*] coordinates
  {(1024,40.7) (2048,40.5) (4096,40.7) (8192,41.8) (16384,49.3)
   (32768,73.7) (65536,128.1) (131072,222.3) (262144,413.0)};
\addplot[blue!65!black, dashed, mark=triangle*] coordinates
  {(1024,19.2) (2048,20.3) (4096,21.7) (8192,22.4) (16384,23.1)
   (32768,25.3) (65536,54.5) (131072,95.6) (262144,271.9)};
\addplot[green!50!black, mark=*] coordinates
  {(1024,19.2) (2048,20.3) (4096,21.7) (8192,22.4) (16384,23.1)
   (32768,25.3) (65536,31.0) (131072,45.4) (262144,73.3)};
\legend{oneCCL $ws{=}4$, \mk{} $P{=}4$, \mk{} best layout}
\end{axis}
\end{tikzpicture}
\caption{\textbf{Node-scale composition on Intel Max GPUs.}
  (a) Non-causal BF16 ring attention uses six packages with one active stack
  each, comparing copy-engine overlap with sequential \mk{} and blocking
  PyTorch schedules.
  (b) Multi-GPU decode time per output token; lower is better. The orange
  and dashed-blue curves compare matched four-GPU paths, where \mk{} is
  $2.91\times$ faster at 32K. The green curve uses four GPUs through 32K,
  six at 64K and 128K, and twelve at 256K. Its longer-context points
  demonstrate the best measured layouts, not matched-resource speedups.}
\Description{Two line charts of node-scale performance. Panel (a) plots
per-GPU ring-attention throughput in TFLOP/s against token count for the
overlapped, sequential, and blocking PyTorch schedules. Panel (b) plots decode
time per output token against context length for matched four-GPU SyclKittens
and baseline paths, together with a best-layout curve that uses four GPUs
through 32K, six at 64K and 128K, and twelve at 256K.}
\label{fig:node-scale}
\end{figure*}
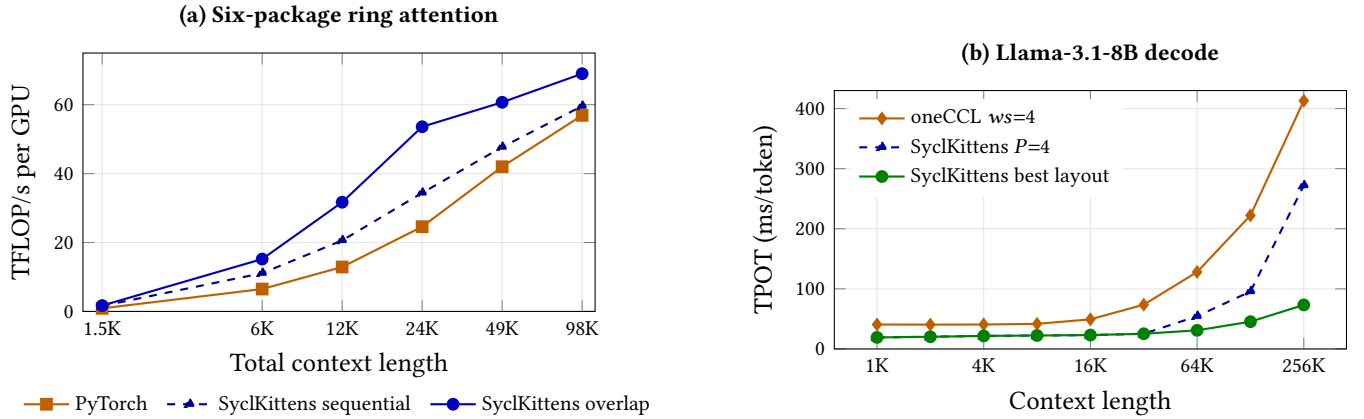

%% file: sections/conclusion.tex
\section{Conclusion}
\label{sec:conclusion}

\mk{} is a tile programming model for Intel GPUs that provides
matrix operand layouts, cooperative cache prefetch, and topology-aware GPU
communication as reusable operations. Controlled experiments show better
single-GPU GEMM endpoints with SK for Opus and GPT under fixed task and
feedback budgets.
Search trajectories also show earlier attainment of high GEMM throughput,
while the GQA results demonstrate that context benefits depend on the
model and task.
Through engineer-agent co-design, \mk{} also supports a kernel
suite that approaches oneDNN on GEMM and outperforms SDPA with IPEX on causal
MHA. In Llama-3.1-8B inference, it outperforms \texttt{torch.compile} on one
GPU and the oneCCL-based decode path on multiple GPUs.

Further evaluation can examine expert development effort and the transfer
of these methods to Intel architectures with different memory and
interconnect systems, such as Crescent Island~\cite{mohan2026cri}.

As coding agents take on kernel development for new accelerators, our results
suggest that the programming interface should carry each architecture's
methods. Agents then start from known-good building blocks and spend their
search on workload-specific policies.

%% file: sections/supp/appx-hardware.tex
\section{Hardware and Programming-Interface Map}
\label{app:hardware-map}

This section defines the terminology and programming interfaces used in the
main paper's \emph{Background and Motivation} and \emph{SyclKittens} sections.
The cross-vendor table aligns functions; implementation costs remain
platform-specific. A node contains six GPU packages,
each with two stacks connected by MDFI, the load/store stack-to-stack
interconnect. XeLink connects the GPU packages all-to-all through two
independent stack planes, even and odd.

\subsection{Measured Fabric Behavior}

We characterize the two communication tiers on the same 12-stack Intel Max GPU
node used in the main evaluation. Direct peer kernels issue contiguous remote
loads or stores through pointers in one shared SYCL context. The collective
measurements use five warmups followed by 15 timed iterations and report the
median; the two-stack all-reduce bandwidth uses
$2(N-1)S/(Nt)$, while distributed-GEMM speedups use raw elapsed-time ratios.
These conventions are reported separately because they are not interchangeable
estimates of a physical link rate.

\begin{table}[t]
\centering
\caption{Measured Intel Max GPU fabric behavior. XeLink rows report direct
large-transfer peer-kernel bandwidth per link. MDFI reports delivered
two-stack all-reduce bus bandwidth at a 96-MB payload. The last two rows report
end-to-end distributed-GEMM speedups with topology-aware XeLink-once plus MDFI
replication over the flat gather pipeline on 12 stacks.}
\label{tab:intel-fabric-measured}
\small
\setlength{\tabcolsep}{4pt}
\begin{tabularx}{\columnwidth}{@{}>{\raggedright\arraybackslash}Xr@{}}
\toprule
\textbf{Measurement} & \textbf{Result} \\
\midrule
XeLink peer write & 18.8 GB/s/link \\
XeLink peer read & 15.7 GB/s/link \\
MDFI two-stack all-reduce & 189.0 GB/s \\
Topology-aware gather, $6144^3$ & $2.15\times$ \\
Topology-aware gather, $12288^3$ & $1.42\times$ \\
\bottomrule
\end{tabularx}
\end{table}

The topology-aware gather sends each source stripe over XeLink once to one
landing stack per remote GPU package, then replicates it to the partner stack
over MDFI. This halves XeLink egress relative to every stack fetching every
remote stripe. Same-plane transfers can occupy both XeLink planes
concurrently; cross-plane transfers additionally consume the six MDFI links,
which lowers the number of independent flows the node can sustain. Kernels can
issue peer loads and stores over both fabrics. For explicit copies, the Intel
Max GPU also exposes per-link copy engines; simply offloading a naive all-pairs
copy does not reproduce the bandwidth of a topology-aware schedule.

\begin{table*}[t]
\centering
\caption{Cross-vendor terminology used in the paper. A dash denotes that the
vendor does not expose a direct counterpart with the same semantics.}
\label{tab:vendor-terms}
\small
\setlength{\tabcolsep}{5pt}
\begin{tabularx}{\textwidth}{@{}lXXX@{}}
\toprule
\textbf{Concept} & \textbf{NVIDIA} & \textbf{AMD} & \textbf{Intel} \\
\midrule
SIMT execution group & warp & wavefront & subgroup \\
Cooperative thread group & thread block / CTA & workgroup & workgroup \\
Compute unit & streaming multiprocessor (SM) & compute unit (CU) & Xe core \\
Programmer-managed local memory & shared memory & LDS & shared local memory (SLM) \\
Matrix engine & Tensor Core / MMA & Matrix Core / MFMA & XMX / DPAS \\
Matrix operand reformat/load & \texttt{ldmatrix} & \texttt{ds\_read\_b64\_tr\_b16} & \texttt{LoadTransformINTEL} \\
Asynchronous movement used by tile kernels & global-to-shared \texttt{cp.async} & global-to-LDS pipeline & 2-D global-to-L1 prefetch \\
Inter-package scale-up network & NVLink / NVSwitch & Infinity Fabric / xGMI & XeLink \\
Hardware multicast collective & NVLS & -- & -- \\
Collective library & NCCL & RCCL & oneCCL \\
\bottomrule
\end{tabularx}
\end{table*}

\begin{table}[t]
\centering
\caption{Architectural facts that determine tile movement and scheduling.}
\label{tab:ampere-pvc-map}
\small
\setlength{\tabcolsep}{4pt}
% Weighted X columns, because plain "Xll" lets the data columns take their
% natural width and starves the label column. Weights must sum to 3.
\begin{tabularx}{\columnwidth}{@{}%
  >{\hsize=1.10\hsize\raggedright\arraybackslash}X%
  >{\hsize=0.95\hsize\raggedright\arraybackslash}X%
  >{\hsize=0.95\hsize\raggedright\arraybackslash}X@{}}
\toprule
 & \textbf{NVIDIA Ampere} & \textbf{Intel Max GPU} \\
\midrule
Execution width & 32 & 16 or 32 \\
Matrix width & warp-level & fixed SIMD16 \\
L1 capacity & 28--192 KB/SM & 64--512 KB/Xe core \\
Local memory & up to 164 KB/SM & up to 448 KB/Xe core \\
Preferred async path & global $\to$ shared & global $\to$ L1 \\
L2 capacity & 40 MB/GPU & 204 MB/stack \\
Matrix operand representation & warp-distributed fragments & SIMD-direction linear vectors \\
\bottomrule
\end{tabularx}
\end{table}

\begin{figure*}[t]
\centering
\input{figures/fig1_overview}
\caption{Capabilities, methods, and policy across three vendors. The right
column reports Opus~4.8's controlled cold-iteration BF16 GEMM on Intel Max GPU
($57.5\%$ of oneDNN with Raw and $82.1\%$ with \mk{}) and the separate
engineer-agent suite's roughly $96\%$ geometric mean across 27 GEMM shapes.
The tile abstractions are shared with ThunderKittens and HipKittens; only
the Intel column is contributed by this work. A new hardware target requires
validating the lowering and adapting workload policies to its execution and
memory system. HipKittens targets a single GPU, so its column excludes
scale-up interconnect.}
\Description{A three-level diagram. A shared tile interface bar spans NVIDIA,
AMD, and Intel columns. Below it, a capability row lists what each vendor's
silicon exposes, a method row lists the architecture-specific operand reformat,
latency-hiding route, and work schedule, and a policy row lists the
workload-specific choices left programmable. A right-hand column reports the
single-shape controlled BF16 GEMM results of 57.5 percent of oneDNN with raw
SYCL and 82.1 percent with SyclKittens. The roughly 96 percent result is the
engineer-agent suite's geometric mean across 27 shapes.}
\label{fig:portability}
\end{figure*}

\subsection{Intel Programming-Interface Taxonomy}

The interfaces available on the same hardware differ in how much schedule
structure they expose. SYCL \texttt{joint\_matrix} is a portable matrix API.
It leaves operand layout and lowering to the implementation, but does
not expose the complete 2-D movement and prefetch schedule used here. ESIMD
provides Intel-specific explicit-SIMD control and is the substrate for larger
template libraries such as XeTLA. Direct SPIR-V built-ins expose individual
XMX and 2-D block operations but leave their legal and profitable composition
to each kernel author.

\mk{} application kernels use a compact tile programming model whose backend lowers to
the same Intel SPIR-V operations. The backend defines the SIMD16/XMX contract,
physical register layout, VNNI operand transform, cache-based 2-D movement, and
partial-tile prefetch operations. Kernels choose prefetch ownership and distance.
The controlled study therefore evaluates a
stable tile programming model with an explicitly architecture-specific lowering.

%% file: figures/fig1_overview.tex
% Figure 1 -- capability / method / policy, and what each level buys an agent.
% Native TikZ (fonts controlled in LaTeX; no external image, no downscaling).
% Rendered as a figure* at \textwidth; all artwork text matches caption size.
\definecolor{intelblue}{HTML}{0068B5}
\begin{tikzpicture}[
  font=\small,
  cell/.style={rounded corners=1.5pt, align=center, text width=2.95cm,
               minimum height=11mm, inner sep=3pt, draw=gray!55, fill=white},
  icell/.style={cell, draw=intelblue, line width=0.9pt, fill=intelblue!12},
  band/.style={rounded corners=3pt, draw=none, fill=gray!8},
  lvl/.style={align=right, text width=2.8cm, font=\bfseries\small},
  sub/.style={align=right, text width=2.8cm, font=\itshape\footnotesize,
              text=black!65},
  res/.style={align=center, text width=2.9cm, font=\footnotesize},
  hd/.style={align=center, text width=3.0cm, font=\bfseries\small},
]
\def\xL{1.9}\def\xN{6.0}\def\xA{9.4}\def\xI{12.8}\def\xM{16.0}
\def\bandL{4.35}\def\bandR{14.45}

% band backdrops (drawn first, so they sit behind the cells)
\fill[band] (\bandL,-0.62) rectangle (\bandR,0.82);
\fill[intelblue!7, rounded corners=3pt] (\bandL,-2.42) rectangle (\bandR,-0.78);
\fill[band] (\bandL,-3.98) rectangle (\bandR,-2.98);

% column headers
\node[hd] at (\xN,1.95) {NVIDIA\\ThunderKittens};
\node[hd] at (\xA,1.95) {AMD\\HipKittens};
\node[hd, text=intelblue!85!black] at (\xI,1.95) {Intel\\\mk{}};
\node[hd, text width=2.9cm] at (\xM,1.95) {Measured\\BF16 GEMM};

% shared tile interface
\node[lvl] at (\xL,1.28) {Tile interface};
\node[sub] at (\xL,0.92) {shared};
\node[draw=gray!60, fill=gray!12, rounded corners=2pt, align=center,
      text width=9.7cm, minimum height=6.5mm] at ({(\bandL+\bandR)/2},1.15)
     {\texttt{rt / st / gl} $\cdot$ mma, load/store, reductions $\cdot$ warp / group / gang};

% capability
\node[lvl] at (\xL,0.28) {Capability};
\node[sub] at (\xL,-0.14) {what the silicon exposes};
\node[cell] at (\xN,0.1) {Tensor Core,\\shared memory,\\NVLink};
\node[cell] at (\xA,0.1) {MFMA,\\shared memory};
\node[icell] at (\xI,0.1) {XMX (DPAS),\\2-D block load, L1,\\MDFI and XeLink};
\node[res] at (\xM,0.1) {raw SYCL agent\\$57.5\%$ of oneDNN};

% method
\node[lvl] at (\xL,-1.42) {Method};
\node[sub] at (\xL,-1.84) {how to use it well};
\node[cell] at (\xN,-1.6) {\texttt{ldmatrix} reformat,\\shared-memory staging,\\warp specialization};
\node[cell] at (\xA,-1.6) {\texttt{ds\_read\_b64\_tr},\\shared-memory staging,\\ping-pong waves};
\node[icell] at (\xI,-1.6) {\texttt{LoadTransformINTEL},\\L1 cache prefetch,\\cooperative partial tiles};
\node[res] at (\xM,-1.6) {\mk{} agent\\$82.1\%$};

% policy
\node[lvl] at (\xL,-3.30) {Policy};
\node[sub] at (\xL,-3.70) {left programmable};
\node[draw=gray!60, fill=white, rounded corners=2pt, align=center,
      text width=9.7cm, minimum height=6.5mm] at ({(\bandL+\bandR)/2},-3.48)
     {partial-tile ownership $\cdot$ prefetch distance $\cdot$ shape-dependent schedule $\cdot$ stack and package placement};
\node[res] at (\xM,-3.48) {engineers and agents\\${\approx}96\%$\\27-shape geometric mean};

\end{tikzpicture}

%% file: sections/dsl_ext.tex
\subsection{Realization Details}
\label{sec:dsl}

Prior work describes the initial SYCL port of ThunderKittens~\cite{jiang2026iwocl}.
Our baseline for this port compiled and linked on \pvc{} but passed only
570 of 1032 unit tests (55\%). Six root causes account for every failure.
With these fixes, the library used by the measured kernels passed all 1032 unit tests.
We describe these fixes, the added operations and performance primitives, the prefetch
API behind \S\ref{sec:prefetch}, and the measured SLM limits that motivate the
L1 operand path.

\subsubsection{Critical Correctness Fixes}
\label{sec:dsl:fixes}

We identified six distinct root causes accounting for all 462 test
failures (Table~\ref{tab:six-fixes}). The largest groups are shared-to-register
layout, incomplete vector operations, and row reduction. The failures include
NVIDIA warp-32 and CUDA-shuffle assumptions that do not survive the move to
\pvc{}'s SIMD16 execution model, XMX packing errors, and incomplete conversion
or reduction paths. Each failure mode requires a separate fix.

\begin{table}[t]
\centering
\caption{Six distinct tile-language correctness fixes that brought \mk{} from
  55\% to 100\% unit-test pass. Counts in the tests-fixed column sum to 462,
  the gap between the 570/1032 initial-port baseline and the final 1032/1032.
  Fix~6 carries no count of its own because it enables the wider tile shapes the
  other fixes are tested on.}
\label{tab:six-fixes}
\footnotesize
\setlength{\tabcolsep}{3pt}
\begin{tabularx}{\columnwidth}{>{\raggedright\arraybackslash\hsize=.55\hsize}X>{\raggedright\arraybackslash\hsize=1.45\hsize}Xr}
\toprule
\textbf{Fix} & \textbf{Root cause} & \textbf{Tests} \\
\midrule
1. shared$\to$register & NVIDIA warp-32 register map applied to SIMD16; only 4 of 8 required packed elements written per lane & 192 \\
2. row reduction & Within-lane stub computed column sums; rewritten as XOR butterfly across the 16 SIMD lanes & 76 \\
3. vector inner-dim & Vector ops (align copy, naive$\leftrightarrow$align conversion, shared$\leftrightarrow$register, reductions) touched only \texttt{[pack\_idx][0]}; seven of eight inner elements never visited & 120 \\
4. XMX \texttt{hmma81616} & Per-element \texttt{bit\_cast} scattered the A-operand of the $8{\times}16{\times}16$ BF16 XMX variant; replaced by contiguous \texttt{reinterpret\_cast} and K-split accumulator chaining & 56 \\
5. BF16$\to$float load & Stray \texttt{continue} skipped the type conversion after the 2D block load on \texttt{rt\_16x32} & 18 \\
6. row-map lane overflow & \texttt{data\_iter*2} produced lane indices up to 30 for \texttt{rt\_16x32}; wrapped modularly (unlocks 16$\times$32 tiles) & n/a \\
\bottomrule
\end{tabularx}
\end{table}

Fix~4 also resolves an interaction with the VNNI hardware transform load
(\S\ref{sec:dsl:perf}). Incorrect \texttt{hmma81616} operand marshalling
made transform-loaded operands appear corrupt, prompting a temporary switch
from \texttt{LoadTransformINTEL} to \texttt{LoadINTEL}. Once contiguous
operand packing and K-split accumulator chaining corrected
\texttt{hmma81616}, restoring \texttt{LoadTransformINTEL} retained 1032/1032
correctness and removed software VNNI repacking.

\subsubsection{Performance Primitives and New Operations}
\label{sec:dsl:perf}

Three PVC-specific primitives affect the measured kernels.
\texttt{move\_local<T>} annotates SLM pointers with
\texttt{\_\_attribute\_\_((opencl\_local))} (\S\ref{sec:dsl:slm}). For \texttt{rt\_16x32} float tiles,
the 2D block-load path splits a \texttt{BlockCount=2} request into two
\texttt{BlockCount=1} loads to remain within \pvc{}'s 64-byte hardware load
width. The \texttt{\_\_spirv\_Subgroup2DBlockLoadTransformINTEL} path
performs the GEMM B-operand VNNI repack during the load. Combined with the
corrected \texttt{hmma81616}, restoring this transform raises single-GPU
$8192^3$ GEMM throughput from 106 to 147~TFLOP/s, a 39\% gain over the
regular-load path with software repacking.

The kernels also use three groups of tile-language operations.
\texttt{load\_part} and \texttt{store\_part} with a position parameter (the
split-D strategy that processes $D{=}128$ as two $D{=}64$ passes within
register budget); \texttt{make\_causal} with partial-tile support for
causal masking at non-aligned tile boundaries; and \texttt{mul::reduce}
and \texttt{min::reduce} completions, since upstream ThunderKittens
only exposed \texttt{sum} and \texttt{max} reductions.

\input{sections/prefetch}

\subsubsection{Measured Limits of SLM Operand Staging}
\label{sec:dsl:slm}

Two measured costs keep shared local memory (SLM) off the operand-staging
path on \pvc{}, namely narrow matrix-operand loads and
compiler overhead from generic address-space qualification.

\paragraph{A- and B-operand costs.}
\pvc{} has no SLM matrix load (\texttt{ldmatrix} equivalent). Loading
an XMX B-operand ($16{\times}16$ bf16) from SLM requires $\sim$64
individual reads vs.\ a single \texttt{LoadTransformINTEL} from HBM.
Multiple SLM-based attention-kernel attempts ran an order of magnitude
below the HBM-native path (the measured SLM B-operand variant reached
$2.3$--$3.0$~TFLOPS against tens of TFLOPS for the HBM path) because
that $64{:}1$ read penalty compounds with the compiler overhead below.

\paragraph{Address-space qualification matters.}
ThunderKittens' \texttt{st<>} shared-tile type stores SLM base pointers as
\texttt{uintptr\_t}, which Intel's IGC compiler treats as generic
address space. IGC emits approximately 1,400 runtime disambiguation IR
operations, adding more cost than the SLM staging removes. Typed
\texttt{local\_accessor<T>} or \texttt{multi\_ptr<T,local\_space>}
fixes this, and \mk{}'s \texttt{move\_local<T>} of \S\ref{sec:dsl:perf}
applies the equivalent annotation. Address-space annotations do not remove
the $64{:}1$ read-count difference above, which motivates the direct operand
path used by the measured kernels.

%% file: sections/prefetch.tex
\subsubsection{Prefetch API}
\label{sec:dsl:prefetch}

\mk{} separates cache warming from the register load that consumes a tile.
\begin{lstlisting}[basicstyle=\ttfamily\footnotesize,aboveskip=2pt,belowskip=2pt]
prefetch_load(tile, global, next);
prefetch_load_coop(tile, global, next, {row, col});
load(tile, global, next);
\end{lstlisting}
\texttt{prefetch\_load} emits Intel 2-D block-prefetch instructions for the
base regions implied by the register-tile shape. It warms L1 but does not
write the destination registers; the later \texttt{load} performs the
XMX-compatible 2-D block load. The prefetch is therefore a performance hint,
not an asynchronous copy whose completion is required for correctness.

\texttt{prefetch\_load\_coop} exposes ownership within a workgroup. One call
warms a single 16-column, 8-row, two-block region at the supplied raw row and
column offset. The GEMM schedule assigns disjoint offsets to its computing
subgroups, so every shared operand region is warmed once while all subgroups
continue to issue matrix work. The destination argument supplies tile shape
and coordinate types only; it is not modified.

\paragraph{Cache-based staging.}
This interface is distinct from two shared memory mechanisms often also called
prefetching. SLM staging copies global data into programmer-managed local
memory before a synchronized load, while SLM fetch-ahead combines a load,
local-memory store, and named-barrier handoff. The measured \pvc{} path uses
neither. Hardware 2-D prefetch places data in the configurable 512\,KB L1,
and each consumer later loads its own register tile. This cache-centric split
preserves ordinary load semantics while allowing the cooperative ownership
policy in \S\ref{sec:prefetch} to remove redundant warming.

%% file: sections/supp/appx-agent-study.tex
\section{Controlled Coding-Agent Studies}
\label{app:agent-study}

The controlled studies cover single-GPU BF16 GEMM, causal grouped-query
attention (GQA)~\cite{ainslie2023gqa}, and 12-GPU all-gather BF16 GEMM under
single-shot and independent cold-iterative feedback
protocols. Claude Code evaluates Opus~4.8, Sonnet~5, and Haiku~4.5 on all
three tasks in both protocols. GitHub Copilot evaluates Opus~5 and
GPT-5.6-Sol in cold iteration. Warm refinement uses Opus~4.8 on single-GPU GEMM and
GQA and all eligible model--context sources in the two \mk{} contexts on
distributed GEMM. They seed feedback with
held-out-correct single-shot sources, separating blank-start search from warm
refinement.
The task, programming context, evaluator, and budget were fixed before
measurement. The following subsections specify each controller's protocol
and task-level results.

\subsection{Claude Code Protocol}

Across all three tasks in Claude Code, each model--context pair receives three independent
starts for both single-shot and cold-iterative evaluation. The evaluated models
are Opus~4.8, Sonnet~5, and Haiku~4.5. Each task uses three nested programming
contexts. \textsc{Raw} provides
Intel SYCL/SPIR-V documentation but no \mk{}. The SyclKittens context adds its
headers and API documentation without a target-kernel implementation. That
documentation covers the whole interface, including
\texttt{prefetch\_load\_coop} and its region-offset argument, so every mechanism the
co-designed kernels use is available to the agent. The example
context is \mk{} + examples, whose task-specific contents are specified below. Every
condition permits direct SYCL/SPIR-V intrinsics and forbids oneAPI
\texttt{joint\_matrix}, its matrix headers, and its namespace.

The agent runtime is Claude Code~2.1.214. Our custom harness
shares ParallelKernelBench's edit--evaluate--refine
structure~\cite{chan2026parallelkernelbench,yang2024sweagent}.
ParallelKernelBench uses mini-SWE-agent, a repository-rooted shell, and a budget
of 20 model calls. Our harness
fixes condition-specific staged files and programming context and limits
feedback to 20 evaluator calls. After development, the selected candidate must
pass five unseen correctness seeds; no held-out input or result is exposed to
the agent.

Single-shot agents can read, search, and edit staged task files but have no
shell, compiler, evaluator, benchmark, network, or external repository access.
Cold-iterative agents additionally receive at most 20 calls to one fixed
compile--correctness--performance evaluator. Warm-iterative agents begin from
an eligible held-out-correct single-shot source as checkpoint~0. Replicates
start independently, no source transfers between trials, and the agent never
sees the five held-out seeds. Under feedback, the harness retains the
development-correct candidate with the highest co-measured baseline ratio.

\subsection{Single-GPU Task}

Every trial implements $M{=}N{=}K{=}4096$ BF16 GEMM with FP32 accumulation and output.
\mk{} + examples adds fixed ThunderKittens and HipKittens GEMM
implementations. All conditions target the same GPU, compiler toolchain,
output contract, and evaluator.

\paragraph{Protocol.}
\begin{table}[h]
\centering
\caption{Single-GPU coding-agent protocol.}
\label{tab:agent-single-stack-protocol}
\small
% Ragged right in the label column, because justifying it hyphenates the
% short compound labels mid-word.
\begin{tabularx}{\columnwidth}{@{}>{\raggedright\arraybackslash}p{0.30\columnwidth}X@{}}
\toprule
\textbf{Item} & \textbf{Setting} \\
\midrule
Models and runtime & \texttt{claude-opus-4-8}, \texttt{claude-sonnet-5}, and
	\texttt{claude-haiku-4-5-20251001}; Claude Code~2.1.214 \\
Single-shot budget & 20-minute timeout; \$10 cost backstop; no evaluator calls \\
Cold-iterative budget & 60-minute timeout; \$25 cost backstop; at most 20 evaluator calls \\
Warm-iterative initialization & Eligible held-out-correct single-shot source
retained as checkpoint~0; at most 20 evaluator calls \\
Seeds & Development: 42; held out: 101, 202, 303, 404, 505 \\
Held-out correctness & Relative Frobenius error $<0.02$ on all five seeds \\
Selection under feedback & Development-correct candidate with the highest co-measured
	tuned oneDNN ratio \\
\bottomrule
\end{tabularx}
\end{table}

Final correctness requires all five held-out seeds to pass. Performance is
throughput divided by the fixed 197.5-TFLOP/s BF16-output oneDNN baseline, reached
through the \texttt{torch.mm} dispatch on Intel XPU and measured on the same
GPU. The main paper reports every model separately
for single-shot and cold-iterative trials and summarizes Opus warm-iterative
refinement; the full multi-model distributed warm-iterative results appear
below.
Both performance summaries use only correct trials, because the median and range describe
reliability among held-out-correct trials, whereas \texttt{peak@3} is the
maximum held-out-correct result from the three independent trials.

\subsection{Single-GPU Complete Controlled Results}

Table~\ref{tab:agent-gemm-full} gives the ranges omitted from
Table~\ref{tab:agent-gemm} in the main paper. A dash means that no trial
passed all five held-out seeds.

\begin{table*}[t]
\centering
\caption{Complete controlled BF16 GEMM results. Each cell reports
held-out-correct trials / 3; median (full range) /
\texttt{peak@3}, in percent of the fixed tuned oneDNN baseline. A
singleton is shown without a range, and a dash means no held-out-correct
trial. Cold-iterative replicates are independent starts.}
\label{tab:agent-gemm-full}
\footnotesize
\setlength{\tabcolsep}{3pt}
\begin{tabularx}{\textwidth}{@{}ll>{\centering\arraybackslash}X>{\centering\arraybackslash}X>{\centering\arraybackslash}X@{}}
\toprule
\textbf{Protocol} & \textbf{Model} & \textbf{Raw SYCL} &
\textbf{SyclKittens} & \textbf{+ examples} \\
\midrule
\multirow{3}{*}{Single-shot}
 & Opus   & 1/3; 6.6 / 6.6 & 3/3; 49.1 (42.1--63.4) / 63.4 & 3/3; 49.4 (49.1--81.2) / 81.2 \\
 & Sonnet & 0/3; --  & 3/3; 31.9 (31.7--53.4) / 53.4  & 3/3; 27.5 (20.2--31.0) / 31.0 \\
 & Haiku  & 2/3; 0.7 (0.6--0.8) / 0.8 & 2/3; 24.0 (0.5--47.5) / 47.5 & 0/3; -- \\
\midrule
\multirow{3}{*}{Cold iterative}
 & Opus   & 3/3; 49.5 (14.0--57.5) / 57.5 & 3/3; 81.0 (71.5--82.1) / 82.1 & 3/3; 82.5 (81.1--82.6) / 82.6 \\
 & Sonnet & 3/3; 5.4 (5.2--5.7) / 5.7    & 3/3; 84.7 (84.6--85.2) / 85.2 & 3/3; 71.0 (70.9--84.8) / 84.8 \\
 & Haiku  & 3/3; 1.1 (0.8--1.6) / 1.6    & 3/3; 0.4 (0.4--0.6) / 0.6 & 3/3; 3.3 (0.1--50.3) / 50.3 \\
\bottomrule
\end{tabularx}
\end{table*}

\subsection{Single-GPU Warm-Iterative Refinement}
\label{app:agent-single-stack-endpoint}

The retained Raw Opus source rises from 6.47\% to 14.24\%.
Raw has one eligible source, so its final median and maximum in
Table~\ref{tab:agent-gemm} are both 14.2\%. The matched follow-up begins from
the six held-out-correct Opus
sources in the two SyclKittens contexts. Together with Raw, all seven Opus
sources remain correct and improve under paired grading.

\begin{table*}[t]
\centering
\caption{Matched single-GPU Opus warm-iterative refinement. Performance is
relative to the co-measured oneDNN baseline; gains are percentage points.}
\label{tab:agent-gemm-endpoint}
\small
\setlength{\tabcolsep}{7pt}
\begin{tabular}{@{}llrrrr@{}}
\toprule
\textbf{Condition} & \textbf{Replicate} & \textbf{Checkpoint 0, \% vendor} &
\textbf{Final, \% vendor} & \textbf{Paired gain} & \textbf{Calls} \\
\midrule
Raw SYCL & r1 & 6.47 & 14.24 & +7.77 & 20 \\
\midrule
\multirow{3}{*}{SyclKittens}
 & r1 & 49.80 & 87.83 & +38.03 & 11 \\
 & r2 & 66.84 & 70.46 & +3.62 & 6 \\
 & r3 & 44.67 & 56.64 & +11.97 & 14 \\
\midrule
\multirow{3}{*}{+ examples}
 & r1 & 51.41 & 87.71 & +36.30 & 10 \\
 & r2 & 85.21 & 86.51 & +1.30 & 5 \\
 & r3 & 51.23 & 82.91 & +31.68 & 7 \\
\bottomrule
\end{tabular}
\end{table*}

Warm-iterative feedback improves every source. The two SyclKittens contexts
reach effectively tied frontiers (87.83\% in \mk{} and 87.71\% in
\mk{} + examples), while examples raise the final median from 70.46\% to 86.51\%. The Raw source
improves from 6.47\% to 14.24\%.

\subsection{Distributed All-Gather GEMM Studies}
\label{app:agent-study-ag-gemm}

The distributed task asks each agent to implement a complete 12-GPU,
row-sharded BF16 all-gather GEMM at $M{=}N{=}K{=}6144$. Each GPU initially
owns 512 rows of dense random $A$; $B$ is replicated; and every GPU must
produce the full replicated BF16 $C{=}AB$ with FP32 accumulation. Candidates
implement the complete communication-and-compute path. The timed interval
includes both components, and throughput is reported per GPU as $2MKN/t$.

\paragraph{Task-specific context.}
\mk{} + examples adds a ThunderKittens H100 BF16 GEMM, a HipKittens AMD
BF16 GEMM, and a two-GPU SyclKittens all-reduce. The GEMM examples target
NVIDIA and AMD GPUs. The all-reduce demonstrates Intel peer access, but none of the
three programs implements an all-gather GEMM, a 12-GPU schedule, or the
topology-aware hand baseline, so the agent must compose the communication and
compute path. All other treatment and tool controls follow the shared protocol
above.

\paragraph{Correctness, timing, and anti-amortization.}
Correctness requires finite sampled output, relative L2 error below 0.03 on
all five held-out seeds, and agreement on every GPU. Timing uses a
2-second power warmup, five explicit warmups, and the median of 15 complete
operations. Before every warmup and scored invocation, the harness regenerates
a distinct deterministic $A$ epoch outside the timed interval, then verifies
the final epoch on all 12 outputs. A negative control that deliberately reuses
gathered data fails this gate with post-timing relative L2 error 1.38, whereas
valid raw-SYCL and SyclKittens candidates pass it; generated code therefore cannot
remove communication by retaining gathered scratch across calls.

Performance is normalized to the co-measured topology-aware hand
baseline on the assigned node. This reference gathers $A$ and
then computes the full $C$ on every GPU. It uses the evaluator's deterministic
input distribution and the frozen masked-prefetch build, rather than the
random-normal cooperative-prefetch configuration in
Table~\ref{tab:dist-gemm-production}. In single-shot evaluation, the hand
baseline spans 126.7--131.0~TFLOPS across six nodes, with a 128.9-TFLOPS
median. The tuned vendor library path combines oneCCL all-gather with
PyTorch's oneDNN GEMM. Timed end to end, it spans 128.9--144.3~TFLOPS, with a
134.9-TFLOPS median.
The matched factor sweep in Table~\ref{tab:dist-gemm-rootcause} reproduces both
\mk{} operating points and attributes the difference primarily to the
evaluator-versus-random-normal operand distribution.
Table~\ref{tab:agent-ag-gemm-full} reports only held-out-correct trials and
uses the node-matched hand baseline as its denominator.

\begin{table*}[t]
\centering
\caption{Complete controlled 12-GPU all-gather GEMM single-shot result.
Performance is relative to the co-measured topology-aware hand baseline.
\texttt{peak@3} is the maximum held-out-correct result from the three trials;
a dash means no trial passed all five held-out seeds.}
\label{tab:agent-ag-gemm-full}
\small
\setlength{\tabcolsep}{5pt}
\begin{tabularx}{\textwidth}{@{}llc>{\centering\arraybackslash}Xc@{}}
\toprule
\textbf{Model} & \textbf{Condition} & \textbf{Correct} &
\textbf{Median (range), \% hand} & \textbf{peak@3, \% hand} \\
\midrule
Opus & Raw SYCL & 0/3 & -- & -- \\
Opus & SyclKittens & 2/3 & 33.1 (29.4--36.7) & 36.7 \\
Opus & + examples & 3/3 & 33.2 (24.3--59.4) & 59.4 \\
\midrule
Sonnet & Raw SYCL & 3/3 & 0.3 (0.3--5.7) & 5.7 \\
Sonnet & SyclKittens & 3/3 & 0.3 (0.1--5.4) & 5.4 \\
Sonnet & + examples & 3/3 & 34.2 (16.7--38.8) & 38.8 \\
\midrule
Haiku & Raw SYCL & 2/3 & 0.3 (0.03--0.5) & 0.5 \\
Haiku & SyclKittens & 1/3 & 0.5 & 0.5 \\
Haiku & + examples & 1/3 & 0.5 & 0.5 \\
\bottomrule
\end{tabularx}
\end{table*}

\paragraph{Model-stratified result.}
Single-shot effects depend on the model. Examples raise the Opus frontier from
36.7\% to 59.4\% and the Sonnet median from 0.3\% to 34.2\%, whereas Haiku
remains below 1\% in both SyclKittens contexts. Four of Raw's five correct
kernels also remain below 0.5\%. No generated kernel exceeds the hand
baseline or tuned oneCCL library path.

\paragraph{Cold-iterative protocol.}
The feedback study keeps the task, programming contexts, evaluator, and three
replicates per model and condition fixed. Each replicate begins independently
from its treatment's initial task template, with no candidate code transferred
from the single-shot study or another replicate. Agents receive at most 20
calls to the fixed compile-correctness-performance evaluator. The fastest
development-correct candidate is then graded on the five held-out seeds.

\begin{table*}[t]
\centering
\caption{12-GPU all-gather GEMM cold-iterative results.
The median, range, and \texttt{peak@3} include only held-out-correct trials and
are percentages of the co-measured topology-aware hand baseline.}
\label{tab:agent-ag-gemm-iterative}
\small
\setlength{\tabcolsep}{5pt}
\begin{tabularx}{\textwidth}{@{}llc>{\centering\arraybackslash}Xc@{}}
\toprule
\textbf{Model} & \textbf{Condition} & \textbf{Correct} &
\textbf{Median (range), \% hand} & \textbf{peak@3, \% hand} \\
\midrule
Opus & Raw SYCL & 3/3 & 42.0 (39.8--42.3) & 42.3 \\
 & SyclKittens & 3/3 & 41.7 (38.4--46.8) & 46.8 \\
 & + examples & 3/3 & 42.0 (41.2--44.0) & 44.0 \\
\midrule
Sonnet & Raw SYCL & 3/3 & 12.9 (12.5--25.8) & 25.8 \\
 & SyclKittens & 3/3 & 46.1 (32.1--50.9) & 50.9 \\
 & + examples & 3/3 & 39.6 (36.3--41.1) & 41.1 \\
\midrule
Haiku & Raw SYCL & 3/3 & 1.6 (1.1--17.2) & 17.2 \\
 & SyclKittens & 3/3 & 0.5 (0.1--0.7) & 0.7 \\
 & + examples & 2/3 & 0.8 (0.1--1.5) & 1.5 \\
\bottomrule
\end{tabularx}
\end{table*}

Cold feedback changes Sonnet most, whose two SyclKittens contexts reach 50.9\%
and 41.1\%, compared with its 25.8\% Raw frontier. Opus changes modestly from
Raw, and Haiku remains low-throughput in both SyclKittens contexts.

\paragraph{Distributed warm-iterative refinement.}
Warm-iterative refinement starts from every eligible held-out-correct
single-shot source in the two SyclKittens contexts, namely six in \mk{} and
seven in \mk{} + examples, spanning five Opus, six Sonnet, and two Haiku trials. All 13 sources remain
held-out correct and improve. Table~\ref{tab:agent-ag-gemm-endpoint} reports
each paired result relative to the node-matched hand baseline.

\begin{table*}[t]
\centering
\caption{12-GPU all-gather GEMM warm-iterative refinement. Each eligible
source remains correct. Gains are paired percentage-point changes from the
fixed checkpoint~0 to the final selected checkpoint.}
\label{tab:agent-ag-gemm-endpoint}
\footnotesize
\setlength{\tabcolsep}{5pt}
\begin{tabular}{@{}lllrrrr@{}}
\toprule
\textbf{Condition} & \textbf{Model} & \textbf{Replicate} &
\textbf{Checkpoint 0, \% hand} & \textbf{Final, \% hand} &
\textbf{Paired gain} & \textbf{Calls} \\
\midrule
\multirow{6}{*}{SyclKittens}
 & Opus   & r1 & 29.49 & 106.57 & +77.08 & 15 \\
 & Sonnet & r1 &  0.28 &   6.19 &  +5.91 & 18 \\
 & Haiku  & r1 &  0.47 &   0.72 &  +0.24 & 20 \\
 & Sonnet & r2 &  5.43 &   8.86 &  +3.43 & 11 \\
 & Opus   & r3 & 35.75 &  40.64 &  +4.89 & 13 \\
 & Sonnet & r3 &  0.15 &  36.72 & +36.58 & 15 \\
\midrule
\multirow{7}{*}{+ examples}
 & Opus   & r1 & 24.13 & 38.51 & +14.38 & 16 \\
 & Sonnet & r1 & 35.18 & 66.57 & +31.39 & 10 \\
 & Opus   & r2 & 59.96 & 61.51 &  +1.55 & 10 \\
 & Sonnet & r2 & 39.76 & 41.17 &  +1.41 & 14 \\
 & Opus   & r3 & 33.42 & 38.90 &  +5.48 &  9 \\
 & Sonnet & r3 & 16.60 & 38.85 & +22.24 & 13 \\
 & Haiku  & r3 &  0.49 &  0.70 &  +0.21 & 20 \\
\bottomrule
\end{tabular}
\end{table*}

The $106.6\%$ source improves the distributed algorithm rather than the
dense GEMM microkernel. It computes each GPU's $C$ row shard from the local
$A$ shard and replicated $B$, then performs a topology-aware all-gather of
$C$. Because this task has $N{=}K$, gathering $C\,(M{\times}N)$ moves the same
BF16 tensor volume as gathering $A\,(M{\times}K)$, but the transformation
reduces aggregate GEMM work from 12 full products to one full product
partitioned across 12 GPUs. It remains a valid result, because the fixed
harness regenerates $A$ before every invocation, waits both queue sets, and verifies
the full replicated $C$ on every GPU for all five held-out seeds.

Every eligible source improves. Median paired gains are 5.4 and 5.5 percentage
points in \mk{} and \mk{} + examples, respectively, and one \mk{} Opus
source reaches 106.6\% of hand (136.311 versus 127.91~TFLOP/s). Because
eligibility produces different source populations, these results establish
per-source refinement rather than a balanced comparison between contexts.

\paragraph{Engineer-agent co-design.}
The controlled sources above remain frozen for attribution. A matched
Box--Muller comparison found that the \mk{} + examples kernel reached
131.616~TFLOP/s, exceeding the 128.545 of our initial refinement of a
\mk{} kernel. We restarted from the unchanged \mk{} + examples kernel
and retained it as the incumbent under a monotonic promotion rule.
The selected refinement retains its DPAS core and
cross-package route, then applies cyclic MDFI band order with a two/four split
over the compute/copy queues. It reaches 125.724~TFLOP/s versus 123.839 for the incumbent
in a new four-node random-normal study, with a $1.007\times$ node-first paired
ratio that sits within run-to-run variation
(Table~\ref{tab:dist-gemm-codesign}). In this workflow, agents expand the
algorithm family, engineers retain the strongest validated source as the
incumbent, and joint tuning is accepted only under a common correctness and
performance gate.

\subsection{Single-GPU Mechanism Inspection}

The single-shot Opus frontier in \mk{} + examples is replicate r2. Its selected
source declares a
$32{\times}32$ register base shape, presents B in the VNNI-compatible column
layout, sets the K-tile prefetch distance to three, uses 2-D loads, and reaches
DPAS through \texttt{mma\_AB}. It reaches 160.4~TFLOP/s, or 81.2\% of the
fixed oneDNN baseline. The cold-iterative frontiers in the remaining columns
also use \texttt{mma\_AB}, 2-D loads, and VNNI-compatible B, but select
different schedules.

\begin{table*}[t]
\centering
\caption{Mechanisms in the single-shot Opus frontier in \mk{} + examples and the
fastest single-GPU cold kernel for each model. Percentages use the
fixed tuned oneDNN baseline.}
\label{tab:agent-mechanisms}
\footnotesize
\setlength{\tabcolsep}{4pt}
\begin{tabularx}{\textwidth}{@{}>{\raggedright\arraybackslash}Xllll@{}}
\toprule
\textbf{Technique} & \shortstack{\textbf{Single-shot}\\\textbf{Opus + examples}} &
\shortstack{\textbf{Cold}\\\textbf{Opus + examples}} &
\shortstack{\textbf{Cold}\\\textbf{Sonnet SK}} &
\shortstack{\textbf{Cold}\\\textbf{Haiku + examples}} \\
\midrule
Performance & 160.4 TF, 81.2\% & 163.1 TF, 82.6\% & 168.3 TF, 85.2\% & 99.3 TF, 50.3\% \\
DPAS through \texttt{mma\_AB} & yes & yes & yes & yes \\
2-D loads and VNNI $B$ & yes & yes & yes & yes \\
Register base shape & $32{\times}32$ & $32{\times}32$ & $32{\times}32$ & $16{\times}16$ \\
K-tile prefetch distance & 3 & 2 & 4 & none \\
K-loop unrolling & none & none & none & $16\times$ \\
\bottomrule
\end{tabularx}
\end{table*}

Raw cold frontiers are 57.5\% for Opus, 5.7\% for Sonnet, and 1.6\% for
Haiku. Direct SPIR-V/XMX access is available in every condition. The generated
sources differ in their tiling and operand-delivery implementations.

\input{sections/supp/appx-agent-gqa}

\subsection{GitHub Copilot Protocol and Results}
\label{app:agent-newer}

GitHub Copilot evaluates Opus~5 and GPT-5.6-Sol in cold iteration.
Raw and SK are compared on
$4096^3$ single-GPU GEMM and $6144^3$ 12-GPU all-gather GEMM; GQA additionally
includes SK + examples.

\paragraph{Controller and grading.}
One fixed VS Code Language Model API controller selects
\texttt{claude-opus-5} and \texttt{gpt-5.6-sol} without fallback.
Each start begins from its condition stub, permits at
most 20 evaluator calls within 60 minutes, and restricts edits to the
candidate implementation. Development uses seed 42. The selected
development-correct checkpoint is graded on seeds 101, 202, 303, 404, and
505, whose inputs and outcomes are not exposed to the agent. All evaluator
requests consume budget, including failed builds. Single-GPU GEMM uses BF16
inputs with FP32 accumulation and output; distributed GEMM uses BF16 output.
The numerical gates are the task-specific criteria defined above.

The collector selects the fastest admissible held-out-correct absolute
throughput among three starts. Its reported percentage uses that trial's
baseline, while the median percentage is the median of correct per-trial
ratios. Single-GPU final grading uses 10 warmups and 100 timed operations
with co-measured oneDNN. Distributed grading uses the
co-measured frozen hand reference and fresh input epochs. Each condition
contains three starts, including model-tool policy failures.

\begin{table*}[t]
\centering
\caption{Raw/SK single-GPU GEMM endpoints through GitHub Copilot.
Values are percentages of each trial's co-measured oneDNN.
Rejected denotes a model-tool policy violation.
All ten admissible endpoints pass five held-out seeds.}
\label{tab:agent-newer-endpoints}
\small
\begin{tabular}{@{}llrrrc@{}}
\toprule
\textbf{Model} & \textbf{Context} & \textbf{r1} & \textbf{r2} & \textbf{r3} & \textbf{Correct} \\
\midrule
Opus 5 & Raw & Rejected & 84.5 & Rejected & 1/3 \\
 & SK & 94.2 & 76.1 & 84.7 & 3/3 \\
GPT-5.6-Sol & Raw & 7.8 & 9.9 & 11.0 & 3/3 \\
 & SK & 87.9 & 85.2 & 68.5 & 3/3 \\
\bottomrule
\end{tabular}
\end{table*}

\paragraph{Search efficiency.}
Table~\ref{tab:agent-newer-search} counts the first development-correct
checkpoint reaching 80\% of its co-measured oneDNN. All starts contribute,
including those with rejected final outcomes. A non-reaching trial is
ranked beyond the observed budget when computing the median; a dash
means fewer than two of three starts reach the threshold. Held-out grading
is performed on the selected endpoint. Figure~\ref{fig:progression}(b)
reports a different aggregation for Opus~4.8, the median best-so-far
throughput at each call. Its SK curve passes Raw's final level at call two
and 80\% at call four.

\begin{table}[t]
\centering
\caption{Single-GPU development calls to 80\% of oneDNN through GitHub Copilot.
NR means not reached within the 20-call budget.}
\label{tab:agent-newer-search}
\footnotesize
\setlength{\tabcolsep}{4pt}
\begin{tabular}{@{}llrrrr@{}}
\toprule
\textbf{Model} & \textbf{Context} & \textbf{r1} & \textbf{r2} & \textbf{r3} & \textbf{Median} \\
\midrule
Opus 5 & Raw & 6 & 10 & NR & 10 \\
 & SK & 5 & NR & 2 & 5 \\
GPT-5.6-Sol & Raw & NR & NR & NR & -- \\
 & SK & 4 & 2 & NR & 4 \\
\bottomrule
\end{tabular}
\end{table}

\begin{figure*}[t]
\centering
\includegraphics[width=\textwidth]{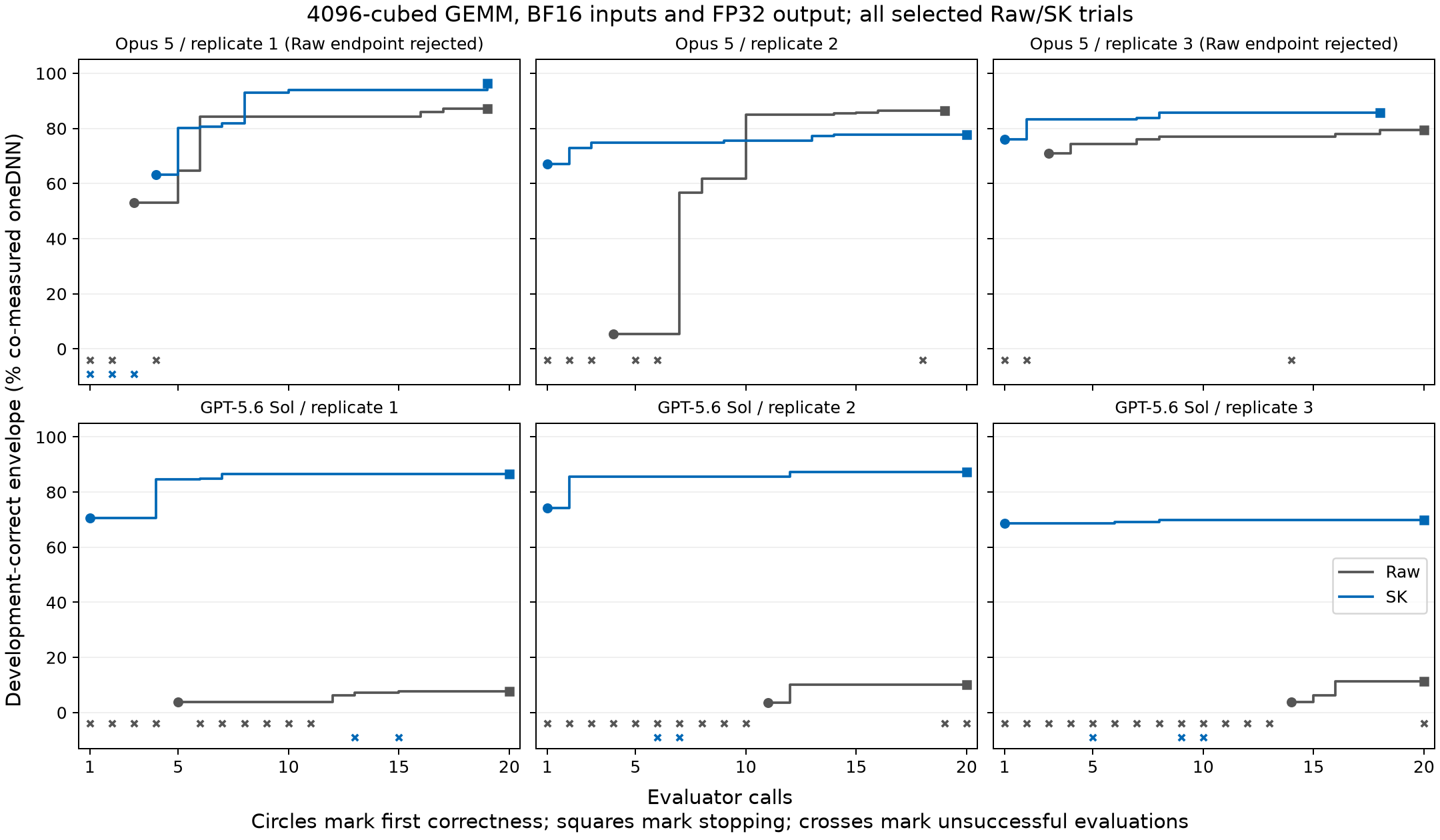}
\caption{All twelve Raw/SK single-GPU GEMM searches through
GitHub Copilot. Each curve is the best development-correct percentage of
co-measured oneDNN observed so far. Traces
stop at their actual last evaluator call; circles mark first correctness,
squares mark stopping, and crosses mark unsuccessful evaluations. The two
Opus~5 Raw searches with rejected final outcomes are identified in their
panel titles.}
\Description{Six panels show the three independent Raw and SK searches
for Opus 5 and GPT-5.6-Sol. Opus has mixed per-replicate ordering, while all
three GPT SK searches substantially exceed their Raw counterparts.}
\label{fig:agent-newer-trajectories}
\end{figure*}

\paragraph{Numerical sensitivity.}
With the selected kernels fixed, we test a tenfold tighter relative
Frobenius limit of 0.002. All 50 held-out seed errors from the ten
admissible FP32-output endpoints pass, with maximum error approximately
$6.98\times10^{-7}$.

\paragraph{Controller resources.}
Table~\ref{tab:agent-newer-resources} includes every Raw/SK single-GPU start,
including failures. Controller time includes model requests and tools.
Counted input tokens sum request histories, including repeated context.
Although both models reach high throughput in fewer evaluator calls with
SK, total controller time increases for Opus~5 and decreases for GPT.

\begin{table}[t]
\centering
\caption{Mean controller resources per single-GPU Copilot start, including failures.}
\label{tab:agent-newer-resources}
\footnotesize
\setlength{\tabcolsep}{4pt}
\begin{tabular}{@{}llrr@{}}
\toprule
\textbf{Model} & \textbf{Context} & \textbf{Minutes} & \shortstack{\textbf{Input tokens}\\\textbf{(millions)}} \\
\midrule
Opus 5 & Raw & 25.5 & 0.712 \\
 & SK & 35.5 & 0.723 \\
GPT-5.6-Sol & Raw & 38.9 & 0.988 \\
 & SK & 33.8 & 1.114 \\
\bottomrule
\end{tabular}
\end{table}

\paragraph{Source observations.}
All three admissible Opus~5 SK-context sources use direct matrix intrinsics,
not \texttt{mma\_AB}. The strongest assigns separate prefetch regions to
subgroups with a guarded three-K-tile lookahead; the other sources use
different ownership and lookahead choices. GPT's strongest SK-context
source uses the tile API and two-tile prefetch, while its other two sources
use direct intrinsics and one has no prefetch. Every GPT Raw source uses
scalar loads and explicit packing without prefetch.

\subsection{Trial Inclusion and Protocol Deviations}

Table~\ref{tab:agent-deviations} records exclusions and reruns in the
Claude Code trials. Warm-iterative analyses are paired because
eligibility determines their source populations.

\begin{table*}[t]
\centering
\caption{Protocol deviations and their effect on the reported studies.}
\label{tab:agent-deviations}
\footnotesize
\begin{tabularx}{\textwidth}{@{}p{0.22\textwidth}XX@{}}
\toprule
\textbf{Deviation} & \textbf{Resolution} & \textbf{Effect} \\
\midrule
Earlier definitions of the two SyclKittens contexts &
Excluded before analysis; the reported dataset uses the final context
definitions. &
All reported comparisons use the same final definitions. \\
Off-workspace edit in a single-GPU single-shot trial &
Discarded before evaluation; repeated from the same initial task and
programming context. &
No execution feedback entered evaluation; the replacement completed the
planned trial. \\
Missing trusted grading record in a distributed single-shot trial &
Repeated from the unchanged initial task before analysis. &
Only the repeat contributes to the reported result. \\
Two distributed pilot protocols &
Datasets that omitted treatment-specific guides or reused one input epoch were
excluded. &
These datasets do not enter the results; the reported protocol includes both guides and
fresh inputs. \\
\bottomrule
\end{tabularx}
\end{table*}

The accompanying artifact retains prompts, programming contexts, generated
sources, agent interactions, development evaluations, selected candidates,
and held-out logs for independent inspection.

\subsection{AI Use in \mk{} development}

The autonomous controlled study is distinct from the engineer-agent suite.
Intel hardware experts helped design the programming model. Engineers and
coding agents collaborated on implementation, call-site integration, variant
generation, and tuning. Engineers owned the experiments and correctness and
performance acceptance. Generated code entered the suite only after
compilation, operation-specific tests, interleaved measurement, and human
review, as described by the main-paper AI disclosure.
For cooperative prefetch, a human expert defined the disjoint regions and
aligned their phases; agents integrated and tuned the primitive. Joint
kernel development used Intel VTune, whereas autonomous controlled trials
received compilation, correctness, and timing feedback.

%% file: sections/supp/appx-agent-gqa.tex
 \subsection{Grouped-Query Attention}
 \label{app:agent-study-gqa}

 The Claude Code trials evaluate Opus~4.8, Sonnet~5, and Haiku~4.5 on causal
 GQA at $B{=}1$, $H{=}32$, $H_{KV}{=}8$, $N{=}2048$, and $D{=}128$.
 The five staged task inputs are byte-identical across models, including the
 raw-pointer harness with SHA-256
 \texttt{bdd6bd8f\allowbreak f88d83c2\allowbreak 9b0f6a46\allowbreak
 55268407\allowbreak 906a2fd6\allowbreak 3e9179cc\allowbreak
 e173935d\allowbreak b143b9c}.
 Raw, \mk{}, and \mk{} + examples use that harness and the same candidate function
 signature. Every candidate owns its launch geometry, work-group and subgroup
 mapping, and submissions. ESIMD is forbidden in every condition because it is
 a separate Intel-specific SIMD programming model that neither raw SYCL nor
 \mk{} builds on. Direct SPIR-V subgroup matrix and 2-D block-I/O intrinsics
 remain available. Raw provides Intel SYCL/SPIR-V documentation, \mk{} adds its
 headers, and \mk{} + examples additionally provides two HipKittens causal-GQA
 kernels for AMD GPUs rather than an Intel implementation.

 Each of the 27 cold-iterative trials independently starts from a no-op
 condition stub and receives at most 20 evaluator calls. The 27 single-shot
 trials use the same frozen inputs without evaluator feedback. Correctness
 requires finite output, relative L2 error below 0.05, and elementwise error
 rate below 0.01 on all five held-out seeds. Performance is normalized by the
 BF16-output native PyTorch XPU SDPA baseline co-measured in each trial, with
 \texttt{enable\_gqa=True} and \texttt{is\_causal=True}. The baseline process
 does not import IPEX. The observed cold-iterative baselines range from 73.0 to
 85.3~TFLOP/s. The kernel-suite sweep in \S\ref{app:attn-fwd} imports IPEX
 and measures its grouped-head fallback.

 \begin{table*}[t]
 \centering
 \caption{Complete equal-launch cold-iterative GQA results. Replicate columns
 report held-out candidate TFLOP/s and the parenthesized percentage of that
 trial's co-measured native PyTorch XPU SDPA baseline, without IPEX imported.
 The final column reports median /
 \texttt{peak@3} among correct trials. All 27 trials are correct.}
 \label{tab:agent-gqa-additional}
 \footnotesize
 \setlength{\tabcolsep}{3pt}
 \begin{tabular}{@{}llcrrrc@{}}
 \toprule
 \textbf{Model} & \textbf{Context} & \textbf{Correct} & \textbf{r1} &
 \textbf{r2} & \textbf{r3} & \textbf{Median / peak, \% SDPA} \\
 \midrule
 \multirow{3}{*}{Opus 4.8}
 & Raw SYCL    & 3/3 & 42.30 (57.7) & 13.81 (18.6) & 26.38 (36.1) & 36.1 / 57.7 \\
 & SyclKittens & 3/3 & 24.84 (29.7) & 34.41 (41.6) & 15.89 (18.6) & 29.7 / 41.6 \\
 & + examples    & 3/3 & 59.95 (79.5) & 38.04 (50.4) & 32.77 (43.2) & 50.4 / 79.5 \\
 \midrule
 \multirow{3}{*}{Sonnet 5}
 & Raw SYCL    & 3/3 & 14.47 (17.6) & 15.36 (18.6) & 14.29 (17.4) & 17.6 / 18.6 \\
 & SyclKittens & 3/3 &  3.48 (4.6)  &  1.38 (1.8)  &  3.33 (4.4)  &  4.4 / 4.6 \\
 & + examples    & 3/3 &  6.82 (8.8)  &  2.96 (3.9)  &  3.41 (4.4)  &  4.4 / 8.8 \\
 \midrule
 \multirow{3}{*}{Haiku 4.5}
 & Raw SYCL    & 3/3 & 0.56 (0.7) & 0.45 (0.6) & 0.36 (0.5) & 0.6 / 0.7 \\
 & SyclKittens & 3/3 & 0.29 (0.4) & 0.73 (1.0) & 0.21 (0.3) & 0.4 / 1.0 \\
 & + examples    & 3/3 & 1.00 (1.3) & 0.40 (0.5) & 0.24 (0.3) & 0.5 / 1.3 \\
 \bottomrule
 \end{tabular}
 \end{table*}

 \paragraph{Context effects differ by model.}
 Examples lead for Opus at
 $79.5\%$, while Raw leads for Sonnet at $18.6\%$, followed by examples at
 $8.8\%$ and \mk{} at $4.6\%$. Haiku remains between $0.7\%$ and $1.3\%$
 across all contexts. The best Sonnet cell falls below the weakest Opus cell at
 $41.6\%$. All 27 kernels are correct, so correctness alone would hide this
 model effect.

 \begin{table*}[t]
 \centering
 \caption{Complete equal-launch single-shot GQA results. Replicate columns
 report percentages of the co-measured native PyTorch XPU SDPA baseline,
 without IPEX imported. A dash denotes a
 trial without an admissible held-out-correct endpoint. The final column reports
 \texttt{peak@3} among correct trials.}
 \label{tab:agent-gqa-single-shot}
 \footnotesize
 \setlength{\tabcolsep}{5pt}
 \begin{tabular}{@{}llcrrrc@{}}
 \toprule
 \textbf{Model} & \textbf{Context} & \textbf{Correct} & \textbf{r1} &
 \textbf{r2} & \textbf{r3} & \textbf{peak@3, \% SDPA} \\
 \midrule
 \multirow{3}{*}{Opus 4.8}
 & Raw SYCL    & 1/3 & --  & --  & 1.4  & 1.4 \\
 & SyclKittens & 2/3 & --  & 1.7 & 11.3 & 11.3 \\
 & + examples    & 3/3 & 1.3 & 8.3 & 12.7 & 12.7 \\
 \midrule
 \multirow{3}{*}{Sonnet 5}
 & Raw SYCL    & 3/3 & 0.1 & 2.9 & 2.6 & 2.9 \\
 & SyclKittens & 3/3 & 2.4 & 1.2 & 3.0 & 3.0 \\
 & + examples    & 3/3 & 2.4 & 3.2 & 0.4 & 3.2 \\
 \midrule
 \multirow{3}{*}{Haiku 4.5}
 & Raw SYCL    & 3/3 & 0.2 & 0.01 & 0.01 & 0.2 \\
 & SyclKittens & 2/3 & --  & 0.2 & 0.03 & 0.2 \\
 & + examples    & 3/3 & 0.2 & 0.01 & 0.3 & 0.3 \\
 \bottomrule
 \end{tabular}
 \end{table*}

 The single-shot arms contain 27 infrastructure-valid cells and 23
 held-out-correct endpoints. The only incorrect Sonnet or Haiku endpoint is
 Haiku SyclKittens replicate~1, whose generated candidate fails the build gate.
 This is a correctness outcome rather than a harness failure.

 \paragraph{Opus warm refinement.}
 Warm refinement starts from the six held-out-correct Opus single-shot sources.
 All six remain correct and improve. Paired gains range from
 9.5 to 81.7 percentage points, and the strongest endpoint reaches 71.0~TFLOP/s
 or $83.4\%$ of its co-measured baseline. We report paired gains on these
 eligible sources.

 \begin{table}[t]
 \centering
 \caption{Opus endpoint-seeded warm GQA results. Final values are percentages
 of the co-measured native PyTorch XPU SDPA baseline, without IPEX imported.}
 \label{tab:agent-gqa-warm}
 \footnotesize
 \setlength{\tabcolsep}{4pt}
 \begin{tabular}{@{}lccc@{}}
 \toprule
 \textbf{Context} & \textbf{Eligible} & \textbf{Final values} &
 \textbf{Median / maximum} \\
 \midrule
 Raw SYCL    & 1/1 & 50.1             & 50.1 / 50.1 \\
 SyclKittens & 2/2 & 83.4, 22.0       & 52.7 / 83.4 \\
 + examples    & 3/3 & 28.8, 48.2, 22.1 & 28.8 / 48.2 \\
 \bottomrule
 \end{tabular}
 \end{table}

 \paragraph{Opus mechanisms and variance.}
 All nine selected Opus cold-iterative kernels reach XMX through direct SPIR-V
 subgroup DPAS, and eight also use direct 2-D block I/O. The strongest Raw
 kernel reaches 42.3~TFLOP/s with transformed 2-D loads and prefetch, while the
 strongest example-assisted kernel reaches 59.9~TFLOP/s with direct 2-D loads, transposed
 loads, and prefetch. The selected Opus kernels from \mk{} and \mk{} + examples
 use direct intrinsics rather than SyclKittens register-tile or MMA APIs. The
 treatment therefore changes the schedules Opus discovers even when the final
 implementations use the same low-level mechanism family.

 The HipKittens examples expose tiled causal-attention algebra and pipelining, but
 the generated kernels are Intel implementations. The strongest Opus result
 translates that structure into PVC DPAS and cache-prefetch operations, while
 the other two example-assisted replicates do not reproduce the same schedule.
 Only the strongest of the three example-assisted Opus starts exceeds Raw's
 cold-iteration peak.

\subsubsection{GitHub Copilot Results}

The Copilot trials evaluate Opus~5 and GPT-5.6-Sol across Raw, SK,
and SK + examples, with three independent cold starts per condition,
the causal-GQA shape above, a 20-evaluator-call budget, and
private five-seed grading. All contexts own their launch geometry, permit
direct SPIR-V, and forbid ESIMD and \texttt{joint\_matrix}. Correctness
requires finite output, relative L2 error below 0.05, and elementwise
error rate below 0.01 on every held-out seed.

Seventeen of the 18 starts produce admissible correct endpoints. Opus~5 SK
replicate~3 violates the model-tool policy. Table~\ref{tab:agent-gqa}
reports absolute causal TFLOP/s for both controller groups, totaling 45
starts and 44 admissible endpoints. Programming-context comparisons are
within each model and controller, as defined in \S\ref{sec:ai:protocol}.

Examples raise Opus~5's peak from Raw's 57.11 to 75.55~TFLOP/s and GPT's
from 8.95 to 44.52~TFLOP/s. Without examples, SK reaches 50.54 for Opus~5
and 24.57 for GPT. Examples therefore give the highest peak for both
models, while SK alone improves GPT and falls below Raw for Opus~5.

%% file: sections/supp/appx-gemm.tex
\section{BF16 GEMM}
\label{app:gemm}

Across the 27-shape evaluation suite, the \mk{} bf16 GEMM with collaborative
de-duplicated prefetch achieves $0.963\times$ oneDNN in geometric-mean
throughput. Measurements use bf16 output, shape-adaptive launch
policies, within-trial co-measurement, and six-node aggregation.

\subsection{Configuration}
\label{app:gemm-config}

\begin{table}[t]
\centering
\small
\caption{GEMM configuration.}
\begin{tabular}{@{}p{0.24\linewidth}p{0.68\linewidth}@{}}
\toprule
Parameter & Value \\
\midrule
Problem (square) & $M{=}K{=}N$, with
  $\begin{array}{@{}c@{}}N \in \{1024, 2048, 3072, 4096, 5120, 6144,\\
  7168, 8192, 9216, 10240, 12288, 14336, 16384\}\end{array}$ \\
Problem (rectangular) & 14 shapes drawn from Llama-3/BERT-style inference
  workloads (QKV, output projection, FFN up/down, decode with
  $M \in \{256, 512\}$) \\
Dtype & bf16 $A$/$B$/$C$, fp32 DPAS accumulation \\
Hardware & \pvc{}, one active stack per GPU package, sibling stack idle \\
Nodes & Six nodes; results are aggregated by cross-node median \\
Baselines & oneDNN, reached through the \texttt{torch.mm} dispatch used on
  this platform \\
Tiling & $\mathrm{BM}{=}32$, $\mathrm{BN}{=}64$, $\mathrm{BK}{=}32$
  with a $32{\times}32$ register tile and 32-way K-loop unrolling; per-shape snake
  width, shape-adaptive root-sync launch, and collaborative de-duplicated
  prefetch (distance 2 for squares; shape-selected distances 1--3 for rectangles) \\
\bottomrule
\end{tabular}
\end{table}

GEMM FLOPs are $2MKN$ for both implementations, so their TFLOPS figures are
directly comparable.

\subsection{Kernel Implementation and Key Techniques}
\label{app:gemm-kernel}

The GEMM kernel is a single compile-time-templated translation unit built
into a callable per-shape module; each shape in the 27-shape evaluation suite
is compiled separately.

\textbf{Data flow (per output tile).} Each subgroup owns a
$\mathrm{BM}\times\mathrm{BN}$ ($32{\times}64$) fp32 accumulator register
tile \texttt{c\_reg} and streams the $K$ dimension in
$\mathrm{BK}{=}32$ chunks. For each K-step it loads a
$\mathrm{BM}\times\mathrm{BK}$ bf16 $A$ tile (row-major) and a
$\mathrm{BK}\times\mathrm{BN}$ bf16 $B$ tile (column-major) via
2D block loads, with the VNNI transform applied to B. It issues the DPAS
contraction into \texttt{c\_reg} and prefetches future $A$/$B$ panels at the
configured lookahead distance. After the K loop,
\texttt{c\_reg} is scaled and stored.

\paragraph{Key techniques}

\begin{itemize}
\item \textbf{Register-resident DPAS tiling.} A $32{\times}64$ fp32
  accumulator per subgroup (falling back to $16{\times}16$ register tiles for
  sub-32 output tiles), bf16 operand tiles, and up to 32
  subgroups per work-group keep the systolic array fed. Profiling shows the
  kernel is DPAS-bound (XVE about 94.5\% array-active, L3 bandwidth about
  1.3\% of peak), so large, well-populated work-groups matter more than
  additional cache optimization.
\item \textbf{VNNI 2D block loads.} Hardware block-load instructions with
  the VNNI operand transform deliver bf16 operands to the DPAS unit in
  their native packed layout, avoiding a local-memory staging round trip.
\item \textbf{Masked prefetch baseline.} Before de-duplication, full-tile A
  prefetches are issued when \texttt{sg\_id \% 5 != 0}, and full-tile B
  prefetches are issued when \texttt{sg\_id \& 8 != 0}. This selective
  schedule reaches $88.8\%$ of oneDNN. Loop unrolling and snake traversal
  then raise the geometric mean to $91.0\%$.
\item \textbf{Snake-curve group swizzle.} Work-group tiles are traversed in
  an L2-friendly super-stack order, with the swizzle width fixed per shape
  (4 at $2048^3$, 16 at $3072^3$/$8192^3$, 4 at $12288^3$/$16384^3$, 8 for
  other square shapes, 8 for all rectangular shapes) from a full
  snake-width sweep. A guard falls back to linear indexing for non-aligned
  grids, and grid padding with an early-return bounds check (returning
  when a work-group's row or column tile index exceeds the grid) prevents
  tile skips and output corruption on non-divisible snake grids.
\item \textbf{Collaborative de-duplicated prefetch.} The partial-tile policy
  assigns every subgroup a disjoint
  slice of the shared $256{\times}32$ A and $32{\times}256$ B workgroup
  panels. Each panel byte is warmed once per K-step, rather than once per
  consuming subgroup; the prologue uses the same assignment.
\item \textbf{Shape-adaptive lookahead.} De-duplication reduces explicit
  prefetch traffic by about $3.6\times$ and enables deeper
  lookahead, namely distance 2 for all squares and mostly distance 3 for rectangles,
  with measured distance-1 and distance-2 exceptions. The retained masked
  baseline instead uses distance 1 for $K>4096$ and distance 2 otherwise.
\item \textbf{K-loop unroll (32-way).} Exposing 32 K-iterations to the
  compiler overlaps loads with MMA, measured $+7\%$ at $8192^3$ and
  $+16\%$ at $10240^3$ over a 16-way unroll.
\item \textbf{bf16-output store.} The fp32 accumulator is down-converted
  in-register by a 2-wide round-to-nearest-even convertor, then written
  with a 2-byte block store, which halves the store traffic of the earlier fp32
  path and removes the fp32 global round trip. This matches the oneDNN output dtype;
  the co-measured bf16/fp32 time ratio is 0.95 on a healthy
  node.
\end{itemize}

\subsubsection{Collaborative de-duplicated inner loop}
\label{app:gemm-coop-loop}

The partial-tile policy keeps the same demand
loads and DPAS contraction but replaces full-tile warming with partial-tile
ownership. The 32 subgroups form an $8{\times}4$ grid. For A, the four
subgroup columns partition each $32{\times}32$ tile into row slices; for B,
the subgroup-row indices partition its K extent and select one 32-column
half. These offsets cover the distinct workgroup panel exactly once. Unlike
the earlier full-tile cooperative-prefetch experiment, this path has no split
barrier.

\begin{lstlisting}
// Collaborative partial-tile ownership arithmetic.
const int coop_h_idx         = sg_id % (256 / BN);
const int coop_v_idx         = sg_id / (256 / BN);
const int pref_a_row_offset  = BM / (256 / BN);
const int pref_b_sg_row_size = 256 / (BM * 2);
const int pref_b_row_offset  = BK / (256 / BN);
const int coop_b_row_idx     = coop_v_idx % pref_b_sg_row_size;
const int coop_b_col_idx     = coop_v_idx / pref_b_sg_row_size;
const int coop_a_row_offset  = coop_h_idx * pref_a_row_offset;
const int coop_b_row_offset  = coop_b_row_idx * pref_b_row_offset;
const int coop_b_col_offset  = coop_b_col_idx * 32;

int prefetch_a = GEMM_PF_A;
int prefetch_b = GEMM_PF_B;

// De-duplicated prologue: use the same disjoint slices as steady state.
#pragma unroll
for (int k = 0; k < prefetch_a; k++)
  prefetch_load_coop(a_reg, g_a, {0, 0, wg_m, k},
                     {coop_a_row_offset, 0});
#pragma unroll
for (int k = 0; k < prefetch_b; k++)
  prefetch_load_coop(b_reg, g_b, {0, 0, k, wg_n},
                     {coop_b_row_offset, coop_b_col_offset});

#pragma unroll GEMM_UNROLL
for (int k = 0; k < GK / BK; k++, prefetch_a++, prefetch_b++) {
  load(a_reg, g_a, {0, 0, wg_m, k});
  load(b_reg, g_b, {0, 0, k, wg_n});
  if (prefetch_a < GK / BK)
    prefetch_load_coop(a_reg, g_a, {0, 0, wg_m, prefetch_a},
                       {coop_a_row_offset, 0});
  if (prefetch_b < GK / BK)
    prefetch_load_coop(b_reg, g_b, {0, 0, prefetch_b, wg_n},
                       {coop_b_row_offset, coop_b_col_offset});
  mma_AB(c_reg, a_reg, b_reg, c_reg);
}
\end{lstlisting}

The loop needs no producer-consumer split, SLM staging, multi-stage buffer
rotation, or barrier. ThunderKittens instead splits its pipeline across
producer and consumer warpgroups and waits on a semaphore at every K step,
while the HipKittens loop issues two workgroup barriers per matrix instruction.

Each \texttt{prefetch\_load\_coop} call lowers its assigned offset to one
Intel 2-D block-prefetch instruction with a $16{\times}8$ element block and
count two. Across the workgroup, the calls warm 32,768 bytes per K-step,
equal to the distinct A+B footprint, versus approximately 118,784 bytes for
the masked full-tile path. The same assignment is used in the prologue;
de-duplicated priming adds a 0.8\% geometric-mean gain on small and medium
squares.

\subsection{Methodology}
\label{app:gemm-methodology}

Measurements use the shared single-GPU isolation protocol. oneDNN and \mk{}
are loaded into one process and timed in an interleaved round robin. Each
trial uses 10 warmup and 30 measured iterations and rotates the implementation
that starts first. Interleaving reduces clock drift in the \mk{}/oneDNN ratio.

Within a node, each value is the per-shape median of per-trial ratios. The
reported ratios are medians of the six node medians. Before a shape enters the
results tables, its output is checked against \texttt{torch.mm} by relative
Frobenius norm. All measured configurations pass, with maximum error
$1.663\times10^{-3}$.

\subsection{De-Duplicated-Prefetch Performance}
\label{app:gemm-gold-results}

\paragraph{Matched cumulative tail.}
The final three cumulative stages share shapes, output type, launch policy,
correctness gate, and six-node co-measurement. They isolate snake traversal,
partial-tile ownership de-duplication, and shape-specific prefetch distance.

\begin{table}[t]
\centering
\small
\caption{Matched cumulative prefetch tail. The $8192^3$ row reports cross-node
medians; the 27-shape row reports geometric means.}
\label{tab:gemm-prefetch-ab}
\begin{tabular}{@{}lrrr@{}}
\toprule
Suite & Snake & De-duplicated & Shape-specific PF \\
\midrule
$8192^3$ & 94.3\% & 95.8\% & 95.9\% \\
All 27 & 91.0\% & 96.2\% & 96.3\% \\
\bottomrule
\end{tabular}
\end{table}

Across all shapes, de-duplication improves the geometric mean by $1.056\times$.
Shape-specific prefetch distances produce the final $96.3\%$ configuration.

Every value is a six-node cross-node median from co-measured trials using
the loop in \S\ref{app:gemm-coop-loop}. Ratios use measured oneDNN
throughput as the denominator.

\begin{table}[t]
\begin{minipage}{\columnwidth}
\centering
\small
\caption{Shape-specific-PF square GEMM
(\mbox{\mk{}/oneDNN} geometric mean of 0.970\texttimes).}
\label{tab:gemm-gold-square}
\begin{tabular}{@{}lr@{}}
\toprule
Size & \mk{}/oneDNN \\
\midrule
$1024^3$  & \textbf{1.165\texttimes} \\
$2048^3$  & 0.881\texttimes \\
$3072^3$  & 0.999\texttimes \\
$4096^3$  & 0.962\texttimes \\
$5120^3$  & 0.940\texttimes \\
$6144^3$  & 0.947\texttimes \\
$7168^3$  & 0.944\texttimes \\
$8192^3$  & 0.959\texttimes \\
$9216^3$  & 0.989\texttimes \\
$10240^3$ & 0.948\texttimes \\
$12288^3$ & 0.96\texttimes \\
$14336^3$ & 0.972\texttimes \\
$16384^3$ & 0.964\texttimes \\
\bottomrule
\end{tabular}
\end{minipage}
\end{table}

\begin{table}[t]
\begin{minipage}{\columnwidth}
\centering
\small
\caption{Shape-specific-PF rectangular GEMM
(\mbox{\mk{}/oneDNN} geometric mean of 0.953\texttimes).}
\label{tab:gemm-gold-rect}
\begin{tabular}{@{}llr@{}}
\toprule
Shape ($M{\times}K{\times}N$) & Workload & \mk{}/oneDNN \\
\midrule
$2048{\times}4096{\times}12288$ & QKV $M{=}2$K & 0.977\texttimes \\
$4096{\times}4096{\times}12288$ & QKV $M{=}4$K & 0.960\texttimes \\
$8192{\times}4096{\times}12288$ & QKV $M{=}8$K & 0.953\texttimes \\
$2048{\times}4096{\times}4096$ & Out $M{=}2$K & 0.924\texttimes \\
$8192{\times}4096{\times}4096$ & Out $M{=}8$K & 0.926\texttimes \\
$2048{\times}4096{\times}11008$ & FFN up $M{=}2$K & 0.925\texttimes \\
$4096{\times}4096{\times}11008$ & FFN up $M{=}4$K & 0.935\texttimes \\
$2048{\times}11008{\times}4096$ & FFN down $M{=}2$K & 0.884\texttimes \\
$4096{\times}11008{\times}4096$ & FFN down $M{=}4$K & 0.941\texttimes \\
$2048{\times}1024{\times}4096$ & BERT FFN up & 0.909\texttimes \\
$2048{\times}4096{\times}1024$ & BERT FFN down & \textbf{1.042\texttimes} \\
$256{\times}4096{\times}4096$ & Inference $M{=}256$ & 0.962\texttimes \\
$512{\times}4096{\times}4096$ & Inference $M{=}512$ & \textbf{1.039\texttimes} \\
$512{\times}4096{\times}11008$ & Inference FFN & 0.976\texttimes \\
\bottomrule
\end{tabular}
\end{minipage}
\end{table}

\mk{} exceeds oneDNN on three of the 27 shapes.

\subsection{Correctness}
\label{app:gemm-correctness}

All 27 shapes pass the relative-Frobenius-norm gate
$\lVert C-\mathrm{ref}\rVert_F / \lVert \mathrm{ref}\rVert_F < 0.02$.
Every shape reports a uniform relative error of $1.66\times10^{-3}$, the
expected bf16 rounding floor for these accumulation depths, with no
per-shape outlier. The bf16-output store retains the fp32 output path's
accuracy.

\subsection{Triton-XPU Comparison}

One sweep co-measures oneDNN, the final cooperative \mk{}
revision, and Intel Triton-XPU~3.6 over the same 27 BF16 input/output shapes.
For each shape on each node, we independently autotune Triton and record its
best configuration for comparison. Each trial rotates their execution order, with 10 warmup iterations
and 30 timed iterations per implementation; host timing includes an explicit
XPU completion barrier. Each implementation is checked against the same fp32
\texttt{torch.mm} reference.

Ratios are reduced first by the median across 10 trials within each node and
then by the median across six nodes. The final cooperative \mk{} reaches
$1.087\times$ geometric-mean and $1.030\times$ median throughput over Triton.
The cross-node median exceeds Triton on all 27 shapes; the smallest margin is
$1.0125\times$. Table~\ref{tab:gemm-triton-final} shows that the advantage holds
for both square and rectangular workloads.

\begin{table}[t]
\centering
\small
\caption{Direct final-cooperative \mk{}/Triton-XPU comparison. Geometric means
and medians are over cross-node-median per-shape throughput ratios.}
\label{tab:gemm-triton-final}
\begin{tabular}{@{}lrrrr@{}}
\toprule
Shape set & Shapes & Geomean & Median & \mk{} $>$ Triton \\
\midrule
All & 27 & $1.087\times$ & $1.030\times$ & 27/27 \\
Square & 13 & $1.095\times$ & $1.031\times$ & 13/13 \\
Rectangular & 14 & $1.079\times$ & $1.019\times$ & 14/14 \\
\bottomrule
\end{tabular}
\end{table}

Across the six nodes, the 27-shape \mk{}/Triton geometric mean ranges
from $1.084\times$ to $1.095\times$, so the advantage does not depend on which
node measures it. The maximum \mk{} and Triton relative Frobenius errors
against the reference are both $1.663\times10^{-3}$, so both implementations
clear the same accuracy bar.

\subsection{SYCL-TLA Comparison}
\label{app:gemm-tla}

We compare the cooperative SK implementation with SYCL-TLA and oneDNN
on 27 shapes across three nodes. The SK binaries are identical to those
used in the six-node ablation above; each experiment co-measures its
baselines on its own node set.

\paragraph{Implementations and measurement.}
The comparison uses SYCL-TLA revision
\texttt{6601377d25d3}, icpx~2025.3.2, compute-runtime~25.18.33578,
PyTorch~2.10, and IPEX~2.10.10. A per-shape, per-node tuner selects among
correctness-qualified documented legacy PVC configurations, including the
legacy GEMM example. Compilation, tuning, and initialization finish before
measurement.

All three implementations receive the same row-major BF16 operands and
produce BF16 output with FP32 accumulation, $\alpha=1$, and $\beta=0$.
One active stack per package is used, with its sibling idle. Each node
runs six independent shape workers across packages. Outputs must be
finite and pass relative Frobenius error below 0.02 against FP32
\texttt{torch.mm}; SK and TLA destinations are prefilled with NaN for
the untimed check. All 81 node-shape records pass; maximum error is
$1.663\times10^{-3}$ for all three implementations.

After power conditioning, ten interleaved trials rotate implementation
order. Each implementation uses ten warmups and 30 synchronized host-wall
timings per trial. Inputs are reused without cache flushing. oneDNN uses
\texttt{torch.mm(A,B)}, including its output allocation; SK and TLA write
preallocated output buffers. Timings cover the complete software call
paths. Section~\ref{app:gemm-diagnosis} separates allocation, submission,
and device effects.

Aggregation first takes each node's median of paired within-trial ratios,
then the cross-node median per shape, then the geometric mean across
shapes. Each pair of implementations is aggregated independently.

\begin{table*}[t]
\centering
\caption{Three-node GEMM comparison over 27 BF16-input/BF16-output
shapes, with 95\% confidence intervals from bootstrapping the 27 shapes.}
\label{tab:gemm-tla-summary}
\small
\setlength{\tabcolsep}{6pt}
\begin{tabular}{@{}lrrrr@{}}
\toprule
\textbf{Ratio} & \textbf{Geomean} & \textbf{Median} & \textbf{Shapes $>1$} & \textbf{95\% interval} \\
\midrule
SK / oneDNN & 0.999 & 0.959 & 6/27 & [0.9617, 1.0594] \\
TLA / oneDNN & 0.982 & 0.961 & 6/27 & [0.9426, 1.0405] \\
SK / TLA & 1.016 & 0.998 & 10/27 & [0.9989, 1.0366] \\
\bottomrule
\end{tabular}
\end{table*}

SK and the tested TLA configurations are comparable overall. SK's
geometric-mean advantage is 1.6\%, while the median is close to parity
and the shape-bootstrap interval includes it. SK's square-shape geometric mean relative to TLA
is 1.038, compared with 0.995 on rectangular shapes. The three largest
squares favor SK by 1.079--1.109 times. Table~\ref{tab:gemm-tla-shapes}
reports all 27 shapes.

\begin{table*}[t]
\centering
\caption{Per-shape ratios in the matched three-node comparison.
Each entry is the cross-node median of within-node paired ratios.
Shapes are $M\times K\times N$; displayed values are rounded.}
\label{tab:gemm-tla-shapes}
\small
\setlength{\tabcolsep}{9pt}
\begin{tabular}{@{}lrrr@{}}
\toprule
\textbf{Shape} & \textbf{SK / oneDNN} & \textbf{TLA / oneDNN} & \textbf{SK / TLA} \\
\midrule
$1024^3$ & 1.850 & 1.775 & 1.039 \\
$2048^3$ & 1.014 & 0.819 & 1.234 \\
$3072^3$ & 1.015 & 1.026 & 0.985 \\
$4096^3$ & 0.977 & 0.982 & 0.998 \\
$5120^3$ & 0.944 & 0.944 & 0.998 \\
$6144^3$ & 0.952 & 0.963 & 0.991 \\
$7168^3$ & 0.946 & 0.947 & 1.000 \\
$8192^3$ & 0.959 & 0.975 & 0.984 \\
$9216^3$ & 0.990 & 0.980 & 1.005 \\
$10240^3$ & 0.947 & 0.943 & 1.005 \\
$12288^3$ & 0.959 & 0.889 & 1.079 \\
$14336^3$ & 0.971 & 0.875 & 1.109 \\
$16384^3$ & 0.961 & 0.874 & 1.100 \\
\midrule
$2048\times4096\times12288$ & 0.976 & 0.989 & 0.993 \\
$4096\times4096\times12288$ & 0.950 & 0.952 & 0.998 \\
$8192\times4096\times12288$ & 0.946 & 0.959 & 0.987 \\
$2048\times4096\times4096$ & 0.916 & 0.920 & 0.985 \\
$8192\times4096\times4096$ & 0.953 & 0.953 & 0.995 \\
$2048\times4096\times11008$ & 0.958 & 0.961 & 0.994 \\
$4096\times4096\times11008$ & 0.930 & 0.940 & 0.990 \\
$2048\times11008\times4096$ & 0.916 & 0.930 & 0.990 \\
$4096\times11008\times4096$ & 0.940 & 0.963 & 0.977 \\
$2048\times1024\times4096$ & 0.974 & 0.963 & 1.012 \\
$2048\times4096\times1024$ & 1.113 & 1.060 & 1.026 \\
$256\times4096\times4096$ & 1.066 & 1.056 & 0.996 \\
$512\times4096\times4096$ & 1.143 & 1.114 & 1.000 \\
$512\times4096\times11008$ & 0.996 & 1.025 & 0.987 \\
\bottomrule
\end{tabular}
\end{table*}

Differences in SK/oneDNN ratios relative to the six-node ablation are
concentrated in short shapes, where oneDNN's host call-path timing varies
most across the measurements. The paired SK/TLA ratio uses the two
implementations' directly measured times.

\subsection{Host and Device Performance}
\label{app:gemm-diagnosis}

To identify the costs that determine GEMM throughput, we measure
$1024^3$, $4096^3$, $16384^3$, and $2048\times4096\times11008$ on one
stack of an additional node, using the same SK binaries and BF16
input/output contract. Unprofiled host timing and VTune device profiling
run in separate processes.

\paragraph{Host and device time.}
At $1024^3$, oneDNN's device kernel takes about 19~\us{}, versus 27~\us{}
for SK. Nevertheless, its host submission through \texttt{torch.mm}
takes 34--39~\us{} per call, versus 8--9~\us{} for SK. Both measured
paths launch one kernel per call. Removing output allocation changes
oneDNN by only 1--3~\us{}, and importing IPEX has no material effect on
this node. The remaining submission cost is measured as an aggregate.

\begin{table*}[t]
\centering
\caption{Host and device timing with IPEX imported and warm
operand reuse, in microseconds. Sync wall brackets one call with completion
waits; submission ends when the host call returns; pipelined timing divides
200 queued calls followed by one wait. Host timings are the fastest
of ten trial medians. Device time is a separate VTune per-kernel average,
so these columns are not additive latency components.}
\label{tab:gemm-host-device}
\small
\setlength{\tabcolsep}{5pt}
\begin{tabular}{@{}llrrrr@{}}
\toprule
\textbf{Shape} & \textbf{Path} & \textbf{Sync wall} & \textbf{Submission} & \textbf{Pipelined} & \textbf{Device} \\
\midrule
$1024^3$ & SK & 46.9 & 7.9 & 27.9 & 27.0 \\
 & oneDNN & 68.7 & 37.3 & 35.5 & 19.0 \\
 & oneDNN, preallocated output & 65.2 & 34.0 & 33.3 & 19.0 \\
\midrule
$4096^3$ & SK & 893.1 & 4.0 & 916.4 & 910.0 \\
 & oneDNN & 869.3 & 24.4 & 870.4 & 865.0 \\
 & oneDNN, preallocated output & 867.6 & 22.7 & 870.7 & 865.0 \\
\bottomrule
\end{tabular}
\end{table*}

Lower host submission cost gives SK the small-shape software-path
advantage despite its longer device time. At $4096^3$,
submission is hidden and both paths are device-bound, with oneDNN still
faster on the device.

\paragraph{Clock-scaled matrix ceiling.}
The architectural compute ceiling assumes 448 XVEs per active stack and
512 BF16 FLOP per XVE per clock, scaled by the loaded clock sampled during
each implementation's run. On the three large warm-reuse shapes, SK runs
at 700--800~MHz and achieves 80--96\% of its clock-scaled ceiling; oneDNN
achieves 86--94\%. Relative to the nominal 1600-MHz ceiling, both are
near 40\%. No register spills are reported for these kernels.

\begin{figure*}[t]
\centering
\includegraphics[width=\textwidth]{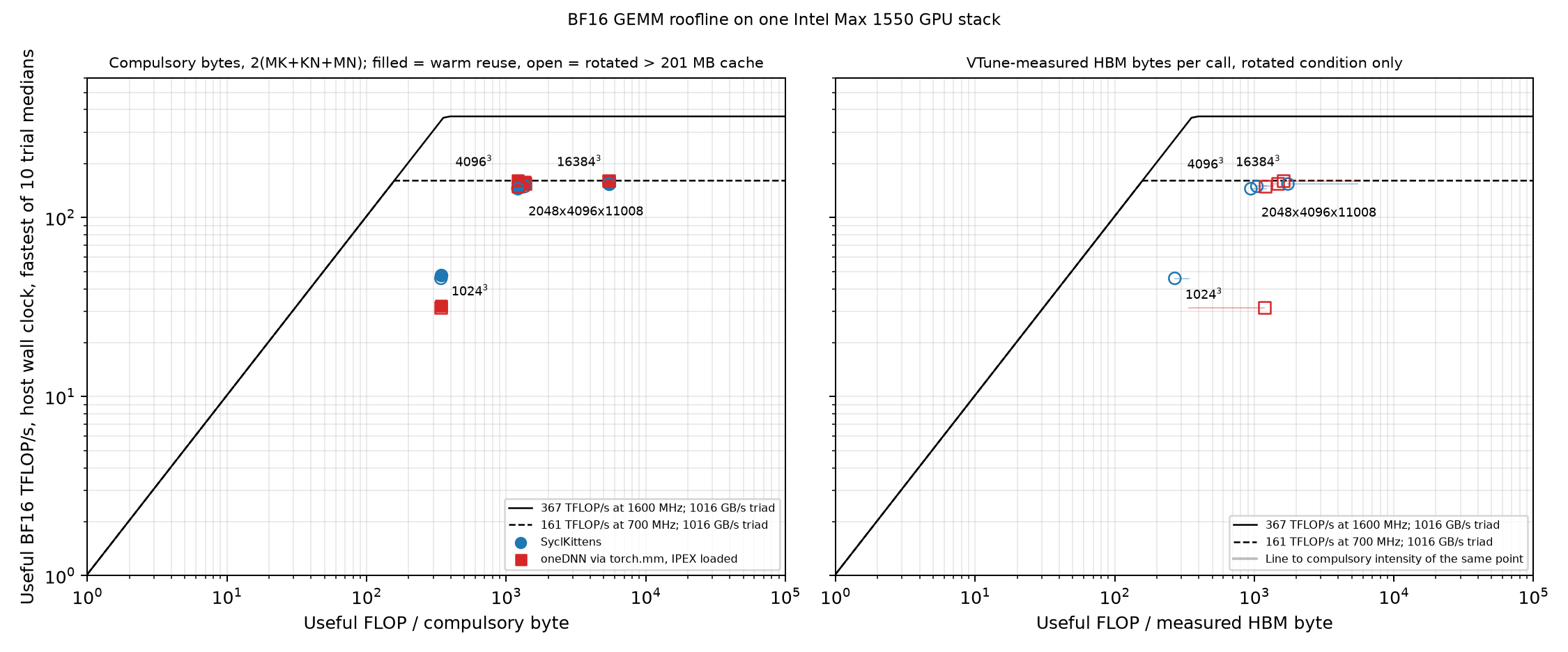}
\caption{Single-stack BF16 GEMM roofline. Host-wall throughput uses the fastest of ten
trial medians. Left, compulsory bytes are $2(MK+KN+MN)$ and points distinguish
warm reuse from a rotated working set larger than the reported 201-MB
cache. Right, rotated-condition intensity uses VTune-measured HBM traffic.
Lines use the architectural matrix rate at nominal 1600~MHz and the
minimum observed 700~MHz, with 1016~GB/s from the fastest of three stream
triads. Per-run utilization uses each run's own loaded clock.}
\Description{Two logarithmic roofline panels compare SK and oneDNN for
four shapes. Large GEMMs approach the loaded-clock matrix ceiling;
the small GEMM remains limited by software-call overhead.}
\label{fig:gemm-clock-roofline}
\end{figure*}

Warm-reuse HBM traffic is near zero for the cache-resident shapes, so
the measured-traffic panel includes only the rotated condition. Measured
bytes are VTune per-kernel read-plus-write bandwidth multiplied by average
kernel time. At $1024^3$, the profiler's
1-ms sampling interval is longer than the kernel, making traffic
attribution approximate.

\subsection{Split-K / Stream-K Design}
\label{app:gemm-splitk}

The \mk{} atomic-combine split-K path targets occupancy-starved decode,
QKV-projection, and MoE-expert shapes with small $M,N$, large $K$, and an
L2-resident output. Eight compile-time K planes reuse the base mainloop and
atomically combine fp32 accumulators before one bf16 output round.

\subsubsection{Delivered performance ($M{=}N{=}512$, K sweep)}

The oneDNN baseline is \texttt{torch.mm} (bf16 in, bf16 out); ``auto''
denotes the auto-selected split ($S{=}8$). All results pass
relative-Frobenius $<2\times10^{-6}$.

\begin{table}[t]
\centering
\small
\caption{Split-K delivered performance, $M{=}N{=}512$, K sweep;
throughput columns are TFLOPS.}
\begin{tabular}{@{}rrrrrr@{}}
\toprule
$K$ & oneDNN & Default & Auto ($S{=}8$) & Auto/oneDNN & Auto/default \\
\midrule
2048  & 46.8  & 20.8 & \textbf{50.0}  & \textbf{1.07\texttimes} & 2.4\texttimes \\
4096  & 65.0  & 22.4 & \textbf{78.4}  & \textbf{1.21\texttimes} & 3.5\texttimes \\
8192  & 100.5 & 23.5 & \textbf{110.5} & \textbf{1.10\texttimes} & 4.7\texttimes \\
16384 & 130.2 & 24.0 & \textbf{143.3} & \textbf{1.10\texttimes} & 6.0\texttimes \\
32768 & 169.9 & 23.8 & \textbf{163.7} & 0.96\texttimes & 6.9\texttimes \\
\bottomrule
\end{tabular}
\end{table}

The default kernel remains at 21--24\,\tflops{} because 128 workgroups
under-fill the machine while the serial mainloop grows with $K$. Split-K is
2.4--6.9\texttimes{} faster than that default and exceeds oneDNN by up to
1.21\texttimes{} for $K\le16384$. At $K=32768$, oneDNN also uses
K-parallelism and reaches 169.9\,\tflops{} versus 163.7\,\tflops{} for
split-K.

\subsubsection{Correctness}

The atomic-add epilogue is bit-exact against the plain-store epilogue in
all 10 unit-test cases (relative-Frobenius error zero).

%% file: sections/supp/appx-attention-fwd.tex
\section{Attention Forward (MHA and GQA)}\label{app:attn-fwd}

The fence-free adaptive forward-attention kernel reaches 71.7~TFLOPS for MHA
and 75.7~TFLOPS for GQA on one GPU. Across non-causal and causal
modes, its geometric-mean speedups over IPEX SDPA are $1.09$--$1.21\times$
for MHA and $1.70$--$4.0\times$ for GQA.

\subsection{Configuration}
\label{app:attn-fwd-config}

\begin{table}[t]
\centering
\caption{Attention-forward configuration (MHA and GQA).}
\label{tab:attn-fwd-config}
\small
\begin{tabularx}{\columnwidth}{@{}>{\raggedright\arraybackslash}p{0.25\columnwidth}>{\raggedright\arraybackslash}X@{}}
\toprule
\textbf{Parameter} & \textbf{Value} \\
\midrule
Problem (MHA) & $B=4,\ H=H_{kv}=16,\ D=128$ (multi-head, no grouping) \\
Problem (GQA) & $B=1,\ H=32,\ H_{kv}=8,\ D=128$ (group size 4; full-sequence queries) \\
Sequence lengths & $N \in \{512, 1024, 2048, 4096, 6144, 8192\}$ \\
Dtype & bf16 inputs/outputs, fp32 accumulation \\
Hardware & \pvc{}, one GPU \\
Peak reference & 157.5~TFLOPS bf16 (oneDNN $512{\times}4096{\times}11008$); MFU quoted against this \\
Baselines & IPEX SDPA (Flash backend), Triton Flash (fp16 operand) \\
Kernel (MHA) & Fence-free adaptive MHA dispatch \\
Kernel (GQA) & Fence-free adaptive dispatch with a 2-D $(N,B)$ GQA branch \\
\bottomrule
\end{tabularx}
\end{table}

MHA is characterized at $B=4$ (throughput regime); the GQA performance tables
in Sec.~\ref{app:attn-fwd-perf} measure batch-one full-sequence queries at $B=1$.
The GQA runtime \emph{selection policy} is a 2-D function of
$(N, B)$ (Sec.~\ref{app:attn-fwd-autotune}), retuned across
$B \in \{1,2,4,8\}$ for full-sequence attention. GQA is compared
against SDPA with IPEX imported and \texttt{enable\_gqa=True}. This suite
measurement uses its grouped-head fallback, which repeats KV heads, rather
than the separately measured pre-expanded-KV fast path. The original Claude Code
agent study uses native PyTorch XPU SDPA without importing IPEX. There is no
Triton grouped-attention baseline.

\subsection{Kernel Implementation and Key Techniques}
\label{app:attn-fwd-kernel}

The MHA implementation is a single-pass, register-resident
FlashAttention-style kernel~\cite{dao2022flashattention} built on
ThunderKittens tile primitives and specialized for
the \pvc{} (Xe) DPAS matrix engine. It is templated on head dimension,
work-group size, query-tile length, and causality, so one compiled module holds
every tuned variant and runtime dispatch selects the measured work-group size
and query-tile length (co-loading separately compiled variants is not possible,
since their generated kernel symbols collide).

\paragraph{Data flow (per query tile).} Each subgroup owns a $\text{seq\_q}
\times D$ query tile and streams the KV sequence in $\text{kv\_stride}=32$
chunks. For every KV tile, the kernel computes $S = QK^\top$ with
\texttt{mma<N,T>}, accumulating $\text{inner\_loop} = D/32 = 4$ DPAS
contractions into an $S[\text{seq\_q}\times 32]$ register tile. It then
applies the optional causal mask, updates the online softmax's running
maximum and normalizer, rescales the output accumulator, and
accumulates $O \mathrel{+}= PV$ with \texttt{mma<N,N>}. After the KV loop,
$O$ is normalized by the running sum and stored.

\paragraph{Key techniques.}
\begin{itemize}
\item \textbf{Online softmax and scaling.} The $N\times N$ score matrix is
  never materialized. The running max (\texttt{max\_vec}), normalizer
  (\texttt{norm\_vec}), and output accumulator remain in registers, with the
  accumulator rescaled by $e^{(m_{\text{old}}-m_{\text{new}})}$ at each KV
  step. Folding $1/\sqrt{D}$ with $\log_2 e=1.44269504$ maps
  $p=\text{exp2}(S\cdot\text{scale}-m)$ directly to the hardware
  \texttt{exp2} unit.
\item \textbf{Register reuse and pipelining.} The
  $\text{seq\_q}\times D$ query tile is loaded once before the KV loop,
  avoiding a reload for every KV tile. Software prefetch issues the
  next K/V loads while DPAS consumes the current tile. Hardware dependency
  tokens order load-to-DPAS and DPAS-to-consumer hazards without software
  scoreboard drains; a 4-seed $\times$ 20-repeat ablation measured up to a
  $\sim$15\% cost from those drains at large-$N$ causal, with no correctness
  benefit.
\item \textbf{DPAS accumulator and operand layouts.} For $D>64$,
  split-D accumulation carries the output in two $D/2$-wide accumulators with
  independent V loads and DPAS, preserving the $32\times32$ DPAS shape and
  register budget. The block loads map row-major A, transposed B for
  \texttt{mma<N,T>}, and VNNI-packed column-major B for \texttt{mma<N,N>}; a
  mismatched role corrupts $S$ or $O$. The transposed-B load lets the matrix
  engine consume $K^\top$ directly without explicit lane-shuffle repacking.
\item \textbf{Query tiling and causal scheduling.} Setting
  $\text{seq\_q}=8$ halves rows per subgroup and doubles concurrent query tiles;
  it is selected at small $N$, with the crossover to $\text{seq\_q}=16$
  determined by the MHA or GQA dispatch table below. The causal loop omits fully masked upper-triangle
  KV tiles and masks only the diagonal tile across
  $\text{kv\_stride}/\text{seq\_q}$ sub-blocks, namely two for
  $\text{seq\_q}=16$ and four for $\text{seq\_q}=8$. Reverse linear query
  ordering dispatches the heaviest causal blocks first to balance workgroups.
\item \textbf{Head mapping and grid residency.}
  Query head $h$ maps to KV head $\lfloor h/(H/H_{kv})\rfloor$; MHA is the
  group-size-one specialization,
  with separate tuned dispatch parameters
  (Sec.~\ref{app:attn-fwd-autotune}). Avoiding forced root synchronization
  lets the full grid run concurrently instead of limiting residency to
  approximately one workgroup per compute core, improving large-$N$
  throughput by up to $\sim$16$\times$.
\item \textbf{Split-K forward.} When KV depth per GPU leaves execution
  resources idle, workers process separate KV partitions and write partial
  $O$/max/norm states. A small log-sum-exp kernel merges them, improving
  performance by 0.6--2.9\% at $N\ge4096$.
\end{itemize}

\subsection{Methodology}
\label{app:attn-fwd-method}

Each $(N,\text{config})$ runs in a separate subprocess pinned to one
GPU, with at most one heavy task per GPU. Baseline, \mk{}, and Triton are
interleaved within each trial; GQA compares only the baseline and \mk{}.
Each of 10 independent trials uses 10 warmup and 30 measured iterations.
TFLOPS is the trial mean, with standard deviation shown where applicable;
GQA values are additionally averaged over three independent runs.

Non-causal forward attention counts $4\,BHN^2D$ FLOPs. The causal count uses
the same coefficient over lower-triangular work, so each path is directly
comparable with its own baseline. Every configuration is checked against a
PyTorch reference before inclusion (Sec.~\ref{app:attn-fwd-correctness}).

\subsection{Delivered Performance}
\label{app:attn-fwd-perf}

Numbers use the validated adaptive runtime policy rather than separately
compiled per-shape kernels.

\subsubsection{MHA (adaptive kernel, $B=4$, $H=16$, $D=128$)}

The model output contract and SDPA baseline require bf16 output, so the reported
measurements include the fp32-to-bf16 epilogue conversion. At 50\% occupancy,
this conversion spills the $\text{seq\_q}=16$ epilogue; omitting it and
returning fp32 is 12\% faster at small $N$ and 4\% at large $N$.
Store-payload reshaping and a compiler register-pressure cap both increased
runtime.

\begin{table}[t]
\centering
\caption{MHA forward, non-causal ($B=4$, $H=H_{kv}=16$, $D=128$).
geomean $1.09\times$ over SDPA; peak 71.7~TFLOPS (45\% MFU) at $N=6144$.
All points average two runs except $N=6144$, which uses one valid run.}
\label{tab:attn-fwd-mha-noncausal}
\small
\begin{tabular}{rrrrrrr}
\toprule
$N$ & \mk{} (T) & IPEX (T) & \mk{}/IPEX & Triton (T) & \mk{}/Triton & MFU \\
\midrule
512  & 49.3 & 48.8 & 1.01$\times$ & 45.7 & 1.08$\times$ & 31\% \\
1024 & 59.2 & 57.1 & 1.04$\times$ & 64.3 & 0.92$\times$ & 38\% \\
2048 & 67.2 & 59.6 & 1.13$\times$ & 68.5 & 0.98$\times$ & 43\% \\
4096 & 69.0 & 61.3 & 1.12$\times$ & 64.8 & 1.06$\times$ & 44\% \\
6144 & 71.7 & 62.6 & \textbf{1.14$\times$} & 69.1 & 1.04$\times$ & 45\% \\
8192 & 70.0 & 62.7 & 1.11$\times$ & 64.6 & 1.08$\times$ & 44\% \\
\bottomrule
\end{tabular}
\end{table}

\begin{table}[t]
\centering
\caption{MHA forward, causal ($B=4$, $H=H_{kv}=16$, $D=128$). Geomean
$1.21\times$ over SDPA; peak 68.7~TFLOPS (44\% MFU) at $N=8192$;
$1.42\times$ at $N=512$. All points average two runs except $N=6144$,
which uses one valid run.}
\label{tab:attn-fwd-mha-causal}
\small
\begin{tabular}{rrrrrrr}
\toprule
$N$ & \mk{} (T) & IPEX (T) & \mk{}/IPEX & Triton (T) & \mk{}/Triton & MFU \\
\midrule
512  & 46.9 & 32.9 & \textbf{1.42$\times$} & 25.1 & \textbf{1.87$\times$} & 30\% \\
1024 & 56.5 & 48.8 & 1.16$\times$ & 42.5 & 1.33$\times$ & 36\% \\
2048 & 61.3 & 52.5 & 1.17$\times$ & 52.4 & 1.17$\times$ & 39\% \\
4096 & 66.4 & 56.6 & 1.17$\times$ & 55.9 & 1.19$\times$ & 42\% \\
6144 & 68.5 & 58.0 & 1.18$\times$ & 58.3 & 1.18$\times$ & 43\% \\
8192 & 68.7 & 58.4 & 1.18$\times$ & 60.7 & 1.13$\times$ & 44\% \\
\bottomrule
\end{tabular}
\end{table}

At $N=512$, the selected causal variant reaches $1.42\times$ IPEX and
$1.87\times$ Triton. Its $\text{seq\_q}=8$ tile doubles concurrent query
tiles, skips fully masked upper-triangle KV blocks, and applies masking only
to the diagonal block, reducing executed $QK^\top$, softmax, and $PV$ work.

\subsubsection{GQA (adaptive kernel, $B=1$, $H=32$, $H_{kv}=8$, $D=128$)}

The tables below characterize full-sequence GQA at $B=1$,
compared against SDPA with grouped-query attention enabled; numbers are the
mean of three independent single-GPU runs. The runtime selection policy
also covers $B \in \{2,4,8\}$ (Sec.~\ref{app:attn-fwd-autotune}), where the
selected variant differs from $B=1$.

\begin{table}[t]
\centering
\caption{GQA forward, non-causal ($B=1$, $H=32$, $H_{kv}=8$, $D=128$).
geomean $1.70\times$ over SDPA; peak 75.7~TFLOPS (48\% MFU) at $N=6144$.}
\label{tab:attn-fwd-gqa-noncausal}
\small
\begin{tabular}{rrrrr}
\toprule
$N$ & \mk{} (T) & IPEX (T) & \mk{}/IPEX & MFU \\
\midrule
512  & $46.3 \pm 0.2$ & 27.2 & \textbf{1.70$\times$} & 29\% \\
1024 & $52.5 \pm 0.0$ & 45.9 & 1.14$\times$ & 33\% \\
2048 & $67.8 \pm 1.0$ & 42.7 & 1.59$\times$ & 43\% \\
4096 & $72.2 \pm 0.6$ & 41.8 & 1.73$\times$ & 46\% \\
6144 & $75.7 \pm 0.3$ & 38.4 & 1.97$\times$ & 48\% \\
8192 & $72.3 \pm 0.1$ & 31.1 & 2.33$\times$ & 46\% \\
\bottomrule
\end{tabular}
\end{table}

\begin{table}[t]
\centering
\caption{GQA forward, causal ($B=1$, $H=32$, $H_{kv}=8$, $D=128$). Geomean
$4.0\times$ over SDPA; peak 72.1~TFLOPS (46\% MFU) at $N=8192$.}
\label{tab:attn-fwd-gqa-causal}
\small
\begin{tabular}{rrrrr}
\toprule
$N$ & \mk{} (T) & IPEX (T) & \mk{}/IPEX & MFU \\
\midrule
512  & $37.1 \pm 0.2$ & 9.5  & \textbf{3.91$\times$} & 24\% \\
1024 & $66.0 \pm 0.1$ & 17.6 & \textbf{3.74$\times$} & 42\% \\
2048 & $69.8 \pm 1.0$ & 19.1 & 3.65$\times$ & 44\% \\
4096 & $69.8 \pm 0.8$ & 19.2 & 3.64$\times$ & 44\% \\
6144 & $71.6 \pm 0.3$ & 17.4 & 4.11$\times$ & 45\% \\
8192 & $72.1 \pm 0.3$ & 13.7 & 5.27$\times$ & 46\% \\
\bottomrule
\end{tabular}
\end{table}

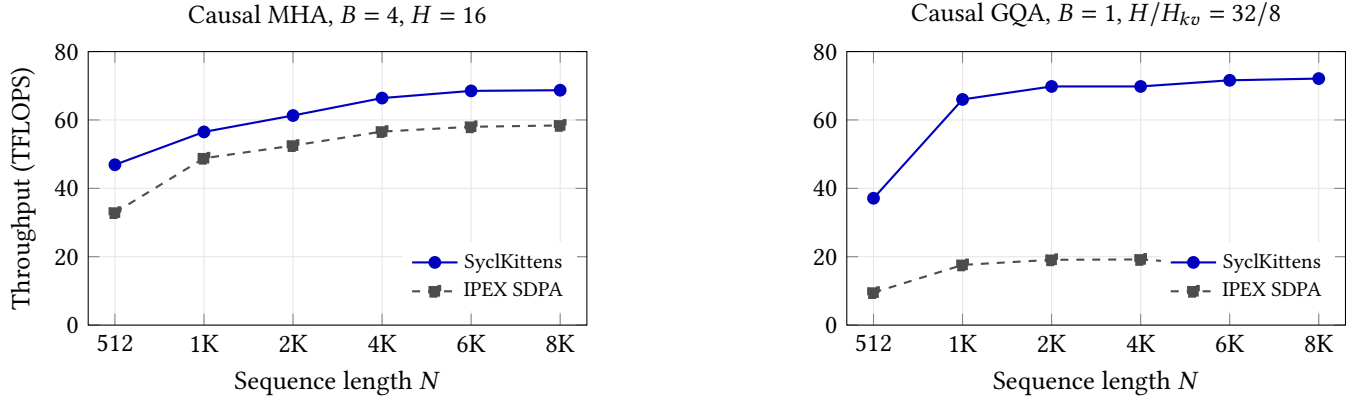
\begin{figure*}[t]
\centering
\begin{tikzpicture}
\begin{axis}[
  width=0.46\textwidth,
  height=5.2cm,
  title={Causal MHA, $B=4$, $H=16$},
  xlabel={Sequence length $N$},
  ylabel={Throughput (TFLOPS)},
  xmin=0.7, xmax=6.3,
  ymin=0, ymax=80,
  xtick={1,2,3,4,5,6},
  xticklabels={512,1K,2K,4K,6K,8K},
  grid=major,
  grid style={gray!20},
  legend style={font=\footnotesize,draw=none,at={(0.98,0.05)},anchor=south east}
]
\addplot[blue!75!black,thick,mark=*] coordinates
  {(1,46.9) (2,56.5) (3,61.3) (4,66.4) (5,68.5) (6,68.7)};
\addlegendentry{\mk{}}
\addplot[black!70,dashed,thick,mark=square*] coordinates
  {(1,32.9) (2,48.8) (3,52.5) (4,56.6) (5,58.0) (6,58.4)};
\addlegendentry{IPEX SDPA}
\end{axis}
\end{tikzpicture}
\hfill
\begin{tikzpicture}
\begin{axis}[
  width=0.46\textwidth,
  height=5.2cm,
  title={Causal GQA, $B=1$, $H/H_{kv}=32/8$},
  xlabel={Sequence length $N$},
  xmin=0.7, xmax=6.3,
  ymin=0, ymax=80,
  xtick={1,2,3,4,5,6},
  xticklabels={512,1K,2K,4K,6K,8K},
  grid=major,
  grid style={gray!20},
  legend style={font=\footnotesize,draw=none,at={(0.98,0.05)},anchor=south east}
]
\addplot[blue!75!black,thick,mark=*] coordinates
  {(1,37.1) (2,66.0) (3,69.8) (4,69.8) (5,71.6) (6,72.1)};
\addlegendentry{\mk{}}
\addplot[black!70,dashed,thick,mark=square*] coordinates
  {(1,9.5) (2,17.6) (3,19.1) (4,19.2) (5,17.4) (6,13.7)};
\addlegendentry{IPEX SDPA}
\end{axis}
\end{tikzpicture}
\caption{Causal BF16 attention scaling on one Intel Max GPU with $D=128$.
  MHA uses $B=4$ and $H=16$; GQA uses $B=1$, $H=32$, and $H_{kv}=8$.
  The baseline is IPEX SDPA, with its grouped-head fallback for GQA.
  \mk{} sustains 66--72 TFLOPS after the
  shortest GQA point, while the single-batch grouped baseline peaks near
  19 TFLOPS and declines at long sequence lengths.}
\Description{Line charts of causal attention throughput in TFLOPS against
sequence lengths from 512 to 8192, comparing SyclKittens with the baseline
paths.}
\label{fig:attn-fwd-causal-scaling}
\end{figure*}

The causal separation follows from the baseline's loss of parallelism at the
single-batch grouped shape beyond $N=4096$. The adaptive GQA kernel retains
parallelism, and its $N\le1024$ causal configurations select
$\text{seq\_q}=8$ (Sec.~\ref{app:attn-fwd-autotune}).

\subsection{Autotune Grid and Selection Policy}
\label{app:attn-fwd-autotune}

The grid sweep covers workgroup size $\in \{64, 128, 256\} \times \text{seq\_q} \in \{8, 16\}$,
swept over all $N$ for both masks.

\begin{table}[t]
\centering
\caption{Selected MHA configurations (compile-time MHA branch, group size 1).}
\label{tab:attn-fwd-mha-champions}
\small
\begin{tabular}{lccc}
\toprule
\textbf{Regime} & $N \le 512$ & $512 < N \le 2048$ & $N \ge 4096$ \\
\midrule
Non-causal & $\text{seq\_q}=8$, wg=256 & $\text{seq\_q}=16$, wg=256 & $\text{seq\_q}=16$, wg=256 \\
Causal     & $\text{seq\_q}=8$, wg=128 & $\text{seq\_q}=16$, wg=128 & $\text{seq\_q}=16$, wg=256 \\
\bottomrule
\end{tabular}
\end{table}

The dispatch function encodes this table in a compile-time MHA branch,
independent of the GQA policy. Handling the narrow $\text{seq\_q}=8$
register tile in the causal masking routine makes this configuration
available and yields the $1.42\times$ result at $N=512$.

\paragraph{GQA 2-D $(N,B)$ policy (compile-time else branch, group
size 4).} Unlike MHA, the selected GQA configuration depends on both sequence
length $N$ and batch $B$. At low grid occupancy, $\text{seq\_q}=8$ doubles
the number of concurrent query tiles. As $B$ fills the machine, the selected
$\text{seq\_q}=16$ crossover moves to smaller $N$, and larger workgroups are
selected more often.

The batch-aware sweep spans $B \in \{1,2,4,8\}$,
$N \in \{256,512,1024,2048\}$, workgroup size $\in \{64,128,256\}$, and
$\text{seq\_q} \in \{8,16\}$ at decoder-model dimensions. Each cell has 24
repetitions and must pass a $\ge 200$~TFLOPS calibration gate with coefficient
of variation $\le 7\%$. The tables report the selected workgroup size and
$\text{seq\_q}$ with median latency in parentheses.

\begin{table}[t]
\centering
\caption{GQA 2-D $(N,B)$ autotune selections, causal. Cells give wg/seq\_q
  (median latency, \textmu s).}
\label{tab:attn-fwd-gqa-causal-grid}
\small
\begin{tabular}{rrrrr}
\toprule
$N \backslash B$ & 1 & 2 & 4 & 8 \\
\midrule
256  & 256/8 (48)   & 128/8 (53)    & 256/8 (71)    & 128/8 (111)   \\
512  & 64/8 (72)    & 128/8 (96)    & 64/16 (159)   & 128/16 (332)  \\
1024 & 128/8 (145)  & 64/16 (266)   & 128/16 (569)  & 256/16 (1168) \\
2048 & 128/16 (473) & 256/16 (1045) & 256/16 (2100) & 256/16 (4249) \\
\bottomrule
\end{tabular}
\end{table}

\begin{table}[t]
\centering
\caption{GQA 2-D $(N,B)$ autotune selections, non-causal. Cells give
  work-items/query rows (median latency, \textmu s).}
\label{tab:attn-fwd-gqa-noncausal-grid}
\small
\begin{tabular}{rrrrr}
\toprule
$N \backslash B$ & 1 & 2 & 4 & 8 \\
\midrule
256  & 256/8 (53)    & 128/8 (73)    & 256/8 (101)   & 128/16 (171)  \\
512  & 128/8 (107)   & 128/8 (162)   & 256/16 (307)  & 256/16 (536)  \\
1024 & 64/8 (296)    & 256/16 (578)  & 256/16 (1017) & 256/16 (2088) \\
2048 & 256/16 (1088) & 256/16 (2007) & 256/16 (3988) & 256/16 (7828) \\
\bottomrule
\end{tabular}
\end{table}

Indexing the policy only by the $B=1$ measurements selects the wrong
configuration after the crossover at $B\ge2$. For causal $N=1024$, applying
the $B=1$ selection (128 work-items and eight query rows) at $B=4$ takes
681~\textmu s, 20\% longer than 128 work-items and 16 query rows at
569~\textmu s; at $B=8$, the difference is 38\%. The 2-D dispatch function
encodes these selections using pre-instantiated variants. Workgroup size
depends on both batch and sequence length, even for a fixed query-tile size.
The $B=1$
selections and the $N>2048$ long-context selection (256 work-items and 16
query rows) are preserved.

\subsection{Correctness}
\label{app:attn-fwd-correctness}

MHA (adaptive kernel, $B=4$, $H=16$, $D=128$, vs.\ a PyTorch SDPA reference).
The table reports maximum absolute error with bf16 inputs and fp32 accumulation.

\begin{table}[t]
\centering
\caption{MHA forward correctness, maximum absolute error vs.\ a PyTorch
  reference.}
\label{tab:attn-fwd-correctness-mha}
\small
\begin{tabular}{lrrrrr}
\toprule
\textbf{Mask} & $N=512$ & 1024 & 2048 & 4096 & 8192 \\
\midrule
Causal & 0.0080 & 0.0078 & 0.0100 & 0.0109 & 0.0103 \\
Non-causal & \multicolumn{5}{c}{$\le 0.0019$ across all $N$} \\
\bottomrule
\end{tabular}
\end{table}

All configurations pass. The causal $\text{seq\_q}=8$ tile passes at
$\le 0.011$.

GQA (adaptive kernel, $B=1$, $H=32$, $H_{kv}=8$, vs.\ a KV-replicated SDPA
reference).

\begin{table}[t]
\centering
\caption{GQA forward correctness, maximum absolute error vs.\ a
  KV-replicated PyTorch reference.}
\label{tab:attn-fwd-correctness-gqa}
\small
\begin{tabular}{lrrrrrr}
\toprule
\textbf{Mask} & $N=512$ & 1024 & 2048 & 4096 & 6144 & 8192 \\
\midrule
Causal     & 0.0134 & 0.0078 & 0.0079 & 0.0091 & 0.0098 & 0.0097 \\
Non-causal & 0.0019 & 0.0010 & 0.0008 & 0.0008 & 0.0005 & 0.0005 \\
\bottomrule
\end{tabular}
\end{table}

All GQA configurations pass (tolerance 0.10); the causal $\text{seq\_q}=8$
path ($N \le 1024$) passes at $\le 0.0134$.

%% file: sections/supp/appx-norms.tex
\section{Memory-Bound Norms, RoPE, and Decode GEMV}
\label{app:norms}

Root mean square layer normalization (RMSNorm)~\cite{zhang2019rmsnorm},
layer normalization (LayerNorm)~\cite{ba2016layernorm}, rotary position
embedding (RoPE)~\cite{su2021roformer}, and single-token GEMV all depend on wide coalesced
memory access, with subgroup cooperation added where a row reduction is
required. Norms and RoPE are compared with eager PyTorch and
\texttt{torch.compile} (TorchInductor to Triton on \pvc{}); decode GEMV is
compared with the tuned vendor library path (oneMKL).

\subsection{Configuration}

RMSNorm uses $D=2048$ and LayerNorm $D=4096$, with bf16 inputs/outputs,
fp32 accumulation, and $\epsilon=10^{-5}$. We test
$(B,N)\in\{(1,1),(1,64),(1,512),(4,1024),(16,2048),(64,2048)\}$ on one
GPU. Plain norms read and write $2BND$ elements; fused residual-add, norm, and
residual writeback read and write $4BND$ elements. The measured kernels assign eight
cooperating subgroups to each row and use 128-bit
\texttt{load\_wide}/\texttt{store\_wide} transactions.

Reported useful traffic is $2\,BND\cdot2$ bytes for plain norm and
$4\,BND\cdot2$ bytes for fused norm at 2 B/elem. This common accounting
convention makes the GB/s columns comparable; it excludes extra intermediate
traffic in implementations that launch multiple kernels.

\subsection{Kernel Implementation and Key Techniques}

Each norm is a single translation unit templated on row width, subgroups per
row, and whether to add and write back the residual. It is built into a
per-width pybind11 module with separate plain and fused entry points.
The kernel is written entirely in \mk{}, with no hand-rolled
intrinsics, on top of the \texttt{load\_wide}/\texttt{store\_wide} 128-bit
vector-memory ops.

\textbf{Data flow (per row).} The grid is $(1,B,N)$ with
$\mathrm{WPR}\cdot16$ threads per block. One work-group owns each row, and
its $\mathrm{WPR}$ subgroups cooperate. Subgroup $w$ owns the contiguous
slice $x[\mathrm{row}, w{\cdot}DC : (w{+}1){\cdot}DC]$ in an
\texttt{rv\_fl<DC, naive>} register vector. This uses $DC/16$ fp32 values
per lane, 16 at $D=2048$ and 32 at $D=4096$, both within the register file.

Each subgroup loads its chunk with \texttt{load\_wide}. In the fused path it
adds the residual chunk and writes the residual sum with
\texttt{store\_wide}. The subgroup then
computes its partial moment or moments and publishes them to SLM. After one
\texttt{group\_barrier}, every subgroup reads all partials, forms the row
statistic, applies the affine transform, and writes its output chunk with
\texttt{store\_wide}.

\textbf{Key techniques.}

\begin{itemize}
\item \textbf{Cooperative multi-subgroup row decomposition.} Splitting the
  row across $\mathrm{WPR}=8$ subgroups reduces per-lane accumulators from
  $D/16$ (single-subgroup layout with 128 at $D{=}2048$, 256 at $D{=}4096$, which
  spills) to $D/(16\,\mathrm{WPR})$ (16 / 32, no spill). This decomposition
  increases measured LayerNorm bandwidth from $\sim$570 to
  $\sim$970--1100 GB/s.
\item \textbf{128-bit coalesced \texttt{load\_wide}/\texttt{store\_wide}.}
  Each lane issues a full \texttt{uint4} (16 B) per transaction; 16 lanes
  touch 256 contiguous bytes in one cache-line-aligned burst (interleaved-128
  mapping), the widest transaction the \pvc{} memory pipe accepts. The old
  scalar \texttt{rv\_fl} fallback issued 32-bit strided loads at
  $\le$half the effective bandwidth. Requires $DC$ a multiple of 128 and a
  16 B-aligned base (satisfied for $D \in \{1024, 2048, 4096\}$).
\item \textbf{SLM moment reduction.} RMSNorm reduces one moment
  ($\sum x^2$) through a $\mathrm{WPR}$-float SLM scratch; LayerNorm reduces
  \emph{two} ($\sum x$, $\sum x^2$) through $2\cdot\mathrm{WPR}$ floats,
  then forms $\mu = \tfrac{1}{D}\sum x$, $\sigma^2 = \tfrac{1}{D}\sum x^2 -
  \mu^2$. A single \texttt{group\_barrier} separates publish from read-back.
\item \textbf{Single-launch fused residual path.} The kernel folds the
  $x + \mathrm{residual}$ add into the load and emits the residual sum, so
  the pre-norm transformer's reused hidden state $h = x + \mathrm{sub}(x)$
  is materialized without a second kernel or a re-read of $h$.
\item \textbf{Register-fused affine + \texttt{rsqrt}.} $\mathrm{inv\_std} =
  \mathrm{rsqrt}(\mathrm{var} + \mathrm{eps})$ maps to the hardware
  reciprocal-sqrt unit; the normalize, scale ($\gamma$), and shift
  ($\beta$, LayerNorm) are fused in registers before the store, with
  $\gamma$/$\beta$ themselves streamed via \texttt{load\_wide}.
\end{itemize}

\subsection{Methodology}

Measurements follow the shared isolation protocol on one selected GPU. Eager
\texttt{F.rms\_norm}/\texttt{F.layer\_norm} decomposes RMSNorm into
multiple kernels; \texttt{torch.compile} lowers the same operations through
Inductor to Triton on XPU.

For the fused comparison, \mk{} writes both the normalized output and
residual sum ($4\times$ traffic). The expression
\texttt{torch.compile(F.rms\_norm(x+res))} discards the sum
($3\times$) and is approximately 30\% faster at identical bandwidth. All
three measured implementations therefore return \texttt{(norm, x+res)} and
use the same useful-byte denominator. An unfused add-then-norm that retains the residual
moves $5\times$ traffic; the fused both-output kernel moves $4\times$, a
20\% traffic reduction.

The three implementations are interleaved round-robin within each of 10
trials, each with 10 warmup and 30 measured iterations; tables report the
per-configuration median. A 2 s GEMM power-condition burn-in precedes
measurement, and the Inductor graph-cache limit is 64 to prevent eager
fallback. All cold-HBM tables use one cache protocol, in which a $\sim$256 MB
buffer is
zeroed before every timed call to evict L2, following Triton's convention, so
all reads miss to HBM. Without eviction, small-$B$ working sets below
the $\sim$204 MB per-GPU L2 report cache bandwidth above the HBM roofline.
Warm steady-state latency is recorded separately for decode.

Traffic is $2\,BND\cdot2$ B for plain norm and $4\,BND\cdot2$ B for fused
norm; GB/s is traffic divided by median time. Before timing, each
configuration checks the plain norm, fused norm, and residual store against
a PyTorch reference with relative error below $2\mathrm{e}{-2}$.

\subsection{Delivered Performance}

The norm tables below report per-configuration cold-HBM medians.
SyclKittens/eager is the speedup over eager PyTorch, and
SyclKittens/compile is the speedup over \texttt{torch.compile}.

\subsubsection{RMSNorm ($D = 2048$)}

\textbf{Plain}, geomean \textbf{1.12$\times$ over torch.compile},
\textbf{15.6$\times$ over eager}.

\begin{table*}[t]
\centering
\small
\caption{RMSNorm, plain, $D=2048$.}
\footnotesize
\begin{tabular}{@{}rrrrrrrr@{}}
\toprule
$B$ & $N$ & SyclKittens (\us{}) & SyclKittens GB/s & eager (\us{}) & SyclKittens/eager & compile (\us{}) & compile GB/s / SyclKittens/compile \\
\midrule
4  & 1024 & 61.8 & 543 & 735   & 11.9$\times$ & 82.6 & 406 / \textbf{1.33$\times$} \\
16 & 2048 & 378  & 711 & 5810  & 15.4$\times$ & 395  & 680 / 1.05$\times$ \\
64 & 2048 & 1090 & 989 & 22400 & 20.7$\times$ & 1090 & 981 / 1.01$\times$ \\
\bottomrule
\end{tabular}

\end{table*}

\textbf{Fused} (residual + norm, $4\times$), geomean \textbf{1.04$\times$
over torch.compile}.

\begin{table*}[t]
\centering
\small
\caption{RMSNorm, fused, $D=2048$.}
\footnotesize
\begin{tabular}{@{}rrrrrrrr@{}}
\toprule
$B$ & $N$ & SyclKittens (\us{}) & SyclKittens GB/s & eager (\us{}) & SyclKittens/eager & compile (\us{}) & compile GB/s / SyclKittens/compile \\
\midrule
4  & 1024 & 112  & 600 & 815   & 7.28$\times$ & 126  & 534 / \textbf{1.13$\times$} \\
16 & 2048 & 655  & 820 & 6700  & 10.2$\times$ & 669  & 803 / 1.02$\times$ \\
64 & 2048 & 2190 & 980 & 25900 & 11.8$\times$ & 2170 & 992 / 0.99$\times$ \\
\bottomrule
\end{tabular}

\end{table*}

\subsubsection{LayerNorm ($D = 4096$)}

\textbf{Plain}, geomean \textbf{1.42$\times$ over torch.compile},
\textbf{2.78$\times$ over eager}.

\begin{table*}[t]
\centering
\small
\caption{LayerNorm, plain, $D=4096$.}
\footnotesize
\begin{tabular}{@{}rrrrrrrr@{}}
\toprule
$B$ & $N$ & SyclKittens (\us{}) & SyclKittens GB/s & eager (\us{}) & SyclKittens/eager & compile (\us{}) & compile GB/s / SyclKittens/compile \\
\midrule
4  & 1024 & 108  & 623 & 255  & 2.37$\times$ & 130  & 517 / \textbf{1.21$\times$} \\
16 & 2048 & 641  & 838 & 1790 & 2.80$\times$ & 921  & 583 / 1.44$\times$ \\
64 & 2048 & 2160 & 993 & 7000 & 3.24$\times$ & 3530 & 608 / \textbf{1.63$\times$} \\
\bottomrule
\end{tabular}

\end{table*}

\textbf{Fused} (residual + norm, $4\times$), geomean \textbf{1.12$\times$
over torch.compile}.

\begin{table*}[t]
\centering
\small
\caption{LayerNorm, fused, $D=4096$.}
\footnotesize
\begin{tabular}{@{}rrrrrrrr@{}}
\toprule
$B$ & $N$ & SyclKittens (\us{}) & SyclKittens GB/s & eager (\us{}) & SyclKittens/eager & compile (\us{}) & compile GB/s / SyclKittens/compile \\
\midrule
4  & 1024 & 210  & 639 & 424   & 2.02$\times$ & 228  & 589 / \textbf{1.08$\times$} \\
16 & 2048 & 1160 & 922 & 3570  & 3.07$\times$ & 1320 & 812 / 1.14$\times$ \\
64 & 2048 & 4310 & 997 & 13700 & 3.18$\times$ & 4900 & 877 / 1.14$\times$ \\
\bottomrule
\end{tabular}

\end{table*}

For the HBM-bound shapes ($B \ge 16$), eager PyTorch sustains
$\sim$45--320 GB/s because its decomposition traverses the data in several
kernels. Triton fusion raises \texttt{torch.compile} to 517--877 GB/s on
LayerNorm, while \mk{} sustains 623--997 GB/s (1.08--1.63$\times$). At the two
large RMSNorm cells, both paths approach the $\sim$1.3 TB/s per-GPU roofline
and differ by 0.99--1.05$\times$; at $B=4$, lower launch overhead gives \mk{}
a 1.13--1.33$\times$ advantage.

\subsection{Decode/Dispatch-Bound Latency}

For few-token decode shapes, each norm touches at most 512 rows and fixed
launch overhead dominates. \mk{} uses one Level-Zero submission;
\texttt{torch.compile} incurs a Dynamo guard and Triton launch, while eager
uses multiple kernels. The preceding matmul or attention leaves the decode
norm input L2-resident, so the table reports warm steady-state latency for
that cache state.

For fused residual-add and norm, \mk{} holds a warm median
$\sim$24 \us{} dispatch floor across the sweep while \texttt{torch.compile}
sits at $\sim$55--61 \us{}.

\begin{table*}[t]
\centering
\small
\caption{Decode-regime latency, fused norm, warm.}
\footnotesize
\begin{tabular}{@{}lrrrrrr@{}}
\toprule
op & $B$ & $N$ & SyclKittens (\us{}) & compile (\us{}) & eager (\us{}) & SyclKittens/compile / SyclKittens/eager \\
\midrule
RMSNorm   & 1 & 1   & 24.3 & 58.0 & 87.6  & \textbf{2.30$\times$} / 3.60$\times$ \\
RMSNorm   & 1 & 64  & 24.6 & 61.3 & 105 & \textbf{2.49$\times$} / 4.25$\times$ \\
RMSNorm   & 1 & 512 & 24.2 & 61.3 & 169 & \textbf{2.53$\times$} / 7.00$\times$ \\
LayerNorm & 1 & 1   & 23.6 & 54.8 & 46.7  & \textbf{2.31$\times$} / 1.98$\times$ \\
LayerNorm & 1 & 64  & 23.7 & 60.9 & 46.5  & \textbf{2.57$\times$} / 1.96$\times$ \\
LayerNorm & 1 & 512 & 23.6 & 61.3 & 66.2  & \textbf{2.60$\times$} / 2.80$\times$ \\
\bottomrule
\end{tabular}

\end{table*}

\mk{}'s near-flat latency (24.2--24.6 \us{} for RMSNorm and 23.6--23.7
\us{} for LayerNorm) reflects one fused submission. The
\texttt{torch.compile} floor is $\sim$2.3--2.6$\times$ higher, while eager
latency grows with $N$ as its decomposed operations re-traverse each row.

\subsection{Correctness}

Every configuration is validated before timing, namely the plain norm and fused
norm against a PyTorch fp32-reference (relative error), and the residual
sum $x+\mathrm{res}$ byte-exactly. Representative maxima appear below.

\begin{table*}[t]
\centering
\small
\caption{Correctness, RMSNorm/LayerNorm.}
\footnotesize
\begin{tabular}{@{}lccc@{}}
\toprule
Check & RMSNorm ($D{=}2048$) & LayerNorm ($D{=}4096$) & Tolerance \\
\midrule
SyclKittens plain rel-err & 2.8e-3 & 1.4e-5 & 2e-2 \\
torch.compile plain rel-err & 2.8e-3 & 1.2e-5 & 2e-2 \\
residual-store rel-err & 0.0 & 0.0 & 2e-2 \\
\bottomrule
\end{tabular}

\end{table*}

All configurations pass; the residual store is exact (both \mk{} and
Inductor write the same $x+\mathrm{res}$). The larger RMSNorm rel-err
(2.8e-3 versus 1.4e-5 for LayerNorm) reflects bf16 rounding of the
un-centered second moment.

\subsection{Rotary Position Embedding (RoPE) with 128-bit Chunk-Pair Mapping}

RoPE is the transformer block's other memory-bound elementwise op. The
128-bit coalesced kernel reaches 1020 GB/s. Relative to
\texttt{torch.compile}, its geometric-mean speedups are $1.09\times$ for cold
bandwidth-bound shapes and $1.98\times$ for warm decode shapes.

\subsubsection{Configuration}

We evaluate halved rotary with $D=128$, $H=32$, bf16 inputs/outputs, and
fp32 computation on the same six $(B,N)$ cells as the norms. Baselines are
eager PyTorch and \texttt{torch.compile}; useful traffic is
$2BNHD\cdot2$ bytes for one input read and output write.

\subsubsection{Wide Chunk-Pair Mapping}

RoPE couples element $d$ with $d+64$ across the two halves of a head. A
contiguous wide load groups eight adjacent bf16 elements, so it never places
a rotary pair on the same lane. The \mk{} \texttt{load\_wide} also requires a
128-element multiple and is designed for reductions, so it cannot be used
unchanged. The kernel instead re-tiles the raw 128-bit packing from the
earlier hand-tuned norm kernel for rotary pairs.

\begin{itemize}
\item \textbf{One work-item owns one 8-element chunk-pair.} Work-item
  $(\mathrm{row}, c)$ takes the $c$-th 8-wide chunk of the first half,
  $x_1[c{\cdot}8 : c{\cdot}8{+}8]$, and its partner chunk in the second half,
  $x_2[c{\cdot}8 : c{\cdot}8{+}8]$, 64 bf16 elements away. Since $D=128$, there are
  8 chunk-pairs per head; consecutive work-items read consecutive 16-byte
  words, so each subgroup's transaction is a fully coalesced 256-byte
  burst.
\item \textbf{Four 16-byte loads, two 16-byte stores, all coalesced.} Each
  item issues one \texttt{uint4} load apiece for its $x_1$, $x_2$,
  \texttt{cos}, and \texttt{sin} chunks. Each load is 16 B, or eight bf16
  elements. The item computes the eight rotary pairs in
  fp32 registers and stores the two output chunks as two \texttt{uint4}
  writes. Every $x$/\texttt{out} transaction is the widest the \pvc{} pipe
  accepts.
\item \textbf{Local pair computation.} RoPE requires no SLM, barrier, or
  shuffle because each chunk-pair is local to one work-item. This mapping
  exposes independent work at high occupancy. It requires $D$ to be a
  multiple of 16 and the base to be 16 B aligned, as it is for contiguous
  $[B,N,H,D]$ bf16 with $D=128$.
\end{itemize}

\subsubsection{Delivered Performance}

RoPE follows the shared interleaved-trial and power-conditioning methodology.
The bandwidth table uses the cold-HBM protocol; the decode table uses warm,
L2-resident inputs. SyclKittens/eager is the speedup over eager, and
SyclKittens/compile is the speedup over \texttt{torch.compile}.

\textbf{Bandwidth} (HBM-bound, L2 flushed before every call, cold), geomean
\textbf{1.09$\times$ over torch.compile}.

\begin{table*}[t]
\centering
\small
\caption{RoPE bandwidth, cold.}
\footnotesize
\begin{tabular}{@{}rrrrrrrr@{}}
\toprule
$B$ & $N$ & SyclKittens (\us{}) & SyclKittens GB/s & eager (\us{}) & SyclKittens/eager & compile (\us{}) & compile GB/s / SyclKittens/compile \\
\midrule
4  & 1024 & 115  & 589  & 370   & 3.24$\times$ & 124  & 544 / \textbf{1.08$\times$} \\
16 & 2048 & 646  & 832  & 3960  & 6.14$\times$ & 678  & 792 / 1.05$\times$ \\
64 & 2048 & 2110 & 1020 & 15800 & 7.46$\times$ & 2440 & 882 / \textbf{1.15$\times$} \\
\bottomrule
\end{tabular}

\end{table*}

At $B{=}64$, \mk{} measures 1020 GB/s and \texttt{torch.compile} measures
882 GB/s; the large-$B$ norm kernels measure 980--997 GB/s.

\textbf{Decode/dispatch-bound} (warm, L2-resident input), latency (\us{}),
geomean \textbf{1.98$\times$ over torch.compile}.

\begin{table}[t]
\centering
\small
\caption{RoPE decode-regime latency, warm.}
\footnotesize
\begin{tabular}{@{}rrrrrr@{}}
\toprule
$B$ & $N$ & SyclKittens (\us{}) & compile (\us{}) & eager (\us{}) & SyclKittens/compile / SyclKittens/eager \\
\midrule
1 & 1   & 26.9 & 51.5 & 90.6 & \textbf{1.92$\times$} / 3.37$\times$ \\
1 & 64  & 27.3 & 58.5 & 90.4 & \textbf{2.15$\times$} / 3.32$\times$ \\
1 & 512 & 30.7 & 57.9 & 99.6 & \textbf{1.89$\times$} / 3.25$\times$ \\
\bottomrule
\end{tabular}

\end{table}

As with the norms, \mk{} holds a near-flat $\sim$27--31 \us{}
single-launch floor across the few-token sweep while \texttt{torch.compile}
sits $\sim$1.9--2.1$\times$ higher. The difference is one Level-Zero
submission versus a Dynamo guard followed by a Triton launch.

\subsubsection{Correctness}

Every config is validated before timing against an fp32 reference
(relative-error tolerance 2e-2). \mk{} computes the halved rotary in fp32 and
rounds to bf16 identically to the reference, making the result bit-exact. The
measured relative error is $0.0$ at all six shapes, matching
\texttt{torch.compile}.

\subsection{Single-Token Decode GEMV ($M=1$)}

The decode-time weight projections ($W_{qkv}$, $W_o$, $W_{gate}$, $W_{up}$,
$W_{down}$) are matrix-vector products $\mathrm{out}[N] = W[N,K] @ x[K]$.
They process one token, so $M=1$. Their arithmetic intensity is approximately
one FLOP per byte of streamed weights. The kernel streams all of $W$
($N{\cdot}K{\cdot}2$ B) once and does $\sim$2 FLOP/element. These five projections account for approximately
$64\%$ of per-token decode latency, and their throughput is determined by bf16
weight-streaming bandwidth.

The DPAS systolic array requires $M\ge8$, so $M=1$ uses at most one eighth
of that matrix shape. For this GEMV comparison, we call oneMKL
\texttt{oneapi::mkl::blas::gemm} directly. Eager \texttt{F.linear} and
\texttt{torch.compile} instead reach the corresponding oneDNN GEMM path through
PyTorch; on this platform the two entry points share the underlying GEMM
implementation.

\subsubsection{Configuration}

We evaluate the five Llama-3.1-8B~\cite{grattafiori2024llama3} $M=1$
projections at TP$=1$, namely $W_{qkv}$
($6144\times4096$), $W_o$ ($4096\times4096$), $W_{gate}$ and $W_{up}$
($14336\times4096$), and $W_{down}$ ($4096\times14336$). Inputs and outputs
are bf16 with fp32 accumulation. The tuned vendor library path is oneMKL GEMM
plus its fp32-to-bf16 cast; \mk{} uses eight split-K subgroups per output
row, 256-element software-prefetched chunks, and a direct bf16 store.

\subsubsection{Split-K Streaming Reduction}

A GEMV row is a dot product, structurally the RMSNorm per-row reduction
with the weight row in place of $x$. The kernel reuses the same \mk{}
wide-load reduction, restructured for the $M=1$ occupancy problem.

\begin{itemize}
\item \textbf{Split-K occupancy.} One workgroup owns each output row and
  $\mathrm{WPR}$ subgroups divide its $K$ reduction. The norms use the same
  decomposition to fit registers; here it also increases parallel work.
  Total launched work is $N{\cdot}\mathrm{WPR}{\cdot}16$ threads,
  so it still scales with $N$, but each row supplies multiple cooperating
  subgroups. This reduces sensitivity to the output-row count relative to
  oneMKL, as the tensor-parallel sweep below shows.
\item \textbf{128-bit coalesced \texttt{load\_wide} over the $K$-slice.}
  Each subgroup streams its contiguous slice $W[n,
  w{\cdot}DC:(w{+}1){\cdot}DC]$ ($DC = K/\mathrm{WPR}$) in
  $\mathrm{CHUNK}$-wide pieces via the same \texttt{load\_wide}
  (\texttt{uint4}/lane, 256 B/subgroup-transaction) that took the norms to
  line rate. A narrow 16-bit, one-workgroup-per-row implementation reaches
  approximately 35\% of that rate.
\item \textbf{Software-prefetched chunks, fp32 accumulate.} Chunk $c{+}1$'s
  \texttt{load\_wide} is issued before the fp32 FMA of chunk $c$ consumes
  chunk $c$, following the GEMM pipeline in Appendix~\ref{app:gemm}. $x$ is
  reused across all rows and stays L2-resident.
\item \textbf{Direct bf16 output.} A single-round
  subgroup sum plus a tiny $\mathrm{WPR}$-float SLM combine (one
  \texttt{group\_barrier}) yields $\mathrm{out}[n]$, written straight to
  bf16. The oneMKL path uses an fp32 scratch buffer and a separate cast
  kernel, adding one launch and $N{\cdot}(4{+}2)$ B of traffic.
\end{itemize}

$\mathrm{WPR}=8, \mathrm{CHUNK}=256$ requires $K \bmod 2048 = 0$ (holds for
all five unsharded shapes); TP-sharded variants must also satisfy this
constraint, and other $K$ fall back to
oneMKL. The kernel is written on the \mk{} \texttt{load\_wide}/\texttt{rv\_fl}/\texttt{sum}
operations, no hand-rolled intrinsics. A hardware L1-prefetch variant of
\texttt{load\_wide} also exists in \mk{} but is not used here. With only
2--7 chunks per slice, the register double-buffer already covers the
latency; measurements showed no benefit and occasional regression from the
extra prefetch.

\subsubsection{Delivered Performance}

Each shape reports the median (p50) over 300 individually event-timed
iterations; the harness also records p95/p99. The table uses the shared
cold-HBM cache protocol, so $W$ is read from HBM on every call.
SyclKittens/oneMKL is the speedup over the oneMKL decode path;
\texttt{\%roof} uses the measured $\sim$1.3 TB/s per-GPU streaming-HBM
reference. The specification is 1.6 TB/s, and the large-operand kernels
above approach a $\sim$990 GB/s asymptote.

\textbf{Cold HBM}, geomean \textbf{1.36$\times$ over oneMKL}.

\begin{table*}[t]
\centering
\small
\caption{Decode GEMV, cold HBM.}
\footnotesize
\begin{tabular}{@{}lrrrrrrr@{}}
\toprule
shape & $N{\times}K$ & oneMKL (\us{}) & oneMKL GB/s & SyclKittens (\us{}) & SyclKittens GB/s & \%roof & SyclKittens/oneMKL \\
\midrule
$W_{qkv}$  & 6144$\times$4096  & 140 & 360 & 99.5 & 506 & 39\% & \textbf{1.40$\times$} \\
$W_o$    & 4096$\times$4096  & 106 & 317 & 69.0 & 486 & 37\% & \textbf{1.54$\times$} \\
$W_{gate}$ & 14336$\times$4096 & 273 & 431 & 221 & 531 & 41\% & 1.23$\times$ \\
$W_{up}$   & 14336$\times$4096 & 274 & 429 & 222 & 530 & 41\% & 1.24$\times$ \\
$W_{down}$ & 4096$\times$14336 & 342 & 344 & 240 & 490 & 38\% & \textbf{1.42$\times$} \\
\bottomrule
\end{tabular}

\end{table*}

Summed per-token GEMM latency (5 GEMVs $\times$ 32 layers, TP$=1$) falls from oneMKL 36.3 ms to
\mk{} 27.2 ms (1.33$\times$). A naive one-work-group-per-row bf16 stream
(narrow 16-bit loads plus a 256-wide reduce) is slower than oneMKL
(42.2 ms, $0.86\times$), establishing that the wide-load and split-K
structure produces the measured improvement.

The kernel is latency-stable, because cold p99 is within $\le$5\% of p50 on every
shape, so the p99 per-token GEMM is 28.4 ms (\mk{}) vs 38.1 ms (oneMKL), the
$1.33\times$ holds at the tail. The L2-resident regime (per-op timed) gives
14.7 ms to 10.7 ms per token, $1.37\times$.

\textbf{Cold and L2-resident regimes.} A single cold GEMV reads 33--117 MB
from an idle memory pipeline and sustains 490--531 GB/s; the 0.5--1 GB norm
streams approach the 990 GB/s long-stream asymptote. With weights resident
in the 204 MB L2 and no flush, \mk{} reaches 1060--1380 GB/s, 82--106\% of
the 1.3 TB/s HBM reference, and 10.7 ms/token. Values above 100\% reflect L2
bandwidth.

The approximately 16-GB model weights do not fit in L2, while the 160
GEMVs per token execute consecutively, placing decode between these two
measurements. Across both regimes, \mk{} exceeds oneMKL, with ratios of
1.24--1.53$\times$ cold and 1.02--1.90$\times$ resident, with an
approximately 1.36$\times$ geometric mean in each. Split-K occupancy and
removal of the cast and scratch path apply in both cache states.

\subsubsection{TP Sharding and Output-Dimension Occupancy}

Inference runs tensor-parallel, so the column-parallel projections
($W_{qkv}$, $W_{gate}$,
$W_{up}$) shard the output dimension $N$ to $N/n$ at TP$=n$. We emulate
TP$=n$ by running the per-shard $(N/n,K)$ GEMV on a single GPU with the same
algorithm. The SyclKittens/oneMKL cold-HBM ratio
rises monotonically with $n$. Values are per-shape cold p50, reproducible
across two runs, with fixed $\mathrm{WPR}=8$.

\begin{table*}[t]
\centering
\small
\caption{TP-sharding SyclKittens/oneMKL cold-HBM ratio (cells are oneMKL /
SyclKittens GB/s, then the ratio).}
\footnotesize
\begin{tabular}{@{}lrllll@{}}
\toprule
proj & $K$ & TP$=1$ & TP$=2$ & TP$=4$ & TP$=8$ \\
\midrule
$W_{qkv}$  ($N{=}6144/n$)  & 4096 & 359 / 503, \textbf{1.40$\times$} & 279 / 496, \textbf{1.78$\times$} & 183 / 410, \textbf{2.24$\times$} & 135 / 323, \textbf{2.39$\times$} \\
$W_{gate}$ ($N{=}14336/n$) & 4096 & 429 / 530, \textbf{1.24$\times$} & 365 / 513, \textbf{1.40$\times$} & 273 / 480, \textbf{1.76$\times$} & 186 / 411, \textbf{2.21$\times$} \\
\bottomrule
\end{tabular}

\end{table*}

oneMKL's $M{=}1$ GEMM parallelism is its output-tile count
$\lceil N/B_N \rceil$, so as $N$ shrinks it loses tiles and its bandwidth
falls from 360 to 135 GB/s for $W_{qkv}$ as $N$ goes from 6144 to 768. \mk{}'s
split-K makes launched work $N\cdot\mathrm{WPR}$ (still $\ge 6144$
subgroups at $N=768$), so it declines only gently (503 to 323). The
ratio widens because oneMKL loses output-tile parallelism faster; both paths
slow at smaller $N$. Ratios separate this occupancy effect from the latency
reduction caused by smaller shards.

Row-parallel $W_o$/$W_{down}$ instead shard $K$ with $N$ fixed at 4096.
Their parallelism and ratio remain unchanged, isolating output-dimension
occupancy as the cause of the widening column-parallel ratio. The ratio
reaches 2.4$\times$ at TP$=8$. A fixed $\mathrm{WPR}=8$ captures this
effect; a per-shape $\mathrm{WPR}(N)$ sweep
found no further gain, since even the smallest shard already saturates at
$\mathrm{WPR}=8$ and larger $\mathrm{WPR}$ only adds combine overhead.

\subsubsection{Correctness}

Each shape is checked against an fp32 reference $\mathrm{out\_ref} =
(W.\mathrm{float}() @ x.\mathrm{float}())$ before timing. \mk{} accumulates
in fp32 and applies one bf16 round on store, so its max-abs error is
identical to the oneMKL path (both round an $O(\sqrt{K})$-magnitude sum
once). The error is in $\{0.50, 0.50, 0.80, 0.50, 0.99\}$ for the five
shapes, reflecting bf16 output quantization. All shapes
match oneMKL to the last bf16 ULP band.

%% file: sections/supp/appx-op-fusions.tex
\section{Op-Fusion Microbenchmarks}
\label{app:op-fusions}

Fusing Q/K rotary combines two launches and allows cached
\texttt{cos}/\texttt{sin} data to be reused across both tensors. We compare this path with
eager PyTorch and \texttt{torch.compile} (TorchInductor to Triton on \pvc{})
using the cold/warm methodology from
Appendix~\ref{app:norms}; Section~\ref{app:e2e-composition} measures the
complete composed path.

\subsection{Fused Q/K Rotary}

Every decoder layer, before attention, applies RoPE to both the projected
query and key. For Mistral-7B~\cite{jiang2023mistral} and
Llama-3.1-8B~\cite{grattafiori2024llama3} GQA, $H=32$ query heads,
$H_{kv}=8$ key heads, and $D=128$. The baseline launches Q and K rotary
separately over the shared $(N,D/2)$ tables. The fused kernel assigns either
a Q or K head-row to each work-item and uses identical halved-rotary math.

\paragraph{Traffic ledger.} Both implementations read Q, write $Q'$, read K,
and write $K'$ exactly once. Launch overhead controls few-token decode;
fusion reduces two launches to one. The shared $(N, D/2)$
\texttt{cos}/\texttt{sin} tables are loaded by both Q and K work-items and can
be reused through the cache within the same launch. Useful traffic excludes
these shared table reads and counts the Q/K tensors. It is
therefore $2\,B\,N(H+H_{kv})D\cdot2$ bytes (read
and write Q and K); at $B{=}64,N{=}2048$ that is 2.68\,GB moved in a single
launch at 973\,\gbps{}.

\subsubsection{Configuration}

\begin{table}[h]
\centering
\footnotesize
\setlength{\tabcolsep}{3pt}
\begin{tabularx}{\columnwidth}{@{}>{\raggedright\arraybackslash}p{0.24\columnwidth}>{\raggedright\arraybackslash}X@{}}
\toprule
Parameter & Value \\
\midrule
Problem & Mistral-7B / Llama-3.1-8B Q/K geometry: $H=32$, $H_{kv}=8$, $D=128$; shared \texttt{cos}/\texttt{sin} $(N, D/2)$ \\
Sweep & decode/dispatch-bound $(B,N)\in\{(1,1),(1,64),(1,512)\}$; bandwidth/HBM-bound $\{(4,1024),(16,2048),(64,2048)\}$ \\
Dtype & bf16 in/out, fp32-internal rotary math; \texttt{cos}/\texttt{sin} bf16 $(N, D/2)$ \\
Convention & Halved rotary: $o_d = x_{1,d}c_d - x_{2,d}s_d$, $o_{d+64} = x_{2,d}c_d + x_{1,d}s_d$ \\
Hardware & \pvc{}, one GPU, flat device hierarchy \\
Baselines & eager PyTorch (two halved rotaries on Q and K); \texttt{torch.compile} (Inductor to Triton, XPU) of the same op-sequence \\
Launch & One fused Q/K launch with $D=128$ and four workers \\
\bottomrule
\end{tabularx}
\end{table}

\subsubsection{Kernel implementation and key techniques}

The kernel reuses the 128-bit coalesced chunk-pair rotary worker from
Appendix~\ref{app:norms} and fuses Q and K by laying their
work items end to end in one \texttt{nd\_range}, so the linear-id space is the
concatenation of Q's $B\,N\,H\cdot\text{CHUNKS}$ items and K's
$B\,N\,H_{kv}\cdot\text{CHUNKS}$ items. A single \texttt{tid < q\_threads}
test routes each work-item to its tensor with the correct head count; both
branches call the same per-row worker. One work-item owns one 8-element
chunk-pair of one head-row; with $D=128$, $(D/2)/8 = 8$ chunk-pairs per
row. RoPE couples element $d$ with $d{+}64$, so each item loads its
first-half chunk \texttt{x1} and its second-half partner \texttt{x2} (64
bf16 apart) plus the shared \texttt{cos} and \texttt{sin} chunks. That is four
coalesced 16-byte loads, the rotary math in fp32 registers, then two
16-byte stores.

\begin{lstlisting}
// ONE nd_range whose linear-id space concatenates the Q and K work items:
// q_threads = B*N*H*CHUNKS come first, then k_threads = B*N*HKV*CHUNKS.
// CHUNKS = (D/2)/8 chunk-pairs per row (8 at D=128).

// grid dispatch: one branch selects the sub-grid, both call one worker
const long tid = item.get_global_linear_id();
if (tid < q_threads)                                  // Q sub-grid: heads = H
    rotary_row_chunk<D>(q, q_out, cos, sin, /*heads=*/H,   N, tid);
else                                                  // K sub-grid: heads = HKV
    rotary_row_chunk<D>(k, k_out, cos, sin, /*heads=*/HKV, N, tid - q_threads);

// the shared per-row worker (identical math to the unfused rotary)
const long row   = local_tid / CHUNKS;
const int  chunk = local_tid % CHUNKS;
const int  coff  = chunk * 8;                         // bf16 offset within a half
const long row_off = row * D;                         // (.,N,heads,D) row base
const int  n  = (row / heads) % N;                    // seq index for cos/sin
const long cs = n * (D / 2) + coff;                    // shared cos/sin base

const wide8_t ra = xw[(row_off + coff) / 8];
const wide8_t rb = xw[(row_off + D / 2 + coff) / 8];
const wide8_t rc = cw[cs / 8], rs = sw[cs / 8];
float a[8], b[8], c[8], s[8];
unpack8(ra, a); unpack8(rb, b); unpack8(rc, c); unpack8(rs, s);

float o1[8], o2[8];
#pragma unroll
for (int i = 0; i < 8; ++i) {
    o1[i] = a[i] * c[i] - b[i] * s[i];               // rotate first half
    o2[i] = b[i] * c[i] + a[i] * s[i];               // rotate second half
}
ow[(row_off + coff) / 8]         = pack8(o1);        // two coalesced stores
ow[(row_off + D / 2 + coff) / 8] = pack8(o2);
\end{lstlisting}

The concatenated id space reduces two Level-Zero submissions to one and permits
cached reuse of the shared \texttt{cos}/\texttt{sin} data across Q and K rows. Each
work-item uses 128-bit coalesced transactions, the widest accepted by the
\pvc{} memory path, and requires no SLM, barrier, or shuffle because one
work-item owns each rotary pair. Both paths call the same fp32
\texttt{rotary\_row\_chunk} worker.

\subsubsection{Methodology}

One heavy task per GPU package, with one stack active, so there is no
intra-package HBM contention.
The baselines are eager PyTorch and
\texttt{torch.compile(backend="inductor")} applied to the same function.
Interleaved round-robin timing within each trial keeps implementations under
the same clock conditions, controlling drift in the
SyclKittens/eager and SyclKittens/compile ratios. Each of 10 trials uses 10
warmup and 30 measured iterations. Each config is measured twice, once warm
(operands stay
L2-resident, the decode metric) and once cold (a scratch buffer is zeroed to
flush L2 before every timed call, following Triton's cold-cache convention).
The cold panel reports HBM bandwidth for the bandwidth-bound shapes; the warm
panel reports few-token decode latency. Before timing, rotated Q and K from
both \mk{} and \texttt{torch.compile} are checked against an fp32 reference
with relative-error tolerance 2e-2.

\subsubsection{Delivered performance}

Per-config medians come from 10 co-measured trials of eager PyTorch,
\texttt{torch.compile}, and SyclKittens. Bandwidth-bound shapes are reported
as cold, L2-flushed HBM bandwidth; the few-token decode shapes as warm
dispatch latency. SyclKittens/eager is the speedup over eager, and
SyclKittens/compile is the
speedup over \texttt{torch.compile}.

\begin{table}[h]
\centering
\footnotesize
\setlength{\tabcolsep}{2.5pt}
\caption{Fused Q/K rotary, bandwidth-bound shapes, cold HBM bandwidth
(L2 flushed). Geomean $1.41\times$ over \texttt{torch.compile},
$9.03\times$ over eager; SyclKittens peaks at 973\,\gbps{}.}
\begin{tabular}{rrrrrrrrr}
\toprule
$B$ & $N$ & SyclKittens $\us$ & SyclKittens $\gbps$ & eager $\us$ & SyclKittens/eager & compile $\us$ & compile $\gbps$ & SyclKittens/compile \\
\midrule
4  & 1024 & 228  & 369 & 2810  & 12.2$\times$ & 385  & 219 & 1.69$\times$ \\
16 & 2048 & 878  & 765 & 6800  & 7.74$\times$ & 1110 & 605 & 1.26$\times$ \\
64 & 2048 & 2760 & 973 & 21500 & 7.77$\times$ & 3650 & 735 & 1.32$\times$ \\
\bottomrule
\end{tabular}
\end{table}

\begin{table}[h]
\centering
\footnotesize
\setlength{\tabcolsep}{3pt}
\caption{Fused Q/K rotary, decode/dispatch-bound shapes, warm latency.
Geomean $2.06\times$ over \texttt{torch.compile}, $17.2\times$ over eager.}
\begin{tabular}{rrrrrrr}
\toprule
$B$ & $N$ & SyclKittens $\us$ & compile $\us$ & eager $\us$ & SyclKittens/compile & SyclKittens/eager \\
\midrule
1 & 1   & 149 & 299 & 2620 & 2.13$\times$ & 18.3$\times$ \\
1 & 64  & 138 & 283 & 2370 & 2.05$\times$ & 17.3$\times$ \\
1 & 512 & 148 & 293 & 2390 & 1.98$\times$ & 16.2$\times$ \\
\bottomrule
\end{tabular}
\end{table}

Relative to \texttt{torch.compile}, fused Q/K rotary is 1.26--1.69$\times$
faster for the cold bandwidth-bound shapes and 1.98--2.13$\times$ faster for
the warm decode shapes, where its latency remains 138--149\,\us{}. The gain
comes from one joint Q/K launch and reuse of the shared cosine/sine data;
\texttt{torch.compile} emits separate guarded rotary launches, while eager
decomposes each rotary into several elementwise kernels.

\subsubsection{Correctness}

Against the fp32 reference rounded to bf16, fused \mk{} is bit-exact
(relative error $0.0$) at all six benchmark shapes, while
\texttt{torch.compile} matches within tolerance. Its maximum absolute
difference is 0.0 from both the two-launch unfused rotary path and
\texttt{transformers.apply\_rotary\_pos\_emb}, with standalone coverage over
$B\in\{1,2,4,8\}\times N\in\{256,512,1024,2048\}$.

%% file: sections/supp/appx-dist-gemm.tex
\section{Distributed GEMM}
\label{app:dist-gemm}

The operation starts with row-sharded $A$ and replicated $B$ and must return the
full replicated $C=AB$ on every GPU. Within the original hand-designed
input-gather family, the topology-aware path combines an all-gather of $A$ with
the \mk{} collaborative de-duplicated-prefetch GEMM core. Each source stripe
crosses XeLink once per remote GPU package, is replicated to the partner stack
over MDFI, and is then consumed by one local GEMM. On 12 GPUs, this path
reaches 84\% and 93\% of its co-measured GEMM-only baseline at $6144^3$ and
$12288^3$. The controlled agent study also explores computing
local $C$ shards first and gathering the output.

\subsection{Configuration}

A node contains six GPU packages, each with two stacks connected by MDFI,
the load/store stack-to-stack interconnect. XeLink connects the GPU packages
all-to-all through two independent stack planes, even and odd.

\begin{table}[h]
\centering
\small
\begin{tabular}{@{}p{0.22\linewidth}p{0.7\linewidth}@{}}
\toprule
Parameter & Value \\
\midrule
Problem & Row-sharded GEMM, replicated output:
  $A_{\text{local}}[d]: \text{bf16}[M/12,K]$, replicated
  $B[d]:\text{bf16}[K,N]$; every GPU returns the full
  $C[d]=AB:\text{bf16}[M,N]$ \\
Shapes & $M{=}N{=}K \in \{6144,\,12288\}$ (square); shard rows per GPU
  $\text{rpd} = M/12 \in \{512,\,1024\}$ \\
Dtype & bf16 inputs/outputs, fp32 accumulation (in-register RNE
  down-convert on store) \\
Topology & Six GPU packages $\times$ two stacks; intra-package MDFI and
  all-to-all inter-package XeLink over the even and odd stack planes \\
Peak reference & Co-located \mk{} GEMM-only reference (A already local, no
  communication, all 12 GPUs concurrent): 139\,T @$6144^3$, 135\,T
  @$12288^3$. Isolated GEMM on one GPU is approximately 158--160\,T; the
  12-GPU-concurrent baseline is lower because each GPU package's two stacks
  share its power/clock budget. \\
Hand GEMM core & bf16 register-resident $32{\times}64{\times}32$ DPAS tiling
  with snake swizzle, reused verbatim from the single-GPU kernel \\
Method & Hand input-gather family; agent local-$C$/output-gather family;
  engineer-agent refinement of the \mk{} + examples kernel \\
\bottomrule
\end{tabular}
\end{table}

TFLOP/s uses the semantic operation count $2MNK/t$ per GPU because every
GPU returns the full replicated $C$. Input-gather methods physically
recompute those FLOPs on every GPU; the local-$C$ family computes one row
band per GPU and gathers the completed bands.

\subsection{Kernel Implementation and Key Techniques}

For the original hand-designed family, the GEMM kernel is DPAS-bound and
reaches approximately 90\% of its local roofline. Its distributed overhead
therefore comes from moving sharded $A$ across the fabric. These methods use
the same compute kernel and differ in how and when they place $A$ in local HBM.

\paragraph{Input-gather data flow (per GPU).} (1) each GPU owns one $M/12{\times}K$
row-shard of $A$ in local HBM; (2) the missing 11 shards are gathered from
peers into a contiguous $M{\times}K$ landing buffer; (3) one
register-resident GEMM computes the full $C = A_{\text{full}}B$ over all
$M$ rows. The methods trade off how step (2) overlaps step (3).

\paragraph{Key techniques}

\begin{itemize}
\item \textbf{Topology-aware gather.} Cross XeLink once per stripe and
  replicate within each GPU package. The 12 stacks are organized as six
  GPU packages with two stacks each; each pair shares a fast intra-package MDFI link outside
  the XeLink fabric. Flat all-gather has every stack pull all 11 remote
  shards, roughly 10 fabric crossings per stack. The topology-aware path sends each source
  stripe across XeLink exactly once to one landing stack per remote
  GPU package (\texttt{landing = 2*q + (p\&1)}), then the landing stack
  replicates the shard to its partner (\texttt{partner(t) = t\^{}1}) over
  MDFI. This halves XeLink egress on the communication-bound workload,
  giving 1.42\texttimes\ ($12288^3$) to 2.15\texttimes\ ($6144^3$)
  end-to-end distributed-GEMM speedup over the flat gather pipeline in the
  main paper's fixed-core movement ablation.
\item \textbf{Within input-gather schedules, materialize-then-GEMM.} The topology-aware path gathers all shards to local HBM
  and then runs one GEMM.
  Interleaving the gather into the GEMM mainloop or reading remote $A$ per
  DPAS inside the kernel is slower in the measured implementations. Remote
  operand reads stall the DPAS pipeline, while materialization lets the
  compute kernel use its local operand-delivery path.
\item \textbf{Gather topology.} DMA \texttt{memcpy} gather and EU
  \texttt{vec128} copy gather land within 1\% of each other at both shapes;
  both are fabric-bound. The topology-aware path changes the traffic topology.
\item \textbf{Large shapes hide communication.} At $12288^3$ the GEMM is
  big enough that even flat gather reaches 63\% of the baseline and the
  topology-aware path reaches 90\%; at $6144^3$ the
  communication-to-compute ratio is higher, so flat gather collapses to
  38\% while the topology-aware path reaches 82\%.
\item \textbf{Compute-core reuse.} The topology-aware path uses the bf16 GEMM kernel
  (snake-swizzled $32{\times}64{\times}32$ DPAS, per-shape snake width)
  verbatim; no distributed-specific compute changes, so the single-GPU
  tuning carries over.
\end{itemize}

\subsection{Methodology}

All 12 GPUs run concurrently in a single process (12 SYCL queues, one per
GPU), so the measurement includes intra-node fabric and shared power
contention. Each method is timed as the
median of 30 iterations after 10 warmup iterations and a 2\,s turbo
burn-in, identical to the single-GPU GEMM reference protocol. Every method
is co-measured in the same run against the GEMM-only baseline, so all
ratios are within-run. Inputs use Box-Muller $N(0,1)$ random initialisation
(matching \texttt{torch.randn}) for the performance numbers; a separate
constant-fill run gates correctness analytically (see the correctness
discussion below). Per-GPU throughput is $2MNK / t$.

\subsubsection{Controlled coding-agent task}
\label{app:dist-gemm-agent-task}

Appendix~\ref{app:agent-study-ag-gemm} reports the controlled coding-agent
evaluation of the $6144^3$ row-sharded operation, including its task
definition, protocols, and correctness-gated results.

\subsection{Engineer-Agent Co-Design}
\label{app:dist-gemm:codesign}

Our first engineering pass refined a kernel from the \mk{} context.
A subsequent common-harness comparison with Box--Muller $N(0,1)$ inputs
measured the best \mk{} and \mk{} + examples kernels alongside that initial
refinement. The \mk{} + examples kernel reached 131.616~TFLOP/s, compared
with 123.207 for the \mk{} kernel and 128.545 for the initial refinement.

We restarted from the unchanged \mk{} + examples kernel and retained it as
the incumbent. The selected refinement preserves its DPAS compute core and staggered
even/odd-plane cross-package gather. It changes only the MDFI replication
phase, in which package bands are visited cyclically, with the first two copies
assigned to the compute queue and the remaining four to the copy queue. Candidate
promotion requires both a paired ratio above one and an absolute
median-of-node-medians no lower than the incumbent.

\begin{table}[t]
\centering
\small
\caption{Engineer-agent refinement of the \mk{} + examples kernel at $6144^3$.
Values are medians across four nodes of each node's
seven-repetition median; the ratio is paired within repetition.}
\label{tab:dist-gemm-codesign}
\begin{tabular}{@{}lrr@{}}
\toprule
Source & TFLOP/s & vs. incumbent \\
\midrule
Agent kernel (incumbent) & 123.839 & $1.000\times$ \\
\textbf{Refined kernel} & \textbf{125.724} &
  $\mathbf{1.007\times}$ \\
\bottomrule
\end{tabular}
\end{table}

Each repetition uses a 2-second burn-in, 10 warmups, and the median of 30
timed operations, with source order rotated across pairs. Both sources pass
4,096 sampled references on all 12 outputs before and after timing in all 56
executions; maximum post-timing relative L2 error is
$8.67{\times}10^{-7}$. The margin is small next to run-to-run variation. Across
the 28 paired repetitions the median ratio is $1.007\times$ and the $95\%$
confidence interval on that ratio spans $0.93\times$ to $1.04\times$, so the
refined kernel is best read as matching the incumbent rather than beating it. We report it
as the certified engineer-agent source because it wins the promotion
protocol.

The study suggests enumerating collective placement over inputs, partial
results, and outputs before tuning schedules, carrying topology routing across
the chosen placement, and then retuning for the resulting local shard geometry.
Our refinement protocol retains the strongest validated agent source as a
selectable incumbent throughout tuning. An engineered variant is promoted
only after it wins the same input, correctness, timing, and reduction protocol.

\subsection{Topology-Aware All-Gather with Collaborative Prefetch}
\label{app:dist-gemm:coop}

This path uses the same collaborative de-duplicated prefetch core as
Appendix~\ref{app:gemm}. This cooperative-prefetch measurement is separate
from the earlier fixed-core movement ablation's 82\% and 90\% endpoints.
Results are three-repetition medians on one quiet
node; the distributed measurement includes the complete topology-aware
all-gather and is normalized to the co-measured GEMM-only baseline.

\begin{table}[ht]
\centering
\small
\caption{Production distributed-GEMM run with Box--Muller $N(0,1)$ inputs and
the cooperative-prefetch build. ``GEMM-only'' is the local \mk{} compute
reference; ``distributed'' includes topology-aware all-gather of $A$ followed
by the full \mk{} GEMM on every GPU.}
\label{tab:dist-gemm-production}
\begin{tabular}{@{}lrrrr@{}}
\toprule
Shape & \mk{} GEMM-only & \mk{} distributed & Dist./GEMM & Snake width \\
 & (T/GPU) & (T/GPU) & & \\
\midrule
$6144^3$  & 134.9 & 113.1 & 84\% & 2 \\
$12288^3$ & 134.7 & 124.9 & 93\% & 4 \\
\bottomrule
\end{tabular}
\end{table}

The larger shape amortizes communication and reaches 93\% of the local
compute ceiling. The smaller shape is communication-bound but still retains
84\%. Both shapes pass the analytic constant-fill correctness gate.
These measurements use random-normal inputs and the cooperative-prefetch
build. The coding-agent reference in Appendix~\ref{app:agent-study-ag-gemm}
uses the evaluator's deterministic input distribution and a masked-prefetch
configuration.

\paragraph{Input distribution and prefetch scheduling.}
We compare both configurations using the same source revision and compiler,
co-measuring them on four nodes with three repetitions per condition,
a 2-second burn-in, 10 warmups, and 30 timed operations. Table~\ref{tab:dist-gemm-rootcause}
reports the cross-node median of each node's repetition median at $6144^3$.

\begin{table}[t]
\centering
\small
\caption{Controlled decomposition of the two distributed-GEMM operating
points. ``Eval'' draws from 32 values in $[-1,0.9375]$ at $1/16$ spacing;
``randn'' is Box--Muller $N(0,1)$.}
\label{tab:dist-gemm-rootcause}
\begin{tabular}{@{}llrr@{}}
\toprule
Build / snake width & Inputs & GEMM-only & Distributed \\
 & & (T/GPU) & (T/GPU) \\
\midrule
Cooperative / 2 & randn & 137.5 & 113.5 \\
Cooperative / 2 & Eval & 156.2 & 125.3 \\
Masked / 8 & Eval & 159.8 & 128.0 \\
\bottomrule
\end{tabular}
\begin{minipage}{\columnwidth}
\vspace{0.5\baselineskip}
Masked prefetch with snake width 8 and Eval inputs is $1.130\times$ faster
than cooperative prefetch with snake width 2 and randn inputs in paired
measurements across the four nodes.
Holding the cooperative build and snake width fixed, changing only randn to
Eval inputs contributes $1.109\times$; holding Eval inputs fixed, changing to
the masked-prefetch, snake-width-8 configuration contributes another
$1.022\times$. The cooperative build itself is $1.033\times$ faster than the
masked build at matched snake width 2 and randn inputs. The difference between
the two operating points is primarily due to operand-dependent power behavior
on \pvc{}, with a smaller contribution from the schedule difference. The
cooperative/randn result of 113.5~TFLOP/s is consistent with the
113.1~TFLOP/s measurement in Table~\ref{tab:dist-gemm-production}.
\end{minipage}
\end{table}

\subsection{Correctness}

A separate constant-fill run ($A_{\text{local}}[d] = (d+1)\cdot 0.01$,
$B = 0.01$) gives an analytic reference. The cooperative-prefetch
GEMM-only and topology-aware distributed paths pass the max-diff $<0.5$ gate; the
in-register bf16 RNE store is the only rounding source.

%% file: sections/supp/appx-cp-attention.tex
\section{Context-Parallel Attention (Ring and All-Gather)}\label{app:cp-attn}

Context-parallel attention combines the fence-free core from
Appendix~\ref{app:attn-fwd} with ring and all-gather KV distribution. Their
memory footprints differ, so both paths are measured under the same
performance and correctness gates on three clean nodes.

\begin{itemize}
\item \textbf{Ring attention.} Each active GPU holds one KV shard and
  rotates shards around the XeLink ring, so per-GPU memory is
  O($N_{\text{per}}$). It supports contexts beyond one GPU's HBM.
\item \textbf{All-gather attention.} Each active GPU gathers the full
  $N_{\text{total}}$ KV once (fan-out push all-gather) and runs a single
  flash pass, so per-GPU memory is O($N_{\text{total}}$). When the gathered
  context fits in HBM, it pays the inter-package cost once instead of at each
  ring step.
\end{itemize}

Measurements use a clock-gated, three-node, ten-trial harness. The PyTorch
paths pair ring (send/receive) and all-gather communication with both the
autotuned 2-D-block-pointer Triton flash kernel used in
Appendix~\ref{app:attn-fwd} and an IPEX/aten fused flash kernel. Every
functional variant passes the same correctness gate.

\subsection{Configuration}
\label{app:cp-attn-config}

\begin{table}[t]
\centering
\caption{Context-parallel attention configuration.}
\label{tab:cp-attn-config}
\small
\begin{tabularx}{\columnwidth}{@{}>{\raggedright\arraybackslash}p{0.29\columnwidth}>{\raggedright\arraybackslash}X@{}}
\toprule
\textbf{Parameter} & \textbf{Value} \\
\midrule
Problem & $H = H_{\text{kv}} = 16$, $D = 128$, non-causal full attention over the ring; $B \in \{1,4\}$ \\
Per-GPU shard & $N_{\text{per}} \in \{256, 1024, 2048, 4096, 8192, 16384\}$ for the $B=1$ sweep (six GPU packages, one active stack each) \\
Total context & $N_{\text{total}} = 6\,N_{\text{per}} \in \{1536,\dots,98304\}$ \\
Dtype & bf16 Q/K/V, fp32 output/LSE and accumulation \\
Topology & Six GPU packages, one active stack each, inter-package XeLink ring \\
Peak reference & 157.5~TFLOPS bf16 per active GPU (single-GPU attention roofline) \\
\mk{} kernels & O($N_{\text{per}}$)-memory ring attention and O($N_{\text{total}}$)-memory all-gather attention; both use the fence-free 256-work-item flash core with 16-query blocks \\
Baselines & PyTorch ring: \texttt{sendrecv/aten}, \texttt{sendrecv/triton}, \texttt{async/aten}; PyTorch all-gather: \texttt{allgather/aten} (tuned oneCCL library path), \texttt{allgather/triton} \\
\bottomrule
\end{tabularx}
\end{table}

Per-GPU TFLOPS uses $4\,BH\,N_{\text{per}}N_{\text{total}}D$ because each
GPU's query shard attends the full context; this is directly comparable
across methods and baselines.

\subsection{Kernel Implementation and Key Techniques}
\label{app:cp-attn-kernel}

Ring attention shards context across six GPU packages, with one active GPU
stack each. At every step, each GPU computes partial attention for its local
queries against the resident KV shard, rotates that shard one XeLink hop, and
updates the online-softmax state. The ring implementation pairs independently
tuned collective and compute implementations. Both methods use ThunderKittens tile
primitives.

\paragraph{Data flow (per ring step, per active GPU).} Each step has three
stages. First, the fence-free 256-work-item flash core with 16-query blocks from
Appendix~\ref{app:attn-fwd} computes partial $O$/LSE for the local queries and
resident KV shard. Second, the kernel merges the max, normalizer, and output
accumulator into the running online-softmax state. Third, a double-buffered
bulk \texttt{memcpy} moves the next KV shard from the ring neighbor through
the shared SYCL context into the alternate local-HBM buffer. The in-order
queue serializes compute and transfer, so the next step reads local HBM.
The overlapped follow-up below uses copy engines to overlap rotation with
compute.

\paragraph{Key techniques.}
\begin{itemize}
\item \textbf{Bulk-memcpy KV rotation.} KV shards are copied in bulk to local
  HBM and attended locally. A measured zero-copy variant instead reads K/V
  over XeLink for each DPAS operation. Because these reads do not overlap
  asynchronously with the DPAS engine, that variant stalls compute and is
  5--50$\times$ slower.
\item \textbf{Double-buffered rotation.} Two KV buffers (current/next) keep
  the active and incoming shards separate. They also support the overlapped
  follow-up, which hides rotation latency behind compute at large $N_{\text{per}}$.
\item \textbf{Fence-free 256-work-item compute core.} The same
  software-fence-free single-pass flash kernel used for single-GPU attention
  (Q preloaded once, IGC SWSB hazard ordering, split-D output, LPT launch
  order); about 15\% faster at large $N$ than the earlier fence-heavy
  128-work-item core it replaces.
\item \textbf{In-order queue serialization (no explicit fences between
  phases).} The per-GPU in-order SYCL queue orders compute then rotate
  implicitly; no host-side event fence is inserted between the flash kernel
  and the rotation memcpy.
\end{itemize}

\textbf{All-gather attention.} Instead of
rotating one shard per step, each active GPU gathers the entire $N_{\text{total}}$
KV up front and runs one flash pass over the full context. Two techniques
define this path.
\begin{itemize}
\item \textbf{Fan-out push all-gather.} Each active GPU
  reads its own KV shard once and writes it into the full
  $[B, N_{\text{total}}, H_{\text{kv}}, D]$ buffer of all five peers at its
  own slot in one parallel loop over cached peer pointers. The self-shard
  uses a local \texttt{memcpy}. This is the fan-out push pattern from the
  general-purpose all-gather in Appendix~\ref{app:collectives}. It sustains
  approximately 400~GB/s aggregate across the node, compared with the
  approximately 16~GB/s non-adjacent link that limits strict ring rotation
  and approximately 108~GB/s for the earlier fan-in pull gather.
\item \textbf{Single flash pass without online merge.} The gathered KV is
  complete, so the validated 256-work-item core attends all
  $N_{\text{total}}$ keys in one pass. The kernel already reads the KV
  buffer depth for the number of GPUs holding KV shards and uses
  $N_{\text{per}}$ for the query
  grid, so $N_q \neq N_{kv}$ needs only a thin dispatch wrapper, no
  softmax-merge reduction. Ring instead performs six rotations and online
  merges.
\end{itemize}

The trade-off is memory, because all-gather is O($N_{\text{total}}$) per GPU and
runs out of memory once the gathered context exceeds HBM, whereas ring stays
O($N_{\text{per}}$). Section~\ref{app:cp-attn-perf} quantifies the
crossover.

\subsection{Methodology}
\label{app:cp-attn-method}

The ring uses the shared isolation protocol across six GPU packages, avoiding
intra-package HBM contention. Each configuration runs 10 independent trials
of 15 iterations. PyTorch ring trials reduce timing to the slowest rank with
a max all-reduce. Results use the per-node median followed by the cross-node
median and min--max spread over three nodes.

The PyTorch variants run on the same shards in the same harness. Ring
baselines are \texttt{sendrecv/aten}, \texttt{sendrecv/triton}, and
\texttt{async/aten}, the last using overlapped batched non-blocking
send/receive. All-gather baselines are \texttt{allgather/aten}, the tuned
vendor library path combining oneCCL \texttt{all\_gather} with IPEX fused
flash, and \texttt{allgather/triton}, which combines the oneCCL gather with
autotuned Triton flash. These two are the algorithm-matched baselines for
\mk{} all-gather attention.

Every functional method is checked against a full-context attention reference
before inclusion. The \mk{} all-gather path is also verified in-binary against
an online-softmax reference (\S\ref{app:cp-attn-correctness}).

\subsection{Delivered Performance}
\label{app:cp-attn-perf}

Tables report cross-node median TFLOPS per active GPU over three clean nodes,
with the node-level min--max spread in brackets; both methods use the same
harness.

\subsubsection{Ring attention}
\label{app:cp-attn-ring}

SyclKittens/PyTorch ratios compare \mk{} ring attention with the blocking \texttt{sendrecv} ring
baselines. The single-batch ($B=1$) sweep covers 1.5K--98K total context
(256--16384 tokens per GPU).

\begin{table*}[t]
\centering
\caption{Sequential ring attention, $B=1$ context sweep (six GPU packages; one
active stack each). The aten and Triton columns use blocking send/receive.}
\label{tab:cp-attn-ring-b1}
\small
\setlength{\tabcolsep}{4pt}
\begin{tabular}{rrrrrrr}
\toprule
$N_{\text{per}}$ & $N_{\text{total}}$ & \mk{} ring (T) & aten (T) & \mk{}/aten & Triton (T) & \mk{}/Triton \\
\midrule
256   & 1536  & 1.6 [2--2]   & 0.9  & $\mathbf{1.85\times}$ & 0.6  & $2.89\times$ \\
1024  & 6144  & 11.1 [11--11] & 6.5  & $1.71\times$ & 5.2  & $2.13\times$ \\
2048  & 12288 & 20.6 [21--21] & 12.9 & $1.59\times$ & 10.3 & $2.01\times$ \\
4096  & 24576 & 34.4 [34--34] & 24.6 & $1.40\times$ & 17.6 & $1.95\times$ \\
8192  & 49152 & 47.7 [47--48] & 42.0 & $1.13\times$ & 26.8 & $1.78\times$ \\
16384 & 98304 & 59.6 [59--60] & 56.9 & $1.05\times$ & 33.9 & $1.76\times$ \\
\bottomrule
\end{tabular}
\end{table*}

Against \texttt{sendrecv/aten}, the \mk{} ring margin narrows from
1.85$\times$ at $N_{\text{per}}=256$ to 1.05$\times$ at 16384; against
\texttt{sendrecv/triton}, it remains 1.76--2.89$\times$.

Table~\ref{tab:cp-attn-ring-b1} reports the sequential schedule. Overlapping
the shard rotation with compute on the copy engines raises the same six
configurations to 1.7, 15.2, 31.7, 53.6, 60.7, and 69.0~TFLOP/s per GPU,
which is 1.21--2.46$\times$ the \texttt{aten} ring and 1.45$\times$ at 49K
total tokens (Figure~\ref{fig:node-scale}(a)).

The step breakdown
shows how rotation and online-softmax merge diminish relative to compute as
the shard grows.

\begin{table*}[t]
\centering
\caption{Ring attention step breakdown.}
\label{tab:cp-attn-ring-breakdown}
\small
\begin{tabular}{rrrr}
\toprule
$N_{\text{per}}$ & compute & reduce & P2P rotation \\
\midrule
256  & 20\% & 41\% & 40\% \\
4096 & 48\% & 10\% & 43\% \\
8192 & 65\% & 6\%  & 29\% \\
\bottomrule
\end{tabular}
\end{table*}

From $N_{\text{per}}=256$ to 8192, the flash-core share rises from 20\% to
65\%, while reduction falls from 41\% to 6\% and rotation from 40\% to 29\%.
Ring throughput therefore approaches the single-GPU attention efficiency in
Appendix~\ref{app:attn-fwd} as the shard grows.

\subsubsection{All-gather attention}
\label{app:cp-attn-ag}

\mk{} all-gather attention gathers the full context once and runs a single
flash pass. Ratios are \mk{} over each PyTorch all-gather baseline (same
algorithm), plus \mk{} all-gather over \mk{} ring on the identical shard.
Single-batch ($B=1$) query shard swept 256--16384 per GPU (gathered context
$N_{\text{total}} = 6 \times N_{\text{per}}$).

\begin{table*}[t]
\centering
\caption{All-gather attention, $B=1$ context sweep (six GPU packages; one
active stack each). Both baselines use oneCCL all-gather.}
\label{tab:cp-attn-ag-b1}
\small
\setlength{\tabcolsep}{4pt}
\begin{tabular}{rrrrrrrr}
\toprule
$N_{\text{per}}$ & $N_{\text{total}}$ & \mk{} (T) & aten (T) & \mk{}/aten & Triton (T) & \mk{}/Triton & \mk{}/ring \\
\midrule
256   & 1536  & 7.0 [7--7]   & 3.1  & $\mathbf{2.24\times}$ & 2.5  & $2.79\times$ & $4.26\times$ \\
1024  & 6144  & 32.4 [32--32] & 20.6 & $1.57\times$ & 14.4 & $2.24\times$ & $2.93\times$ \\
2048  & 12288 & 41.5 [42--42] & 34.9 & $1.19\times$ & 21.8 & $1.91\times$ & $2.02\times$ \\
4096  & 24576 & 61.3 [61--61] & 50.2 & $1.22\times$ & 29.2 & $2.10\times$ & $1.78\times$ \\
8192  & 49152 & 61.2 [61--63] & 66.3 & $0.92\times$ & 37.1 & $1.65\times$ & $1.28\times$ \\
16384 & 98304 & 69.6 [69--70] & 75.3 & $0.93\times$ & 39.7 & $1.75\times$ & $1.17\times$ \\
\bottomrule
\end{tabular}
\end{table*}

\begin{figure*}[t]
\centering
\begin{tikzpicture}
\begin{axis}[
  width=0.88\textwidth,
  height=6.0cm,
  xlabel={Context tokens per active GPU ($N_{\mathrm{per}}$)},
  ylabel={Throughput per active GPU (TFLOPS)},
  xmode=log,
  log basis x=2,
  xmin=220, xmax=19000,
  ymin=0, ymax=78,
  xtick={256,1024,2048,4096,8192,16384},
  xticklabels={256,1K,2K,4K,8K,16K},
  grid=major,
  grid style={gray!20},
  legend style={at={(0.5,1.02)},anchor=south,legend columns=3,draw=none},
  legend cell align={left}
]
\addplot[black!65,dashed,thick,mark=triangle*] coordinates {
  (256,1.6) (1024,11.1) (2048,20.6) (4096,34.4) (8192,47.7) (16384,59.6)
};
\addlegendentry{Separated ring}
\addplot[blue!75!black,thick,mark=*] coordinates {
  (256,7.0) (1024,32.4) (2048,41.5) (4096,61.3) (8192,61.2) (16384,69.6)
};
\addlegendentry{All-gather attention}
\end{axis}
\end{tikzpicture}
\caption{Non-causal context-parallel attention on six Intel Max GPU packages,
  one active stack each, with $B=1$, $H=H_{kv}=16$, and $D=128$.
  All-gather exceeds the sequential \mk{} ring across this sweep; ring attention
  retains $O(N_{\mathrm{per}})$ memory for longer contexts. Both curves are
  three-node medians.}
\Description{Line chart of context-parallel prefill throughput per active GPU
in TFLOPS against context tokens per active GPU, from 256 to 16K, comparing the
all-gather and ring methods across six GPU packages with one active stack each.}
\label{fig:cp-attn-methods}
\end{figure*}
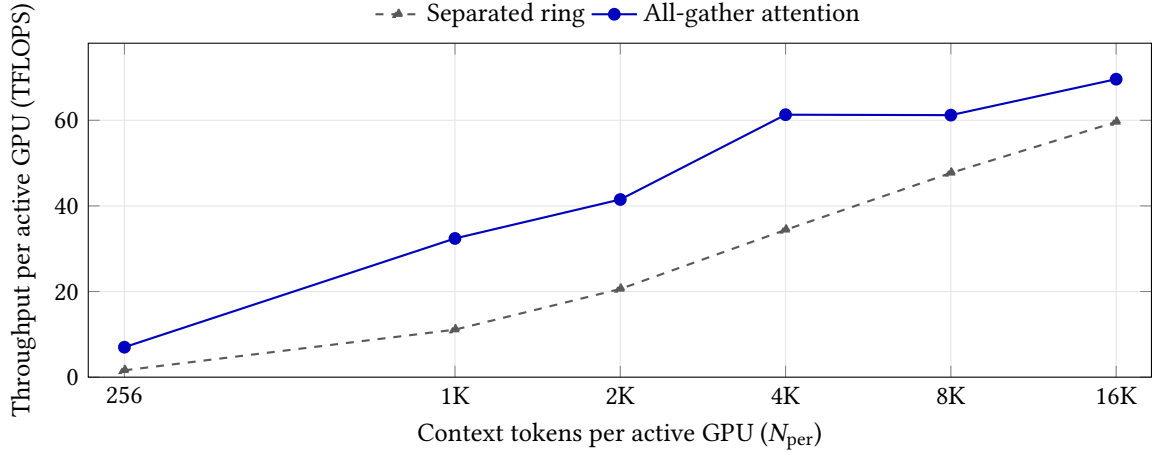

Relative to the sequential \mk{} ring, the all-gather advantage narrows from 4.26$\times$ at
$N_{\text{per}}=256$ to 1.17$\times$ at 16384, while remaining
1.65--2.79$\times$ faster than \texttt{allgather/triton}. The tuned
oneCCL-plus-IPEX path overtakes it only at the two largest shards, where the
\mk{}/aten ratios are 0.92$\times$ and 0.93$\times$. At
$N_{\text{per}}=8192$, the fan-out gather takes 4.7~ms (9\%) and flash takes
46.3~ms (91\%), so the flash pass controls throughput at this shape.

\subsubsection{Memory-capacity trade-off}
\label{app:cp-attn-which}

HBM capacity determines which path applies.
\begin{itemize}
\item \textbf{Gathered context fits in HBM.} All-gather pays the
  inter-package cost once and is $1.17$--$4.26\times$ faster than the
  sequential ring across the measured $B=1$ sweep.
\item \textbf{Context exceeds one GPU's HBM.} Ring retains an
  O($N_{\text{per}}$) footprint and increases supported context with the
  active GPU count.
\end{itemize}
The all-gather buffer holds six KV shards per GPU in this configuration.
Ring stores only a constant number of shards, including its rotation buffers,
so its KV memory per GPU scales with local rather than total context length.

\subsection{Correctness}
\label{app:cp-attn-correctness}

Every functional method is validated against an aten fused-flash
full-context reference (relative error, bf16 in, fp32 accumulate).
Representative maxima across the three nodes appear below.

\begin{table}[t]
\centering
\caption{Ring attention baseline correctness, maximum relative error.}
\label{tab:cp-attn-correctness}
\small
\begin{tabular}{lrrrrrr}
\toprule
& $N_{\text{per}}=256$ & 1024 & 2048 & 4096 & 8192 & 16384 \\
\midrule
\texttt{sendrecv/aten}   & 0.0191 & 0.0126 & 0.0101 & 0.0076 & 0.0063 & 0.0041 \\
\texttt{sendrecv/triton} & 0.0204 & 0.0126 & 0.0102 & 0.0076 & 0.0063 & 0.0041 \\
\bottomrule
\end{tabular}
\end{table}

All functional configurations pass (tolerance 0.05); the error falls with
$N_{\text{per}}$ as the softmax normalizer averages more terms. \mk{} ring
attention is validated by its own in-binary reference check on each shard, and
\mk{} all-gather attention is checked in-binary against an online-softmax
full-context reference (relative error 0.0012 at $N_{\text{per}}=256$, well
under tolerance) before any timing is recorded.

%% file: sections/supp/appx-collectives.tex
\section{Multi-GPU Collectives}
\label{app:collectives}

The \mk{} collectives reduce per-call launch and handshake overhead. In a
six-node matrix, all four win at the smallest standalone payload, 24 KiB, on
every node. The evaluated batch-one decode uses two 8-KiB all-reduces per
layer; this smaller payload is exercised in the end-to-end measurements in
\S\ref{app:e2e-mgpu}, not in the standalone sweep. Reduce-scatter also
wins 0.19 MiB on every node, and destination-split all-to-all wins every
measured payload on every node. Most other medium cells are node-sensitive.
At 96 MiB, all-to-all is 1.20\texttimes{} faster on the median and wins all
six nodes; oneCCL retains the median lead for all-reduce and all-gather, while
reduce-scatter favors \mk{} on four of six nodes. These paths also support
distributed GEMM (Appendix~\ref{app:dist-gemm}).

\subsection{Configuration}

We compare bf16 all-reduce, all-gather, reduce-scatter, and all-to-all on
all 12 GPUs of one node containing six GPU packages. \mk{} uses one process with
shared-context peer pointers; the tuned vendor library path is oneCCL/xCCL with
one process per GPU. Each \mk{} collective dispatches adaptively between a
latency-oriented path and a topology-aware bandwidth path.

\subsection{Kernel Implementation and Key Techniques}

Every \mk{} collective builds on direct peer read/write, so GPU $d$'s kernel
addresses GPU $k$'s HBM through the shared SYCL context, with a single host
barrier per phase and no staging through host memory. Three measured
techniques shape the per-collective kernels.

\begin{itemize}
\item \textbf{Remote-write fan-out.} On the measured peer route, a remote
  write sustains $\sim$18.8 GB/s versus $\sim$15.7 GB/s for a remote read. All-gather
  and large-payload all-reduce use fan-out push. Each GPU reads its local slice
  once and writes it to all $N{-}1$ remote destinations. Local HBM reads cost
  $epd$ rather than $(N{-}1)\,epd$, leaving bandwidth for incoming remote writes
  and avoiding the HBM collapse of large-payload pull kernels.
\item \textbf{Hierarchical reduction.} For all-reduce and
  reduce-scatter the hierarchical path first sums each GPU package's two GPU
  stacks over MDFI, then reduces the partials from the six GPU packages over
  XeLink, halving the XeLink message count (the copy now carries
  already-summed data).
\item \textbf{Destination-split fused EU copy.} For all-to-all, one kernel per
  source GPU pushes all $N$ chunks with by-value dispatch tables, while each
  workgroup owns one destination. This removes the per-message launch floor and
  exposes independent remote endpoints concurrently. The topology-aware
  all-gather uses the corresponding fused latency path.
\end{itemize}

The reduction and gather kernels dispatch adaptively between latency- and
bandwidth-oriented paths. All-to-all uses the destination-split path at every
size because it is neutral at the launch floor and wins as endpoint traffic grows.

\subsubsection{Multi-GPU API}

The collective kernels use the PGL peer-pointer table, \pvc{} topology
planner, and system-scope atomic synchronization introduced in the tile-language
appendix. These replace ThunderKittens' CUDA NVLS multicast lowering while preserving
explicit control over slice ownership, transfer direction, and schedule.

We validated the API conversion of all-gather, reduce-scatter, all-reduce,
and all-to-all against immutable pre-conversion binaries with complementary raw--API--API--raw and
API--raw--raw--API orderings. Equal-weighting the two per-node bracket estimates
gives an API/raw geomean of $1.013\times$ over all 28 operation--payload cells;
the worst cell median is $0.974\times$. Every cell passes the predefined
$0.95\times$ guard and the aggregate passes the $0.99\times$ guard, establishing
no material regression under the declared protocol.

A separate clean current-source recollection rebuilt all four production
kernels and completed the full \mk{}--oneCCL--\mk{} matrix. All 336 \mk{} rows
passed full-buffer validation with zero mismatches, confirming the production
build and oneCCL integration after conversion.

\subsection{Methodology}

Each same-node bracket runs \mk{} and oneCCL on the same contiguous 12 GPUs
and traverses the same physical MDFI and XeLink links under the same topology.
The formal matrix repeats that bracket on six nodes. Each pass reports the
median of 15 timed iterations after five warmups. For oneCCL,
\texttt{all\_reduce(MAX)} selects the slowest rank as the collective
completion time.

All-reduce bandwidth uses
$2(N{-}1)/N\cdot S/t$. All-gather, reduce-scatter, and all-to-all use the
p2p-bytes convention $N(N{-}1)\cdot epd\cdot2/t$. The \mk{} binary and its
oneCCL counterpart print the same formula for aligned size sweeps, making
their GB/s values directly comparable. Every node runs \mk{}--oneCCL--\mk{};
the \mk{} value is the mean of its two bracketing medians, and speedup is paired
on that node before taking the six-node median. Every \mk{} point is full-buffer
verified against a reference sum, gather, or scatter; maximum absolute error
remains within bf16 rounding.

\subsection{Delivered Performance}

Tables report six-node marginal median $\mu$s (12 GPUs), the
matched-convention GB/s for both, and the median paired same-node speedup
(oneCCL divided by \mk{}; greater than 1 means \mk{} is faster). Absolute
performance tiers vary by node, so the paired speedup need not equal the ratio
of the two marginal median times.

\noindent\textbf{All-Reduce} (busbw $2(N{-}1)/N$).

\begin{table}[h]
\centering
\small
\setlength{\tabcolsep}{2pt}
\begin{tabular}{@{}lrrrrr@{}}
\toprule
Payload & \multicolumn{2}{c}{Latency (\us{})} & \multicolumn{2}{c}{Bandwidth (\gbps{})} & Speedup \\
\cmidrule(lr){2-3}\cmidrule(lr){4-5}
& \mk{} & oneCCL & \mk{} & oneCCL & \\
\midrule
24 KiB   & 81.2  & 173.5 & 0.6  & 0.3   & \textbf{2.14\texttimes} \\
0.19 MiB & 131.8 & 172.5 & 2.7  & 2.1   & 1.16\texttimes \\
1.5 MiB  & 208.2 & 274.5 & 13.8 & 10.5  & 1.06\texttimes \\
6 MiB    & 264.5 & 319.0 & 43.6 & 36.2  & 1.04\texttimes \\
12 MiB   & 352.0 & 417.5 & 65.5 & 55.3  & 1.04\texttimes \\
24 MiB   & 516.8 & 596.0 & 89.3 & 77.4  & 1.12\texttimes \\
96 MiB   & 1\,511.8 & 1\,460.5 & 122.1 & 126.4 & 0.94\texttimes \\
\bottomrule
\end{tabular}
\end{table}

\noindent\textbf{All-Gather} (p2p bytes).

\begin{table}[h]
\centering
\small
\setlength{\tabcolsep}{2pt}
\begin{tabular}{@{}lrrrrr@{}}
\toprule
Payload & \multicolumn{2}{c}{Latency (\us{})} & \multicolumn{2}{c}{Bandwidth (\gbps{})} & Speedup \\
\cmidrule(lr){2-3}\cmidrule(lr){4-5}
& \mk{} & oneCCL & \mk{} & oneCCL & \\
\midrule
24 KiB   & 92.8  & 161.5 & 2.9   & 1.7    & \textbf{1.81\texttimes} \\
0.19 MiB & 108.0 & 160.5 & 20.0  & 13.5   & 1.51\texttimes \\
1.5 MiB  & 144.5 & 180.0 & 119.7 & 96.1   & 1.26\texttimes \\
6 MiB    & 227.2 & 245.0 & 304.5 & 282.5  & 1.10\texttimes \\
12 MiB   & 259.2 & 285.0 & 533.9 & 485.7  & 1.12\texttimes \\
24 MiB   & 368.5 & 378.0 & 751.2 & 732.3  & 1.03\texttimes \\
96 MiB   & 987.8 & 903.0 & 1\,121.0 & 1\,226.2 & 0.92\texttimes \\
\bottomrule
\end{tabular}
\end{table}

\noindent\textbf{Reduce-Scatter} (p2p bytes).

\begin{table}[h]
\centering
\small
\setlength{\tabcolsep}{2pt}
\begin{tabular}{@{}lrrrrr@{}}
\toprule
Payload & \multicolumn{2}{c}{Latency (\us{})} & \multicolumn{2}{c}{Bandwidth (\gbps{})} & Speedup \\
\cmidrule(lr){2-3}\cmidrule(lr){4-5}
& \mk{} & oneCCL & \mk{} & oneCCL & \\
\midrule
24 KiB   & 80.8  & 238.5 & 3.3   & 1.1    & \textbf{2.19\texttimes} \\
0.19 MiB & 80.8  & 238.5 & 26.8  & 9.1    & \textbf{2.20\texttimes} \\
1.5 MiB  & 110.5 & 265.5 & 156.6 & 65.2   & 1.97\texttimes \\
6 MiB    & 223.2 & 265.5 & 310.0 & 260.7  & 1.22\texttimes \\
12 MiB   & 249.5 & 283.5 & 554.8 & 488.2  & 1.21\texttimes \\
24 MiB   & 315.0 & 357.5 & 878.8 & 774.3  & 1.19\texttimes \\
96 MiB   & 713.5 & 781.0 & 1\,551.9 & 1\,417.8 & 1.10\texttimes \\
\bottomrule
\end{tabular}
\end{table}

\noindent\textbf{All-to-All} (p2p bytes).

\begin{table}[h]
\centering
\small
\setlength{\tabcolsep}{2pt}
\begin{tabular}{@{}lrrrrr@{}}
\toprule
Payload & \multicolumn{2}{c}{Latency (\us{})} & \multicolumn{2}{c}{Bandwidth (\gbps{})} & Speedup \\
\cmidrule(lr){2-3}\cmidrule(lr){4-5}
& \mk{} & oneCCL & \mk{} & oneCCL & \\
\midrule
24 KiB   & 91.8  & 293.0 & 2.9   & 0.9   & \textbf{3.31\texttimes} \\
0.19 MiB & 91.0  & 302.5 & 23.8  & 7.1   & \textbf{3.46\texttimes} \\
1.5 MiB  & 90.2  & 302.0 & 191.7 & 57.3  & \textbf{3.49\texttimes} \\
6 MiB    & 136.5 & 355.0 & 507.0 & 194.9 & \textbf{2.62\texttimes} \\
12 MiB   & 207.5 & 464.0 & 667.0 & 298.3 & \textbf{2.24\texttimes} \\
24 MiB   & 342.0 & 612.5 & 809.4 & 452.0 & \textbf{1.79\texttimes} \\
96 MiB   & 1\,188.5 & 1\,421.5 & 931.7 & 779.0 & \textbf{1.20\texttimes} \\
\bottomrule
\end{tabular}
\end{table}

\begin{figure*}[t]
\centering
\begin{tikzpicture}
\begin{axis}[
  width=0.90\textwidth,
  height=6.1cm,
  xlabel={Payload per GPU},
  ylabel={Speedup over oneCCL},
  xmode=log,
  log basis x=2,
  xmin=0.018, xmax=125,
  ymin=0.35, ymax=4.2,
  xtick={0.024,0.19,1.5,6,12,24,96},
  xticklabels={24 KiB,0.19 MiB,1.5 MiB,6 MiB,12 MiB,24 MiB,96 MiB},
  x tick label style={rotate=25,anchor=east},
  grid=major,
  grid style={gray!20},
  legend style={at={(0.5,1.02)},anchor=south,legend columns=4,draw=none},
  legend cell align={left}
]
\addplot[blue!75!black,thick,mark=*] coordinates {
  (0.024,2.14) (0.19,1.16) (1.5,1.06) (6,1.04) (12,1.04) (24,1.12) (96,0.94)
};
\addlegendentry{All-reduce}
\addplot[teal!70!black,thick,mark=square*] coordinates {
  (0.024,1.81) (0.19,1.51) (1.5,1.26) (6,1.10) (12,1.12) (24,1.03) (96,0.92)
};
\addlegendentry{All-gather}
\addplot[orange!85!black,thick,mark=triangle*] coordinates {
  (0.024,2.19) (0.19,2.20) (1.5,1.97) (6,1.22) (12,1.21) (24,1.19) (96,1.10)
};
\addlegendentry{Reduce-scatter}
\addplot[red!70!black,thick,mark=diamond*] coordinates {
  (0.024,3.31) (0.19,3.46) (1.5,3.49) (6,2.62) (12,2.24) (24,1.79) (96,1.20)
};
\addlegendentry{All-to-all}
\addplot[black,dashed,thin,forget plot] coordinates {(0.018,1) (125,1)};
\end{axis}
\end{tikzpicture}
\caption{BF16 collective speedup over oneCCL on 12 Intel Max GPUs
  (higher is better). \mk{} uses one process with shared-context peer
  pointers; oneCCL uses one process per GPU.
  Curves are six-node medians of paired same-node brackets. \mk{} wins all six
  nodes for every collective at 24 KiB and for all-to-all through 96 MiB;
  medium AG/RS/AR cells vary by node, and oneCCL leads AG and AR at 96 MiB.}
\Description{Line chart of six-node median collective speedup over oneCCL on 12
GPUs across payloads per GPU ranging from 24 KiB to 96 MiB.}
\label{fig:collectives-speedup}
\end{figure*}
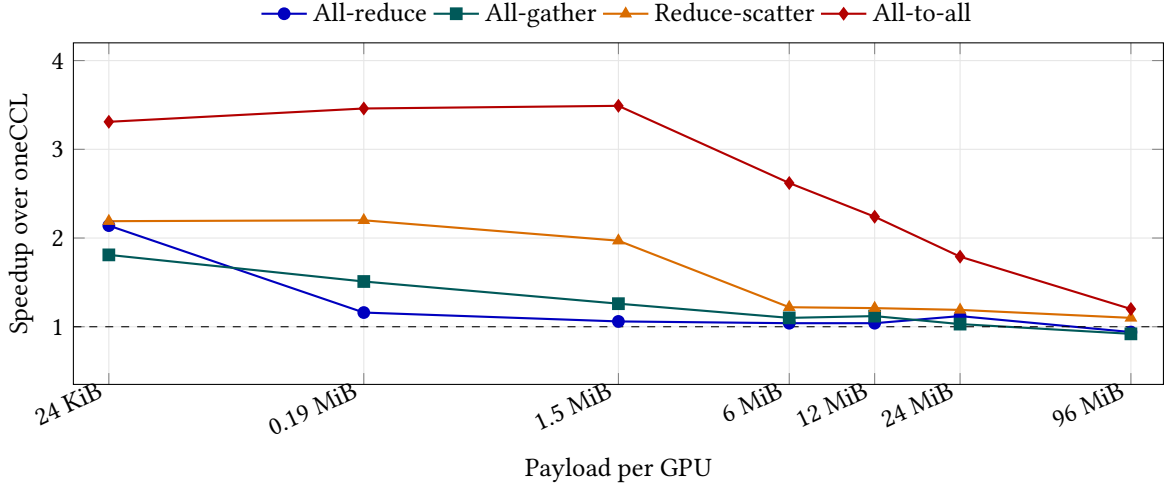

The crossover follows the dominant cost. At 24 KiB, all four \mk{} collectives
win all six nodes; reduce-scatter extends that robust win to 0.19 MiB and
all-to-all extends it through 96 MiB. The medium AG/RS/AR ratios favor \mk{} in
the six-node median but cross parity between nodes. At 96 MiB, oneCCL wins AG,
and AR in the median, RS favors \mk{} on four of six nodes, and A2A favors
\mk{} on all six nodes. Thus the standalone sweep establishes the
per-collective crossover from 24 KiB through 96 MiB; the smaller 8-KiB decode
payload is evaluated in the complete decode path.

The decode orchestrator also replaces 64--96 per-token host waits with a
system-scope atomic peer-counter barrier on the GPUs. This removes host
serialization and improves multi-GPU decode TPOT by 1.1--1.54\texttimes, with
the largest change on the 2D TP$\times$CP layout. The synchronization change
does not alter arithmetic and passes the same correctness checks. It is used
for the decode results in \S\ref{app:e2e-mgpu}.

\subsection{Correctness}

Every \mk{} collective is full-buffer verified on the participating GPUs against a
reference on each run and each size, with all-reduce and reduce-scatter checked
against
an fp32 elementwise sum, all-gather and all-to-all against the exact
gathered/permuted layout. Maximum absolute error is at most 0.01, within bf16
rounding. The oneCCL baseline uses the
\texttt{torch.distributed} collective.

%% file: sections/supp/appx-e2e.tex
\section{End-to-End Llama-3.1-8B}
\label{app:e2e}

On one GPU, composing \mk{} operations with TorchInductor-generated regions
improves Llama-3.1-8B full-request throughput by $1.59\times$ in geometric
mean over \texttt{torch.compile} alone on the 21 configurations that both
paths run. The composed path also serves the complete $N=8192$ grid, including
cells where the baselines run out of memory. The final subsection evaluates
multi-GPU decode separately.

\subsection{Evaluated Kernel Set}
\label{app:e2e-evaluated}

Table~\ref{tab:appxe2e-evaluated} lists the implementations used for the
measured end-to-end path. Each selection is the fastest correct
implementation in its operation-specific study, and every run records the
implementation that executes.

\begin{table}[h]
\centering
\footnotesize
\setlength{\tabcolsep}{2pt}
\caption{Evaluated kernel set for the end-to-end Llama-3.1-8B path.}
\label{tab:appxe2e-evaluated}
\begin{tabularx}{\columnwidth}{@{}>{\raggedright\arraybackslash}p{0.28\columnwidth}>{\raggedright\arraybackslash}p{0.39\columnwidth}>{\raggedright\arraybackslash}X@{}}
\toprule
Block op & Implementation & Notes \\
\midrule
Residual+RMSNorm & Fused \mk{} residual and RMSNorm & $1.04\times$ Inductor \\
GQA ($g{=}4$) & Fence-free \mk{} GQA & shape-tuned \\
MHA & Fence-free \mk{} MHA & non-grouped fallback \\
Q/K rotary & Fused \mk{} Q/K rotary & fusion microbenchmark \\
QKV projection & concatenated $W_{qkv}$ & one prefill GEMM \\
\bottomrule
\end{tabularx}
\end{table}

\subsection{Baselines}
\label{app:e2e-baselines}

Eager PyTorch executes the operator sequence directly. The
\texttt{torch.compile} baseline fuses pointwise and reduction regions but
graph-breaks at SDPA on this backend; all primary ratios compare the
composed path with Inductor alone under identical compilation settings.

\subsection{Operations across the SDPA Graph Break}
\label{app:e2e-fusions}

\mk{} supplies four operations across the attention graph break, namely GQA flash
attention, joint Q/K rotary, a concatenated QKV projection, and residual-add
plus RMSNorm. Inductor retains RMSNorm, SwiGLU, and the remaining pointwise
and epilogue regions.

\subsection{End-to-End Composition with TorchInductor}
\label{app:e2e-composition}

The integration boundary is the SDPA graph break. \mk{} replaces regions
that span this boundary, while TorchInductor continues to generate the
remaining pointwise, reduction, and epilogue regions. Stand-alone norm and
SiLU remain with Inductor because it matches or exceeds the corresponding
stand-alone \mk{} kernels.

\paragraph{Full-request metric.}
Each repeat synchronizes the GPU, times one complete deterministic
\texttt{model.generate()} call with \texttt{max\_new\_tokens=128}, and
synchronizes again after return. For batch size $B$, prompt length $N$,
mean actual generated sequence length $\bar{G}\le128$, and mean synchronized
elapsed time $\bar{t}$ in seconds over the timed repeats, the harness records
$(BN+\bar{G})/\bar{t}$. This convention combines batch-wide prompt tokens
with the mean generated length of one stream. We retain it for the reported
full-request rates and ratios. Total batch-token throughput for equally long
outputs would instead be $B(N+\bar{G})/\bar{t}$. When compared runs generate
the same lengths, the reported ratio equals their inverse elapsed-time ratio.
The timed interval includes prefill and every generated decode step;
compilation occurs during untimed warmup.

\subsubsection{Configuration}

Both composed and Inductor-alone runs use identical compilation settings,
eight untimed warmups, and five synchronized repeats. The only difference is
the four \mk{} operations above; Inductor owns RMSNorm, SwiGLU, and all other
pointwise and epilogue code generation in both runs.

\subsubsection{Delivered performance}

We evaluate bf16 Llama-3.1-8B on one GPU over 24 shapes with
$B\in\{1,2,4,8\}$ and $N\in\{256,512,1024,2048,4096,8192\}$. Composed,
Inductor-alone, and eager runs are node-matched; the primary ratio is
composed over Inductor alone. \textsc{oom} marks a baseline capacity
failure.

\begin{table}[h]
\centering
\small
\caption{Geometric-mean summary for the 24-shape single-GPU grid.
Ratios include only cells where both compared paths run. \emph{comp/ind}
uses 21 cells; Inductor runs out of memory at $N=8192$ with $B=2,4,8$.
The composed path runs all 24 cells, and eager also runs the $B=2$ cell.}
\label{tab:appxe2e-geomean}
\begin{tabular}{@{}lccc@{}}
\toprule
Regime & vs.\ inductor-alone & vs.\ eager & Range (comp/ind) \\
\midrule
Prefill & \textbf{$1.22\times$} & $1.40\times$ & $0.91\times$--$2.03\times$ \\
Decode & \textbf{$1.65\times$} & $1.58\times$ & $1.30\times$--$2.20\times$ \\
End-to-end & \textbf{$1.59\times$} & $1.55\times$ & $1.29\times$--$1.98\times$ \\
\bottomrule
\end{tabular}
\end{table}

\begin{table}[h]
\centering
\footnotesize
\setlength{\tabcolsep}{2.5pt}
\caption{Per-shape prefill throughput (tok/s).}
\label{tab:appxe2e-prefill}
\begin{tabular}{@{}rrrrrccc@{}}
\toprule
BS & N & eager & inductor & composed & comp/ind & comp/eager \\
\midrule
1 & 256  & 5340 & 6720 & 6140 & $0.91\times$ & $1.15\times$ \\
1 & 512  & 6890 & 8450 & 8700 & $1.03\times$ & $1.26\times$ \\
1 & 1024 & 7470 & 8920 & 9490 & $1.06\times$ & $1.27\times$ \\
1 & 2048 & 7160 & 8340 & 9750 & $1.17\times$ & $1.36\times$ \\
1 & 4096 & 5990 & 6790 & 9300 & $1.37\times$ & $1.55\times$ \\
1 & 8192 & 3510 & 3730 & 7590 & $2.03\times$ & $2.16\times$ \\
2 & 256  & 6890 & 8140 & 8350 & $1.03\times$ & $1.21\times$ \\
2 & 512  & 7540 & 8780 & 9210 & $1.05\times$ & $1.22\times$ \\
2 & 1024 & 7720 & 8580 & 9550 & $1.11\times$ & $1.24\times$ \\
2 & 2048 & 7060 & 7510 & 9550 & $1.27\times$ & $1.35\times$ \\
2 & 4096 & 5860 & 6130 & 9460 & $1.54\times$ & $1.61\times$ \\
2 & 8192 & 3700 & \textsc{oom} & 8110 & -- & $2.19\times$ \\
4 & 256  & 7890 & 9520 & 9660 & $1.02\times$ & $1.22\times$ \\
4 & 512  & 8220 & 9800 & 10000 & $1.02\times$ & $1.22\times$ \\
4 & 1024 & 7980 & 9060 & 10400 & $1.14\times$ & $1.30\times$ \\
4 & 2048 & 7090 & 7810 & 10000 & $1.28\times$ & $1.41\times$ \\
4 & 4096 & 5390 & 4930 & 9430 & $1.91\times$ & $1.75\times$ \\
4 & 8192 & \textsc{oom} & \textsc{oom} & 8250 & -- & -- \\
8 & 256  & 8550 & 10100 & 10300 & $1.02\times$ & $1.21\times$ \\
8 & 512  & 8640 & 10100 & 10500 & $1.05\times$ & $1.22\times$ \\
8 & 1024 & 8090 & 9260 & 10500 & $1.14\times$ & $1.30\times$ \\
8 & 2048 & 7090 & 7740 & 10100 & $1.30\times$ & $1.42\times$ \\
8 & 4096 & 5520 & 5040 & 9500 & $1.89\times$ & $1.72\times$ \\
8 & 8192 & \textsc{oom} & \textsc{oom} & 7740 & -- & -- \\
\bottomrule
\end{tabular}
\end{table}

\begin{table}[h]
\centering
\footnotesize
\setlength{\tabcolsep}{2.5pt}
\caption{Decode throughput per stream (tok/s).}
\label{tab:appxe2e-decode}
\begin{tabular}{@{}rrrrrccc@{}}
\toprule
BS & N & eager & inductor & composed & comp/ind & comp/eager \\
\midrule
1 & 256  & 29   & 25   & 49   & $1.96\times$ & $1.67\times$ \\
1 & 512  & 30   & 25   & 48   & $1.90\times$ & $1.62\times$ \\
1 & 1024 & 30   & 25   & 46   & $1.82\times$ & $1.54\times$ \\
1 & 2048 & 30   & 26   & 44   & $1.71\times$ & $1.49\times$ \\
1 & 4096 & 28   & 29   & 39   & $1.37\times$ & $1.38\times$ \\
1 & 8192 & 22.7 & 24.9 & 32.5 & $1.30\times$ & $1.43\times$ \\
2 & 256  & 29   & 25   & 47   & $1.87\times$ & $1.62\times$ \\
2 & 512  & 29   & 25   & 46   & $1.81\times$ & $1.56\times$ \\
2 & 1024 & 29   & 26   & 44   & $1.72\times$ & $1.50\times$ \\
2 & 2048 & 28   & 25   & 40   & $1.58\times$ & $1.43\times$ \\
2 & 4096 & 23   & 25   & 35   & $1.40\times$ & $1.53\times$ \\
2 & 8192 & 15   & \textsc{oom} & 27 & -- & $1.82\times$ \\
4 & 256  & 30   & 26   & 44   & $1.74\times$ & $1.49\times$ \\
4 & 512  & 30   & 26   & 43   & $1.67\times$ & $1.43\times$ \\
4 & 1024 & 28   & 25   & 40   & $1.59\times$ & $1.43\times$ \\
4 & 2048 & 23   & 25   & 36   & $1.44\times$ & $1.57\times$ \\
4 & 4096 & 15   & 17   & 29   & $1.73\times$ & $1.87\times$ \\
4 & 8192 & \textsc{oom} & \textsc{oom} & 24 & -- & -- \\
8 & 256  & 30   & 26   & 42   & $1.64\times$ & $1.43\times$ \\
8 & 512  & 28   & 25   & 40   & $1.58\times$ & $1.43\times$ \\
8 & 1024 & 23   & 25   & 36   & $1.43\times$ & $1.57\times$ \\
8 & 2048 & 15   & 21   & 31   & $1.48\times$ & $2.00\times$ \\
8 & 4096 & 9    & 10   & 22   & $2.20\times$ & $2.32\times$ \\
8 & 8192 & \textsc{oom} & \textsc{oom} & 16 & -- & -- \\
\bottomrule
\end{tabular}
\end{table}

\begin{table}[h]
\centering
\footnotesize
\setlength{\tabcolsep}{2.5pt}
\caption{Per-shape full-request rate, $(BN+\bar{G})/\bar{t}$, using
batch-wide prompt tokens and generated length per stream. All columns are
node-matched.}
\label{tab:appxe2e-e2e}
\begin{tabular}{@{}rrrrrccc@{}}
\toprule
BS & N & eager & inductor & composed & comp/ind & comp/eager \\
\midrule
1 & 256  & 253  & 218  & 413  & $1.89\times$ & $1.64\times$ \\
1 & 512  & 472  & 409  & 749  & $1.83\times$ & $1.59\times$ \\
1 & 1024 & 874  & 768  & 1320 & $1.72\times$ & $1.51\times$ \\
1 & 2048 & 1520 & 1390 & 2210 & $1.59\times$ & $1.46\times$ \\
1 & 4096 & 2280 & 2400 & 3290 & $1.37\times$ & $1.44\times$ \\
1 & 8192 & 2230 & 2360 & 4070 & $1.72\times$ & $1.83\times$ \\
2 & 256  & 465  & 411  & 738  & $1.80\times$ & $1.59\times$ \\
2 & 512  & 863  & 766  & 1310 & $1.70\times$ & $1.51\times$ \\
2 & 1024 & 1530 & 1400 & 2200 & $1.58\times$ & $1.44\times$ \\
2 & 2048 & 2400 & 2290 & 3360 & $1.47\times$ & $1.40\times$ \\
2 & 4096 & 2930 & 3140 & 4610 & $1.47\times$ & $1.57\times$ \\
2 & 8192 & 2480 & \textsc{oom} & 5100 & -- & $2.06\times$ \\
4 & 256  & 880  & 775  & 1280 & $1.65\times$ & $1.45\times$ \\
4 & 512  & 1580 & 1430 & 2180 & $1.53\times$ & $1.39\times$ \\
4 & 1024 & 2500 & 2400 & 3460 & $1.44\times$ & $1.39\times$ \\
4 & 2048 & 3230 & 3540 & 4830 & $1.36\times$ & $1.49\times$ \\
4 & 4096 & 3200 & 3130 & 5760 & $1.84\times$ & $1.80\times$ \\
4 & 8192 & \textsc{oom} & \textsc{oom} & 6190 & -- & -- \\
8 & 256  & 1570 & 1440 & 2170 & $1.51\times$ & $1.38\times$ \\
8 & 512  & 2540 & 2450 & 3450 & $1.41\times$ & $1.36\times$ \\
8 & 1024 & 3410 & 3810 & 4930 & $1.29\times$ & $1.45\times$ \\
8 & 2048 & 3730 & 4480 & 6140 & $1.37\times$ & $1.65\times$ \\
8 & 4096 & 3510 & 3370 & 6670 & $1.98\times$ & $1.90\times$ \\
8 & 8192 & \textsc{oom} & \textsc{oom} & 6270 & -- & -- \\
\bottomrule
\end{tabular}
\end{table}

For prefill at prompt length 256, the composed/Inductor ratio is
$0.91\times$--$1.03\times$. The projection GEMMs are too small to amortize
the fused QKV, rotary, and residual launches, with $0.91\times$ at the
$1\times256$ cell. From prompt length 512 onward, the
ratio increases with context, reaching $1.30\times$ at $8\times2048$,
$1.91\times$ at $4\times4096$, and $2.03\times$ at $1\times8192$ as the
flash-attention path replaces the baseline attention path across the SDPA
graph break.

Decode improves by $1.65\times$ geometrically and reaches $1.96\times$ at
$1\times256$. Decode is dispatch-bound; the SDPA graph break prevents
Inductor from fusing the per-step attention prologue and epilogue, whereas
the composed path uses fewer launches.

The harness retains only last-token logits, preventing a full-vocabulary
all-position logits tensor from determining memory capacity. At $N=8192$,
\texttt{torch.compile} runs out of memory at $BS\ge2$ and eager at $BS\ge4$.
The composed path serves BS 1--8 at the prefill throughputs in
Table~\ref{tab:appxe2e-prefill} because its flash attention has the smaller
memory footprint.

Across the $N\le2048$ subset, the composed path retains a $1.56\times$
end-to-end geometric mean. The gain comes from combining Inductor's
pointwise/norm fusion with the cross-op and attention fusions that span its
SDPA graph break; the flash-attention memory footprint also preserves coverage
at long context.

\subsection{Distributed Multi-GPU Decode (Single-Process TP+CP vs.\ oneCCL)}
\label{app:e2e-mgpu}

We measure Llama-3.1-8B decode on one node with six GPU packages, each with
two stacks connected by MDFI.
XeLink connects the GPU packages all-to-all through independent even and odd
stack planes. A flat device hierarchy exposes all 12 stacks. We compare
two implementations of tensor parallelism~\cite{shoeybi2019megatron}.
\mk{} uses a single-process C++ orchestrator. The tuned Intel oneCCL library
path uses \texttt{torch.distributed} with the \texttt{xccl} backend and one
process per stack.

\subsubsection{Configuration and compared implementations}

The model has hidden size 4096, 32 query and 8 KV heads, head dimension 128,
intermediate size 14336, and 32 layers. The node exposes six GPU packages and
12 stacks. Both systems use the same TP weight and KV sharding; \mk{} uses inline
peer collectives and can additionally shard KV by context.

\subsubsection{Parallelization strategy}

Decode combines a context-invariant weight stream with a KV scan that grows
linearly with context. Tensor parallelism shards the weight stream, whereas
context parallelism shards the growing KV term. The 2-D configurations below
combine these schemes.

\subsubsection{What \mk{} runs per token}

Each layer uses split-K flash decode for the KV scan, 128-bit wide-load GEMV
for the five projections, inline peer all-reduce, and a GPU-side atomic
rendezvous. The entire tile-by-layer loop is one C++ orchestrator call, so
Python dispatch is removed from the inner loop. The C++ orchestrator still
submits GPU work, and its cost is included in the measured decode time.

\subsubsection{TPOT versus context}

Table~\ref{tab:appxe2e-mgpu-tpot} reports steady-state
time-per-output-token (TPOT), the geometric mean over 100 decode steps
after prefilling to length $N$; prefill is excluded. The p95 and p99 remain
within 2\,\ms{} of the geometric mean throughout, and the medians coincide.
Across matched two- and four-GPU configurations, \mk{} is
$1.52$--$2.91\times$ faster than the oneCCL-based decode path over the
contexts where both run.

\begin{table}[h]
\centering
\small
\setlength{\tabcolsep}{4pt}
\caption{Multi-GPU decode TPOT (ms/token, geometric mean over 100 steps;
lower is better). \textsc{oom} marks a baseline capacity failure. All \mk{}
configurations pass the first-new-token logit-vector check against the
single-GPU reference with cosine similarity at least 0.99997.}
\label{tab:appxe2e-mgpu-tpot}
\begin{tabular}{@{}rrrrrr@{}}
\toprule
ctx & \mk{} $P=1$ & \mk{} $P=2$ & \mk{} $P=4$ & xCCL $ws=2$ & xCCL $ws=4$ \\
\midrule
1K   & 28.4  & 22.8  & \textbf{19.2}  & 42.5  & 40.7  \\
2K   & 30.0  & 24.6  & \textbf{20.3}  & 43.0  & 40.5  \\
4K   & 30.6  & 25.7  & \textbf{21.7}  & 43.3  & 40.7  \\
8K   & 31.3  & 26.1  & \textbf{22.4}  & 49.2  & 41.8  \\
16K  & 43.4  & 27.4  & \textbf{23.1}  & 66.6  & 49.3  \\
32K  & 54.1  & 36.2  & \textbf{25.3}  & 104.6 & 73.7  \\
64K  & 117.1 & 73.3  & \textbf{54.5}  & 173.9 & 128.1 \\
128K & 209.7 & 128.0 & \textbf{95.6}  & 310.8 & 222.3 \\
256K & --    & 370.7 & \textbf{271.9} & \textsc{oom} & 413.0 \\
\bottomrule
\end{tabular}
\end{table}

At 32K context \mk{} on 2 GPUs is $2.89\times$ faster than oneCCL on 2
(36.2 vs.\ 104.6\,\ms{}) and on 4 GPUs $2.91\times$ (25.3 vs.\
73.7\,\ms{}); equivalently \mk{} sustains 39.5\,tok/s on 4 GPUs at 32K
versus oneCCL's 13.6. At 128K the speedups are $2.43\times$ ($P=2$) and
$2.33\times$ ($P=4$); at 256K, $P=4$ is $1.52\times$ over xCCL $ws=4$ and
$P=2$ runs at all where xCCL $ws=2$ runs out of memory.

\subsubsection{Per-layer collective cost}

Each layer performs two all-reduces as inline single-process peer pulls over
the shared context. A separate category-level profile assigns all-reduces
about 1.7\,\ms{}/token at $P=2$ and 3.1\,\ms{}/token at $P=4$. In that
instrumented profile, they account for 3.4--3.7\% at short context and less
than 1.5\% after the KV scan dominates. The profiler synchronizes around
categories, so these percentages describe its instrumented step breakdown,
not the final overlapped TPOT table.

At fixed GPU count, this collective cost stays nearly constant as context
grows, so its fraction of TPOT falls. The oneCCL path invokes a cross-process
collective at each of the two reduction sites per layer; its two-GPU TPOT rises from 43\,\ms{} at 1K to
105\,\ms{} at 32K, while the \mk{} path rises from 23 to 36\,\ms{}.

\subsubsection{GPU count and capacity}

Flat TP=4 has the lowest TPOT across the measured TP-only range. The 2-D
measurements below evaluate longer contexts. The 128-KB/token KV cache exceeds one-GPU capacity
at approximately 384K context, after which sharding is required for capacity.

\subsubsection{2D scaling to the full node ($\mathrm{TP}\times\mathrm{CP}$)}

Plain TP is capped at $P=8$ and loses efficiency before that limit. Context
parallelism (CP) instead shards the KV cache by sequence position, computes
a local flash partial on each GPU, and merges the softmax statistics across
GPUs without a head-count limit. A standalone CP-attention prototype accelerates
the 128K KV scan by $6.6\times$ across 4 GPUs and $8.5\times$ across 8,
with maximum absolute error $10^{-4}$.

The full 2-D orchestrator shards weights by TP and replicates them across CP,
so CP leaves per-GPU GEMM work unchanged but replicates that work across CP
groups. It shards the KV cache by sequence, merges the CP
softmax partials within each TP group, and performs the two TP all-reduces
per layer. One process drives all 12 GPUs.

\begin{table}[h]
\centering
\small
\setlength{\tabcolsep}{4pt}
\caption{Measured 2D end-to-end decode TPOT (ms/token, geomean over 20
steps). Full end-to-end $\mathrm{TP}\times\mathrm{CP}$ decode, all layers,
one \pvc{} node. Every configuration passes the first-new-token logit-vector
check against the single-GPU reference with cosine similarity at least 0.9999.
Compared against the best flat-TP point
and oneCCL.}
\label{tab:appxe2e-mgpu-2d}
\begin{tabular}{@{}rcccc@{}}
\toprule
ctx & 6-GPU $\mathrm{TP}2\times\mathrm{CP}3$ & 12-GPU $\mathrm{TP}4\times\mathrm{CP}3$ & best flat TP ($P=4$) & xCCL $ws=4$ \\
\midrule
32K  & 29.4 & 59.8 & \textbf{25.3} & 73.7 \\
64K  & \textbf{31.0} & 59.0 & 54.5 & 128.1 \\
128K & \textbf{45.4} & 60.0 & 95.6 & 222.3 \\
256K & 93.9 & \textbf{73.3} & 271.9 & 413.0 \\
\bottomrule
\end{tabular}
\end{table}

The lowest-TPOT mapping is flat $P=4$ through 32K,
$\mathrm{TP}=2\times\mathrm{CP}=3$ on six GPUs at 64K--128K, and
$\mathrm{TP}=4\times\mathrm{CP}=3$ on 12 GPUs at 256K. The six-GPU
path measures 29.4, 31.0, and 45.4\,\ms{} at 32K, 64K, and 128K. At 64K
and 128K it is respectively $1.76\times$ and $2.11\times$ faster than flat
$P=4$. The 12-GPU path remains at 59.0--60.0\,\ms{} through 128K, then
reaches 73.3\,\ms{} at 256K, compared with 93.9\,\ms{} for six GPUs and
271.9\,\ms{} for flat $P=4$. Both 2-D layouts use $\mathrm{CP}=3$; the
additional TP sharding affects the largest KV scan.

With these additional GPUs, the measured ratio over the oneCCL-based decode
path is $4.90\times$ at
128K (6-GPU 45.4 vs.\ $ws=4$'s 222.3) and $5.63\times$ at 256K (12-GPU
73.3 vs.\ 413.0, where $ws=2$ runs out of memory). These are scaling
comparisons against four GPUs, not matched-resource speedups. Both selected layouts use
the same single-process collective and GPU-side barrier.

\subsubsection{Correctness methodology}

The correctness gate uses first-new-token logit-vector cosine similarity
against the single-GPU ($P=1$) run. Greedy token matching varies across
runs because fused SDPA reduction order changes near-tied argmax logits. In
the broader validation sweep, which also includes eight GPUs,
$P\in\{2,4,8\}$ configurations reproduce $P=1$ to cosine 0.99998 at dense
context and 0.99973 at 128K--256K. The maximum absolute logit difference is
0.125, corresponding to bf16 rounding.

%% file: main.bbl

\begin{thebibliography}{38}

%%% the Web address.

\ifx \showCODEN    \undefined \def \showCODEN     #1{\unskip}     \fi
\ifx \showISBNx    \undefined \def \showISBNx     #1{\unskip}     \fi
\ifx \showISBNxiii \undefined \def \showISBNxiii  #1{\unskip}     \fi
\ifx \showISSN     \undefined \def \showISSN      #1{\unskip}     \fi
\ifx \showLCCN     \undefined \def \showLCCN      #1{\unskip}     \fi
\ifx \shownote     \undefined \def \shownote      #1{#1}          \fi
\ifx \showarticletitle \undefined \def \showarticletitle #1{#1}   \fi
\ifx \showURL      \undefined \def \showURL       {\relax}        \fi
% invisible to TeX
\providecommand\bibfield[2]{#2}
\providecommand\bibinfo[2]{#2}
\providecommand\natexlab[1]{#1}
\providecommand\showeprint[2][]{arXiv:#2}

\bibitem[Ainslie et~al\mbox{.}(2023)]%
        {ainslie2023gqa}
\bibfield{author}{\bibinfo{person}{Joshua Ainslie}, \bibinfo{person}{James
  Lee-Thorp}, \bibinfo{person}{Michiel de Jong}, \bibinfo{person}{Yury
  Zemlyanskiy}, \bibinfo{person}{Federico Lebr\'{o}n}, {and}
  \bibinfo{person}{Sumit Sanghai}.} \bibinfo{year}{2023}\natexlab{}.
\newblock \showarticletitle{{GQA}: Training Generalized Multi-Query Transformer
  Models from Multi-Head Checkpoints}. In \bibinfo{booktitle}{\emph{Proceedings
  of the 2023 Conference on Empirical Methods in Natural Language Processing}}.
  \bibinfo{pages}{4895--4901}.
\newblock
\href{https://doi.org/10.18653/v1/2023.emnlp-main.298}{doi:\nolinkurl{10.18653/v1/2023.emnlp-main.298}}


\bibitem[Allcock et~al\mbox{.}(2025)]%
        {aurora2025}
\bibfield{author}{\bibinfo{person}{William~E. Allcock},
  \bibinfo{person}{Benjamin~S. Allen}, \bibinfo{person}{James Anchell},
  \bibinfo{person}{Victor Anisimov}, \bibinfo{person}{Thomas Applencourt},
  \bibinfo{person}{Abhishek Bagusetty}, \bibinfo{person}{Ramesh Balakrishnan},
  \bibinfo{person}{Riccardo Balin}, \bibinfo{person}{Solomon Bekele},
  \bibinfo{person}{Colleen Bertoni}, \bibinfo{person}{Cyrus Blackworth},
  \bibinfo{person}{Renzo Bustamante}, \bibinfo{person}{Kevin Canada},
  \bibinfo{person}{John Carrier}, \bibinfo{person}{Christopher Chan-nui},
  \bibinfo{person}{Lance~C. Cheney}, \bibinfo{person}{Taylor Childers},
  \bibinfo{person}{Paul Coffman}, \bibinfo{person}{Susan Coghlan},
  \bibinfo{person}{Tanima Dey}, \bibinfo{person}{Michael D'Mello},
  \bibinfo{person}{Ashok Emani}, \bibinfo{person}{Murali Emani},
  \bibinfo{person}{Kyle~G. Felker}, \bibinfo{person}{Sam Foreman},
  \bibinfo{person}{Olivier Franza}, \bibinfo{person}{Longfei Gao},
  \bibinfo{person}{Marta García}, \bibinfo{person}{María Garzarán},
  \bibinfo{person}{Balazs Gerofi}, \bibinfo{person}{Yasaman Ghadar},
  \bibinfo{person}{Subrata Goswami}, \bibinfo{person}{Neha Gupta},
  \bibinfo{person}{Kevin Harms}, \bibinfo{person}{Väinö Hatanpää},
  \bibinfo{person}{Brian Holland}, \bibinfo{person}{Carissa Holohan},
  \bibinfo{person}{Brian Homerding}, \bibinfo{person}{Khalid Hossain},
  \bibinfo{person}{Xue Hu}, \bibinfo{person}{Louise Huot},
  \bibinfo{person}{Huda Ibeid}, \bibinfo{person}{Joseph~A. Insley},
  \bibinfo{person}{Sai Jayanthi}, \bibinfo{person}{Hong Jiang},
  \bibinfo{person}{Wei Jiang}, \bibinfo{person}{Xiao-Yong Jin},
  \bibinfo{person}{Jeongnim Kim}, \bibinfo{person}{Christopher Knight},
  \bibinfo{person}{Panagiotis Kourdis}, \bibinfo{person}{Kalyan Kumaran},
  \bibinfo{person}{JaeHyuk Kwack}, \bibinfo{person}{Janghaeng Lee},
  \bibinfo{person}{Ti Leggett}, \bibinfo{person}{Ben Lenard},
  \bibinfo{person}{Chris Lewis}, \bibinfo{person}{Nevin Liber},
  \bibinfo{person}{Johann Lombardi}, \bibinfo{person}{Raymond~M. Loy},
  \bibinfo{person}{Ye Luo}, \bibinfo{person}{Bethany Lusch},
  \bibinfo{person}{Nilakantan Mahadevan}, \bibinfo{person}{Beth Markey},
  \bibinfo{person}{Victor~A. Mateevitsi}, \bibinfo{person}{Gordon McPheeters},
  \bibinfo{person}{Ryan Milner}, \bibinfo{person}{Jerome Mitchell},
  \bibinfo{person}{Vitali~A. Morozov}, \bibinfo{person}{Servesh Muralidharan},
  \bibinfo{person}{Tom Musta}, \bibinfo{person}{Mrigendra Nagar},
  \bibinfo{person}{Vikram Narayana}, \bibinfo{person}{Marieme Ngom},
  \bibinfo{person}{Anthony-Trung Nguyen}, \bibinfo{person}{Nathan Nichols},
  \bibinfo{person}{Aditya Nishtala}, \bibinfo{person}{James~C. Osborn},
  \bibinfo{person}{Michael~E. Papka}, \bibinfo{person}{Scott Parker},
  \bibinfo{person}{Saumil~S. Patel}, \bibinfo{person}{Julia Piotrowska},
  \bibinfo{person}{Adrian~C. Pope}, \bibinfo{person}{Sucheta Raghunanda},
  \bibinfo{person}{Esteban Rangel}, \bibinfo{person}{Paul~M. Rich},
  \bibinfo{person}{Katherine~M. Riley}, \bibinfo{person}{Silvio Rizzi},
  \bibinfo{person}{Kris Rowe}, \bibinfo{person}{Varuni Sastry},
  \bibinfo{person}{Adam Scovel}, \bibinfo{person}{Filippo Simini},
  \bibinfo{person}{Haritha~Siddabathuni Som}, \bibinfo{person}{Patrick
  Steinbrecher}, \bibinfo{person}{Rick Stevens}, \bibinfo{person}{Xinmin Tian},
  \bibinfo{person}{Peter Upton}, \bibinfo{person}{Thomas Uram},
  \bibinfo{person}{Archit~K. Vasan}, \bibinfo{person}{Álvaro
  Vázquez-Mayagoitia}, \bibinfo{person}{Kaushik Velusamy},
  \bibinfo{person}{Brice Videau}, \bibinfo{person}{Venkatram Vishwanath},
  \bibinfo{person}{Brian Whitney}, \bibinfo{person}{Timothy~J. Williams},
  \bibinfo{person}{Michael Woodacre}, \bibinfo{person}{Sam Zeltner},
  \bibinfo{person}{Chuanjun Zhang}, \bibinfo{person}{Gengbin Zheng}, {and}
  \bibinfo{person}{Huihuo Zheng}.} \bibinfo{year}{2025}\natexlab{}.
\newblock \bibinfo{title}{{Aurora}: Architecting {Argonne}'s First Exascale
  Supercomputer for Accelerated Scientific Discovery}.
\newblock
\showeprint[arxiv]{2509.08207}~[cs.DC]
\urldef\tempurl%
\url{https://arxiv.org/abs/2509.08207}
\showURL{%
\tempurl}


\bibitem[Ba et~al\mbox{.}(2016)]%
        {ba2016layernorm}
\bibfield{author}{\bibinfo{person}{Jimmy~Lei Ba}, \bibinfo{person}{Jamie~Ryan
  Kiros}, {and} \bibinfo{person}{Geoffrey~E. Hinton}.}
  \bibinfo{year}{2016}\natexlab{}.
\newblock \showarticletitle{Layer Normalization}.
\newblock \bibinfo{journal}{\emph{arXiv preprint arXiv:1607.06450}}
  (\bibinfo{year}{2016}).
\newblock
\urldef\tempurl%
\url{https://arxiv.org/abs/1607.06450}
\showURL{%
\tempurl}


\bibitem[Blythe(2021)]%
        {blythe2021pvc}
\bibfield{author}{\bibinfo{person}{David Blythe}.}
  \bibinfo{year}{2021}\natexlab{}.
\newblock \bibinfo{title}{{Intel}'s {Ponte Vecchio} {GPU} Architecture}.
\newblock \bibinfo{howpublished}{Hot Chips 33 Symposium on High-Performance
  Chips}.
\newblock
\urldef\tempurl%
\url{https://hc33.hotchips.org/assets/program/conference/day2/hc2021_pvc_final.pdf}
\showURL{%
\tempurl}


\bibitem[Chan et~al\mbox{.}(2026)]%
        {chan2026parallelkernelbench}
\bibfield{author}{\bibinfo{person}{Willy Chan}, \bibinfo{person}{Nathan Paek},
  \bibinfo{person}{Simon Guo}, \bibinfo{person}{Simran Arora}, {and}
  \bibinfo{person}{Daniel Fu}.} \bibinfo{year}{2026}\natexlab{}.
\newblock \bibinfo{title}{{ParallelKernelBench}: Can {LLM}s Write Fast
  Multi-{GPU} Kernels?}
\newblock
\shownote{Benchmark and evaluation harness for multi-GPU kernel generation}.
\newblock
\urldef\tempurl%
\url{https://github.com/togethercomputer/ParallelKernelBench}
\showURL{%
\tempurl}


\bibitem[Dao et~al\mbox{.}(2022)]%
        {dao2022flashattention}
\bibfield{author}{\bibinfo{person}{Tri Dao}, \bibinfo{person}{Daniel~Y. Fu},
  \bibinfo{person}{Stefano Ermon}, \bibinfo{person}{Atri Rudra}, {and}
  \bibinfo{person}{Christopher R\'{e}}.} \bibinfo{year}{2022}\natexlab{}.
\newblock \showarticletitle{{FlashAttention}: Fast and Memory-Efficient Exact
  Attention with {IO}-Awareness}. In \bibinfo{booktitle}{\emph{Advances in
  Neural Information Processing Systems}}, Vol.~\bibinfo{volume}{35}.
  \bibinfo{pages}{16344--16359}.
\newblock
\href{https://doi.org/10.52202/068431-1189}{doi:\nolinkurl{10.52202/068431-1189}}


\bibitem[Grattafiori et~al\mbox{.}(2024)]%
        {grattafiori2024llama3}
\bibfield{author}{\bibinfo{person}{Aaron Grattafiori} {et~al\mbox{.}}}
  \bibinfo{year}{2024}\natexlab{}.
\newblock \showarticletitle{The {Llama 3} Herd of Models}.
\newblock \bibinfo{journal}{\emph{arXiv preprint arXiv:2407.21783}}
  (\bibinfo{year}{2024}).
\newblock
\urldef\tempurl%
\url{https://arxiv.org/abs/2407.21783}
\showURL{%
\tempurl}


\bibitem[Hu et~al\mbox{.}(2025)]%
        {hu2024hipkittens}
\bibfield{author}{\bibinfo{person}{William Hu}, \bibinfo{person}{Drew
  Wadsworth}, \bibinfo{person}{Sean Siddens}, \bibinfo{person}{Stanley Winata},
  \bibinfo{person}{Daniel~Y. Fu}, \bibinfo{person}{Ryann Swann},
  \bibinfo{person}{Muhammad Osama}, \bibinfo{person}{Christopher R\'{e}}, {and}
  \bibinfo{person}{Simran Arora}.} \bibinfo{year}{2025}\natexlab{}.
\newblock \showarticletitle{{HipKittens}: Fast and Furious {AMD} Kernels}.
\newblock \bibinfo{journal}{\emph{arXiv preprint arXiv:2511.08083}}
  (\bibinfo{year}{2025}).
\newblock
\urldef\tempurl%
\url{https://arxiv.org/abs/2511.08083}
\showURL{%
\tempurl}


\bibitem[{Intel Corporation}(2023a)]%
        {intel_esimd2023}
\bibfield{author}{\bibinfo{person}{{Intel Corporation}}.}
  \bibinfo{year}{2023}\natexlab{a}.
\newblock \bibinfo{title}{Explicit {SIMD} Programming Extension for {DPC++}}.
\newblock
\urldef\tempurl%
\url{https://github.com/intel/llvm/blob/sycl/sycl/doc/extensions/supported/sycl_ext_intel_esimd/sycl_ext_intel_esimd.md}
\showURL{%
\tempurl}


\bibitem[{Intel Corporation}(2023b)]%
        {intel_max_2023}
\bibfield{author}{\bibinfo{person}{{Intel Corporation}}.}
  \bibinfo{year}{2023}\natexlab{b}.
\newblock \bibinfo{title}{{Intel Data Center GPU Max 1550} Specifications}.
\newblock
\shownote{Launched Q1 2023; codename Ponte Vecchio}.
\newblock
\urldef\tempurl%
\url{https://www.intel.com/content/www/us/en/products/sku/232873/intel-data-center-gpu-max-1550/specifications.html}
\showURL{%
\tempurl}


\bibitem[{Intel Corporation}(2023c)]%
        {intel_xetla2024}
\bibfield{author}{\bibinfo{person}{{Intel Corporation}}.}
  \bibinfo{year}{2023}\natexlab{c}.
\newblock \bibinfo{title}{{Intel Xe Templates for Linear Algebra (XeTLA)
  v0.3.7}}.
\newblock
\urldef\tempurl%
\url{https://github.com/intel/xetla/releases/tag/v0.3.7}
\showURL{%
\tempurl}


\bibitem[{Intel Corporation}(2025a)]%
        {ipex}
\bibfield{author}{\bibinfo{person}{{Intel Corporation}}.}
  \bibinfo{year}{2025}\natexlab{a}.
\newblock \bibinfo{title}{{Intel Extension for PyTorch v2.10.10+xpu}}.
\newblock
\shownote{aurora\_frameworks 2025.3.1 (PyTorch 2.10)}.
\newblock
\urldef\tempurl%
\url{https://github.com/intel/intel-extension-for-pytorch/releases/tag/v2.10.10%2Bxpu}
\showURL{%
\tempurl}


\bibitem[{Intel Corporation}(2025b)]%
        {intel_onemkl2025}
\bibfield{author}{\bibinfo{person}{{Intel Corporation}}.}
  \bibinfo{year}{2025}\natexlab{b}.
\newblock \bibinfo{title}{{Intel oneAPI Math Kernel Library} -- Data Parallel
  {C++} Developer Reference: \texttt{gemm}}.
\newblock
\urldef\tempurl%
\url{https://www.intel.com/content/www/us/en/docs/onemkl/developer-reference-dpcpp/2025-0/gemm.html}
\showURL{%
\tempurl}


\bibitem[{Intel Corporation}(2025c)]%
        {sycltla2024}
\bibfield{author}{\bibinfo{person}{{Intel Corporation}}.}
  \bibinfo{year}{2025}\natexlab{c}.
\newblock \bibinfo{title}{{SYCL* Templates for Linear Algebra (SYCL*TLA)
  v0.6}}.
\newblock
\urldef\tempurl%
\url{https://github.com/intel/sycl-tla/releases/tag/v0.6}
\showURL{%
\tempurl}


\bibitem[{Intel Corporation}(2026)]%
        {intel_triton_xpu}
\bibfield{author}{\bibinfo{person}{{Intel Corporation}}.}
  \bibinfo{year}{2026}\natexlab{}.
\newblock \bibinfo{title}{{Intel XPU Backend for Triton v3.6.0}}.
\newblock
\urldef\tempurl%
\url{https://github.com/intel/intel-xpu-backend-for-triton/releases/tag/v3.6.0}
\showURL{%
\tempurl}


\bibitem[Jiang et~al\mbox{.}(2023)]%
        {jiang2023mistral}
\bibfield{author}{\bibinfo{person}{Albert~Q. Jiang}, \bibinfo{person}{Alexandre
  Sablayrolles}, \bibinfo{person}{Arthur Mensch}, \bibinfo{person}{Chris
  Bamford}, \bibinfo{person}{Devendra~Singh Chaplot}, \bibinfo{person}{Diego
  de~las Casas}, \bibinfo{person}{Florian Bressand}, \bibinfo{person}{Gianna
  Lengyel}, \bibinfo{person}{Guillaume Lample}, \bibinfo{person}{Lucile
  Saulnier}, \bibinfo{person}{L\'{e}lio~Renard Lavaud},
  \bibinfo{person}{Marie-Anne Lachaux}, \bibinfo{person}{Pierre Stock},
  \bibinfo{person}{Teven Le~Scao}, \bibinfo{person}{Thibaut Lavril},
  \bibinfo{person}{Thomas Wang}, \bibinfo{person}{Timoth\'{e}e Lacroix}, {and}
  \bibinfo{person}{William El~Sayed}.} \bibinfo{year}{2023}\natexlab{}.
\newblock \showarticletitle{{Mistral 7B}}.
\newblock \bibinfo{journal}{\emph{arXiv preprint arXiv:2310.06825}}
  (\bibinfo{year}{2023}).
\newblock
\urldef\tempurl%
\url{https://arxiv.org/abs/2310.06825}
\showURL{%
\tempurl}


\bibitem[Jiang(2022)]%
        {jiang2022pvc}
\bibfield{author}{\bibinfo{person}{Hong Jiang}.}
  \bibinfo{year}{2022}\natexlab{}.
\newblock \bibinfo{title}{{Intel}'s {Ponte Vecchio} {GPU}: Architecture, System
  and Software}.
\newblock \bibinfo{howpublished}{Hot Chips 34 Symposium on High-Performance
  Chips}.
\newblock
\urldef\tempurl%
\url{https://www.hc34.hotchips.org/assets/program/conference/day1/GPU%20HPC/Intel_s%20Ponte%20Vecchio%20GPU%20-%20Architecture%20Systems%20and%20Software%20FINAL.pdf}
\showURL{%
\tempurl}


\bibitem[Jiang et~al\mbox{.}(2026)]%
        {jiang2026iwocl}
\bibfield{author}{\bibinfo{person}{Yehong Jiang}, \bibinfo{person}{Sheng Chen},
  \bibinfo{person}{Fangwen Fu}, \bibinfo{person}{Vincent Lu},
  \bibinfo{person}{Yen-Kuang Chen}, \bibinfo{person}{Hong Wang}, {and}
  \bibinfo{person}{Xinmin Tian}.} \bibinfo{year}{2026}\natexlab{}.
\newblock \showarticletitle{Porting {ThunderKittens} from {CUDA} to {SYCL} for
  {Intel GPU}: Process, Challenges, and Lessons Learned}. In
  \bibinfo{booktitle}{\emph{Proceedings of the International Workshop on OpenCL
  and SYCL}}. \bibinfo{publisher}{ACM}, \bibinfo{address}{Heilbronn, Germany},
  \bibinfo{pages}{1--5}.
\newblock
\href{https://doi.org/10.1145/3811257.3811269}{doi:\nolinkurl{10.1145/3811257.3811269}}


\bibitem[{Khronos Group}(2026)]%
        {sycl2020}
\bibfield{author}{\bibinfo{person}{{Khronos Group}}.}
  \bibinfo{year}{2026}\natexlab{}.
\newblock \bibinfo{booktitle}{\emph{{SYCL} 2020 Specification}}.
\newblock \bibinfo{type}{{T}echnical {R}eport}. \bibinfo{institution}{Khronos
  Group, Inc}.
\newblock
\shownote{Revision 12, August 6, 2026}.
\newblock
\urldef\tempurl%
\url{https://registry.khronos.org/SYCL/specs/sycl-2020/html/sycl-2020.html}
\showURL{%
\tempurl}


\bibitem[Liu et~al\mbox{.}(2024)]%
        {liu2023ringattention}
\bibfield{author}{\bibinfo{person}{Hao Liu}, \bibinfo{person}{Matei Zaharia},
  {and} \bibinfo{person}{Pieter Abbeel}.} \bibinfo{year}{2024}\natexlab{}.
\newblock \showarticletitle{{RingAttention} with Blockwise Transformers for
  Near-Infinite Context}. In \bibinfo{booktitle}{\emph{The Twelfth
  International Conference on Learning Representations (ICLR)}}.
\newblock
\urldef\tempurl%
\url{https://openreview.net/forum?id=WsRHpHH4s0}
\showURL{%
\tempurl}


\bibitem[Mohan and Jiang(2026)]%
        {mohan2026cri}
\bibfield{author}{\bibinfo{person}{Sumit Mohan} {and} \bibinfo{person}{Hong
  Jiang}.} \bibinfo{year}{2026}\natexlab{}.
\newblock \bibinfo{title}{{Crescent Island}: {GPU} Designed for Agentic {AI}
  Inference}.
\newblock \bibinfo{howpublished}{Hot Chips 38 Symposium on High-Performance
  Chips}.
\newblock
\urldef\tempurl%
\url{https://hc2026.hotchips.org/program/}
\showURL{%
\tempurl}


\bibitem[{NVIDIA}(2024)]%
        {nvidia_cute2024}
\bibfield{author}{\bibinfo{person}{{NVIDIA}}.} \bibinfo{year}{2024}\natexlab{}.
\newblock \bibinfo{title}{{CuTe}: {C++ CUDA} Template Abstractions for Tensors
  and Layouts}.
\newblock
\urldef\tempurl%
\url{https://github.com/NVIDIA/cutlass/blob/main/media/docs/cpp/cute/00_quickstart.md}
\showURL{%
\tempurl}


\bibitem[{NVIDIA}(2025)]%
        {nvidia_cutlass2025}
\bibfield{author}{\bibinfo{person}{{NVIDIA}}.} \bibinfo{year}{2025}\natexlab{}.
\newblock \bibinfo{title}{{CUTLASS}: {CUDA} Templates for Linear Algebra
  Subroutines and Solvers}.
\newblock
\urldef\tempurl%
\url{https://github.com/NVIDIA/cutlass}
\showURL{%
\tempurl}


\bibitem[{oneDNN Project}(2024)]%
        {onednn}
\bibfield{author}{\bibinfo{person}{{oneDNN Project}}.}
  \bibinfo{year}{2024}\natexlab{}.
\newblock \bibinfo{title}{{oneAPI Deep Neural Network Library (oneDNN)}}.
\newblock
\shownote{Open-source project hosted by the UXL Foundation}.
\newblock
\urldef\tempurl%
\url{https://github.com/uxlfoundation/oneDNN}
\showURL{%
\tempurl}


\bibitem[{oneDNN Project}(2025a)]%
        {onednn_gpu_jit}
\bibfield{author}{\bibinfo{person}{{oneDNN Project}}.}
  \bibinfo{year}{2025}\natexlab{a}.
\newblock \bibinfo{title}{{oneDNN v3.10 GPU JIT-compiled Kernels}}.
\newblock
\urldef\tempurl%
\url{https://github.com/uxlfoundation/oneDNN/blob/v3.10/src/gpu/intel/jit/README.md}
\showURL{%
\tempurl}


\bibitem[{oneDNN Project}(2025b)]%
        {onednn_gemm_kernel_catalog}
\bibfield{author}{\bibinfo{person}{{oneDNN Project}}.}
  \bibinfo{year}{2025}\natexlab{b}.
\newblock \bibinfo{title}{{oneDNN v3.10 Intel GPU GEMM JIT Strategy Catalog}}.
\newblock
\urldef\tempurl%
\url{https://github.com/uxlfoundation/oneDNN/blob/v3.10/src/gpu/intel/gemm/jit/selector/db/kernel.db}
\showURL{%
\tempurl}


\bibitem[Ouyang et~al\mbox{.}(2025)]%
        {ouyang2025kernelbench}
\bibfield{author}{\bibinfo{person}{Anne Ouyang}, \bibinfo{person}{Simon Guo},
  \bibinfo{person}{Simran Arora}, \bibinfo{person}{Alex~L. Zhang},
  \bibinfo{person}{William Hu}, \bibinfo{person}{Christopher R\'{e}}, {and}
  \bibinfo{person}{Azalia Mirhoseini}.} \bibinfo{year}{2025}\natexlab{}.
\newblock \showarticletitle{{KernelBench}: Can {LLM}s Write Efficient {GPU}
  Kernels?}. In \bibinfo{booktitle}{\emph{Proceedings of the 42nd International
  Conference on Machine Learning}} \emph{(\bibinfo{series}{Proceedings of
  Machine Learning Research}, Vol.~\bibinfo{volume}{267})}.
  \bibinfo{publisher}{PMLR}, \bibinfo{pages}{47356--47415}.
\newblock
\urldef\tempurl%
\url{https://proceedings.mlr.press/v267/ouyang25a.html}
\showURL{%
\tempurl}


\bibitem[{PyTorch Contributors}(2026)]%
        {pytorch_xpu_mm_210}
\bibfield{author}{\bibinfo{person}{{PyTorch Contributors}}.}
  \bibinfo{year}{2026}\natexlab{}.
\newblock \bibinfo{title}{{PyTorch v2.10.0 XPU ATen Matrix Multiplication
  Implementation}}.
\newblock
\shownote{Release commit 449b176; \texttt{xpu::mm\_out} calls
  \texttt{onednn::matmul}}.
\newblock
\urldef\tempurl%
\url{https://github.com/pytorch/pytorch/blob/449b1768410104d3ed79d3bcfe4ba1d65c7f22c0/aten/src/ATen/native/mkldnn/xpu/Blas.cpp#L182-L206}
\showURL{%
\tempurl}


\bibitem[Shoeybi et~al\mbox{.}(2019)]%
        {shoeybi2019megatron}
\bibfield{author}{\bibinfo{person}{Mohammad Shoeybi}, \bibinfo{person}{Mostofa
  Patwary}, \bibinfo{person}{Raul Puri}, \bibinfo{person}{Patrick LeGresley},
  \bibinfo{person}{Jared Casper}, {and} \bibinfo{person}{Bryan Catanzaro}.}
  \bibinfo{year}{2019}\natexlab{}.
\newblock \showarticletitle{{Megatron-LM}: Training Multi-Billion Parameter
  Language Models Using Model Parallelism}.
\newblock \bibinfo{journal}{\emph{arXiv preprint arXiv:1909.08053}}
  (\bibinfo{year}{2019}).
\newblock
\urldef\tempurl%
\url{https://arxiv.org/abs/1909.08053}
\showURL{%
\tempurl}


\bibitem[Spector et~al\mbox{.}(2025)]%
        {spector2025thunderkittens}
\bibfield{author}{\bibinfo{person}{Benjamin~F. Spector},
  \bibinfo{person}{Simran Arora}, \bibinfo{person}{Aaryan Singhal},
  \bibinfo{person}{Arjun Parthasarathy}, \bibinfo{person}{Daniel~Y. Fu}, {and}
  \bibinfo{person}{Christopher R\'{e}}.} \bibinfo{year}{2025}\natexlab{}.
\newblock \showarticletitle{{ThunderKittens}: Simple, Fast, and
  \textit{Adorable} Kernels}. In \bibinfo{booktitle}{\emph{Proceedings of the
  13th International Conference on Learning Representations}}.
\newblock
\urldef\tempurl%
\url{https://iclr.cc/virtual/2025/poster/31243}
\showURL{%
\tempurl}


\bibitem[Su et~al\mbox{.}(2024)]%
        {su2021roformer}
\bibfield{author}{\bibinfo{person}{Jianlin Su}, \bibinfo{person}{Murtadha
  Ahmed}, \bibinfo{person}{Yu Lu}, \bibinfo{person}{Shengfeng Pan},
  \bibinfo{person}{Wen Bo}, {and} \bibinfo{person}{Yunfeng Liu}.}
  \bibinfo{year}{2024}\natexlab{}.
\newblock \showarticletitle{{RoFormer}: Enhanced Transformer with Rotary
  Position Embedding}.
\newblock \bibinfo{journal}{\emph{Neurocomputing}}  \bibinfo{volume}{568}
  (\bibinfo{year}{2024}), \bibinfo{pages}{127063}.
\newblock
\href{https://doi.org/10.1016/j.neucom.2023.127063}{doi:\nolinkurl{10.1016/j.neucom.2023.127063}}


\bibitem[Sul et~al\mbox{.}(2025)]%
        {sul2025parallelkittens}
\bibfield{author}{\bibinfo{person}{Stuart~H. Sul}, \bibinfo{person}{Simran
  Arora}, \bibinfo{person}{Benjamin~F. Spector}, {and}
  \bibinfo{person}{Christopher R\'{e}}.} \bibinfo{year}{2025}\natexlab{}.
\newblock \showarticletitle{{ParallelKittens}: Systematic and Practical
  Simplification of Multi-GPU {AI} Kernels}.
\newblock \bibinfo{journal}{\emph{arXiv preprint arXiv:2511.13940}}
  (\bibinfo{year}{2025}).
\newblock
\urldef\tempurl%
\url{https://arxiv.org/abs/2511.13940}
\showURL{%
\tempurl}


\bibitem[Tillet et~al\mbox{.}(2019)]%
        {tillet2019triton}
\bibfield{author}{\bibinfo{person}{Philippe Tillet}, \bibinfo{person}{H.~T.
  Kung}, {and} \bibinfo{person}{David Cox}.} \bibinfo{year}{2019}\natexlab{}.
\newblock \showarticletitle{{Triton}: An Intermediate Language and Compiler for
  Tiled Neural Network Computations}. In \bibinfo{booktitle}{\emph{Proceedings
  of the 3rd ACM SIGPLAN International Workshop on Machine Learning and
  Programming Languages}}. \bibinfo{publisher}{ACM}, \bibinfo{pages}{10--19}.
\newblock
\href{https://doi.org/10.1145/3315508.3329973}{doi:\nolinkurl{10.1145/3315508.3329973}}


\bibitem[{UXL Foundation}(2025)]%
        {intel_oneccl2025}
\bibfield{author}{\bibinfo{person}{{UXL Foundation}}.}
  \bibinfo{year}{2025}\natexlab{}.
\newblock \bibinfo{title}{{Intel oneAPI Collective Communications Library
  (oneCCL) 2021.17}}.
\newblock
\urldef\tempurl%
\url{https://github.com/uxlfoundation/oneCCL/releases/tag/2021.17}
\showURL{%
\tempurl}


\bibitem[Vaswani et~al\mbox{.}(2017)]%
        {vaswani2017attention}
\bibfield{author}{\bibinfo{person}{Ashish Vaswani}, \bibinfo{person}{Noam
  Shazeer}, \bibinfo{person}{Niki Parmar}, \bibinfo{person}{Jakob Uszkoreit},
  \bibinfo{person}{Llion Jones}, \bibinfo{person}{Aidan~N. Gomez},
  \bibinfo{person}{Lukasz Kaiser}, {and} \bibinfo{person}{Illia Polosukhin}.}
  \bibinfo{year}{2017}\natexlab{}.
\newblock \showarticletitle{Attention Is All You Need}. In
  \bibinfo{booktitle}{\emph{Advances in Neural Information Processing
  Systems}}, Vol.~\bibinfo{volume}{30}. \bibinfo{pages}{5998--6008}.
\newblock
\urldef\tempurl%
\url{https://proceedings.neurips.cc/paper/2017/hash/3f5ee243547dee91fbd053c1c4a845aa-Abstract.html}
\showURL{%
\tempurl}


\bibitem[Wang et~al\mbox{.}(2025)]%
        {wang2025tilelang}
\bibfield{author}{\bibinfo{person}{Lei Wang}, \bibinfo{person}{Yu Cheng},
  \bibinfo{person}{Yining Shi}, \bibinfo{person}{Zhengju Tang},
  \bibinfo{person}{Zhiwen Mo}, \bibinfo{person}{Wenhao Xie},
  \bibinfo{person}{Lingxiao Ma}, \bibinfo{person}{Yuqing Xia},
  \bibinfo{person}{Jilong Xue}, \bibinfo{person}{Fan Yang}, {and}
  \bibinfo{person}{Zhi Yang}.} \bibinfo{year}{2025}\natexlab{}.
\newblock \bibinfo{title}{{TileLang}: A Composable Tiled Programming Model for
  {AI} Systems}.
\newblock
\showeprint[arxiv]{2504.17577}
\urldef\tempurl%
\url{https://arxiv.org/abs/2504.17577}
\showURL{%
\tempurl}


\bibitem[Yang et~al\mbox{.}(2024)]%
        {yang2024sweagent}
\bibfield{author}{\bibinfo{person}{John Yang}, \bibinfo{person}{Carlos~E.
  Jimenez}, \bibinfo{person}{Alexander Wettig}, \bibinfo{person}{Kilian
  Lieret}, \bibinfo{person}{Shunyu Yao}, \bibinfo{person}{Karthik~R.
  Narasimhan}, {and} \bibinfo{person}{Ofir Press}.}
  \bibinfo{year}{2024}\natexlab{}.
\newblock \showarticletitle{{SWE-agent}: Agent-Computer Interfaces Enable
  Automated Software Engineering}. In \bibinfo{booktitle}{\emph{Advances in
  Neural Information Processing Systems}}, Vol.~\bibinfo{volume}{37}.
  \bibinfo{publisher}{Neural Information Processing Systems Foundation, Inc.},
  \bibinfo{address}{San Diego, California, USA}, \bibinfo{pages}{50528--50652}.
\newblock
\href{https://doi.org/10.52202/079017-1601}{doi:\nolinkurl{10.52202/079017-1601}}


\bibitem[Zhang and Sennrich(2019)]%
        {zhang2019rmsnorm}
\bibfield{author}{\bibinfo{person}{Biao Zhang} {and} \bibinfo{person}{Rico
  Sennrich}.} \bibinfo{year}{2019}\natexlab{}.
\newblock \showarticletitle{Root Mean Square Layer Normalization}. In
  \bibinfo{booktitle}{\emph{Advances in Neural Information Processing
  Systems}}, Vol.~\bibinfo{volume}{32}. \bibinfo{pages}{12376--12385}.
\newblock
\urldef\tempurl%
\url{https://proceedings.neurips.cc/paper/2019/hash/1e8a19426224ca89e83cef47f1e7f53b-Abstract.html}
\showURL{%
\tempurl}


\end{thebibliography}
